\documentclass[fleqn,usenatbib]{mnras}

\usepackage[T1]{fontenc}
\usepackage{ae,aecompl}

\usepackage{graphicx}	% Including figure files
\usepackage{amsmath}	% Advanced maths commands
\usepackage{amssymb}	% Extra maths symbols

\usepackage{newtxtext,newtxmath}
\usepackage[usenames,dvipsnames]{xcolor}
\usepackage{physics}
\usepackage{tikz}
\usetikzlibrary{arrows,babel}
\usepackage[labelfont={bf},labelsep=endash]{caption}
\usetikzlibrary{positioning,arrows}
\usetikzlibrary{decorations.pathmorphing}
\usetikzlibrary{decorations.markings}
\tikzset{
momentum/.style={postaction={decorate},decoration={markings,mark=at position 1 with {\arrow{>}}}},
particle/.style={dashed%, postaction={decorate},
    },
photon/.style={decorate, 
    decoration={snake}},
    math/.style={draw, execute at begin node={$\displaystyle}, execute at end node={$}}
 }
\newcommand{\boxandcomment}[4][]{%
    \tikz[baseline=(#2.base), remember picture]{%
        \node[math, label=below:{#3}, #1] (#2) {#4};}}

\title[Low mass subhaloes for gamma-ray DM searches]{The viability of low-mass subhaloes as targets for gamma-ray dark matter searches}

\author[Aguirre-Santaella \& S\'anchez-Conde]{
Alejandra Aguirre-Santaella$^{1, 2}$\thanks{E-mail: alejandra.aguirre@uam.es} 
\& 
Miguel A. S\'anchez-Conde$^{1, 2}$\thanks{E-mail: miguel.sanchezconde@uam.es}
\\
$^{1}$ Instituto de F\'isica Te\'orica UAM-CSIC, Universidad Aut\'onoma de Madrid, C/ Nicol\'as Cabrera, 13-15, 28049 Madrid, Spain\\
$^{2}$ Departamento de F\'isica Te\'orica, M-15, Universidad Aut\'onoma de Madrid, E-28049 Madrid, Spain
}

\date{Accepted 2024 April 2. Received 2024 March 27; in original form 2023 September 19}

\pubyear{2024}

\begin{document}
\label{firstpage}
\pagerange{\pageref{firstpage}--\pageref{lastpage}}
\maketitle

% Abstract of the paper
\begin{abstract}
In this work, we investigate the discovery potential of low-mass Galactic dark matter (DM) subhaloes for indirect searches of DM. 
We use data from the Via Lactea II (VL-II) N-body cosmological simulation, which resolves subhaloes down to $\mathcal{O}(10^4)$ solar masses and it is thus ideal for this purpose. First, we characterize the abundance, distribution and structural properties of the VL-II subhalo population in terms of both subhalo masses and maximum circular velocities. 
Then, we repopulate the original simulation with millions of subhaloes of masses down to about five orders of magnitude below the minimum VL-II subhalo mass (more than one order of magnitude in velocities).
We compute subhalo DM annihilation astrophysical ``J-factors'' and angular sizes for the entire subhalo population, by placing the Earth at a random position but at the right Galactocentric distance in the simulation. 
Thousands of these realizations are generated in order to obtain statistically meaningful results. 
We find that some nearby low-mass Galactic subhaloes, not massive enough to retain stars or gas, may indeed yield DM annihilation fluxes comparable to those expected from other, more massive and acknowledgeable DM targets like dwarf satellite galaxies. 
Typical angular sizes are of the order of the degree, thus subhaloes potentially appearing as extended sources in gamma-ray telescopes, depending on instrument angular resolution and sensitivity.
Our work shows that low-mass Galactic subhaloes with no visible counterparts are expected to play a relevant role in current and future indirect DM search searches and should indeed be considered as excellent DM targets.

\end{abstract}

% Select between one and six entries from the list of approved keywords.
% Don't make up new ones.

%%%%%%%%%%%%%%%%%%%%%%%%%%%%%%%%%%%%%%%%%%%%%%%%%%

%%%%%%%%%%%%%%%%% BODY OF PAPER %%%%%%%%%%%%%%%%%%

%\tableofcontents

\section{Introduction}

 Plenty of cosmological and astrophysical observations at different scales suggest that, if our theory of gravity is correct, the visible mass in the Universe is not enough and that we need to add a new matter component, the so-called dark matter (DM), to explain these observations~\citep{Bertone+05, Garrett:2010hd, 2012AnP...524..507F, 2018RvMP...90d5002B, 2020A&A...641A...6P}. 
 This DM, despite its nature being still unknown, should constitute more than 80\% of the matter content in the Universe.

Among the plethora of proposed DM particle candidates, the Weakly Interacting Massive Particle \citep[WIMP,][]{Bertone10} has been for decades now the preferred one and the most intensely searched for.
Physicists have three complementary techniques to search for 
this kind of DM: 
direct production at colliders \citep{2018ARNPS..68..429B}, 
direct detection aimed at finding signs of interactions between DM and baryonic matter at the laboratories~\citep{2010arXiv1002.1912C} and indirect detection~\citep{2005MPLA...20.1021B}. The latter, on which our work focuses, aims to observe the outcome, i.e. photons, neutrinos and antimatter, generated by DM annihilation or decay into Standard Model particles. 
These products are expected to contain relevant information which might give a clue about DM properties \citep{2011ARA&A..49..155P} and, among them, photons are specially important because they do not suffer magnetic deviations in its path to the observer, therefore their origin can be tracked. Same happens to neutrinos, but these are much harder to detect. The energy of these photons depends mainly on the DM particle mass. The gamma-ray regime is relevant, since 
WIMPs
are expected to
have GeV-TeV masses \citep{Bertone+05, Bertone10}.
Besides, since DM was proposed in the first place to explain astrophysical findings, indirect searches also have the potential to find the DM distribution in the Universe, which cannot be done using the other methods.
These very energetic photons could be detected through gamma-ray experiments, both spatial and terrestrial, such as \textit{Fermi}-LAT~\citep{fermiglast}, VERITAS~\citep{Weekes:2001pd}, H.E.S.S. \citep{hinton2004} and MAGIC \citep{Lorenz:2004ah}.

Structure formation is pictured as hierarchical by the most supported cosmological framework, $\Lambda$CDM, with low-mass virialized structures or {\it haloes} being the first ones to form. These would eventually merge, originating larger haloes~\citep{2006Natur.440.1137S, 2012AnP...524..507F, 2019Galax...7...81Z}.
Hence, a large number of low-mass {\it subhaloes} are expected inside larger haloes like our own galaxy, the Milky Way (MW). It is believed that the most massive of these subhaloes would host dwarf satellite galaxies, while so-called dark satellites, i.e. less massive subhaloes with no stars or gas at all, would exist in a much larger number and would not hold any visible counterparts.

%%%%%%%%%%%%%%%%%%%%%%%%%%%%%%%%%%%%%%%%%%

Using N-body cosmological simulations with a large number of particles per virialized object and both a high time and force resolution makes it possible to study the formation of cold DM haloes and their substructure in the non-linear regime in great detail \citep{jurg1, 2020NatRP...2...42V, 2022LRCA....8....1A}.
DM-only simulations are done assuming that all the matter is dark, that is, baryons are not included. Therefore, they are collisionless N-body simulations, and even though they are not so accurate near the centre of haloes, where baryons are expected to play a significant role, they  
are one of the best tools we have to understand structure formation and halo structural properties at present. More recently, hydrodynamical simulations including baryons have also been successfully developed~\citep{2014MNRAS.444.1518V, 2016MNRAS.457..844F, 2016MNRAS.457.1931S}. However, basic properties of subhaloes such as their abundance, distribution and structure remain unclear for the less massive subhaloes due to the limited resolution in the simulations \citep{2014MNRAS.442.3256A}. 
Indeed, there exists a hot debate within the community about the survival probability of low-mass subhaloes. Some authors \citep{2017MNRAS.469.1997D, 2017MNRAS.471.1709G, Kelley2019, 2021MNRAS.501.3558G, 2021MNRAS.507.4953G} state that tidal forces and the impact of baryons inside host haloes lead to subhalo disruption, while others \citep{2004MNRAS.348..333D, 2008MNRAS.386.2135G, 2010MNRAS.404..502G, 2018MNRAS.474.3043V, Ogiya2019, 2019MNRAS.490.2091G, 2020MNRAS.491.4591E, 2022MNRAS.509.2624G, 2021arXiv211101148A, 2023MNRAS.518...93A, 2022arXiv220700604S} affirm that the inner cusp of a subhalo should always survive, even in the less massive ones. Following our own findings on this matter, in this work we will assume that the lack of subhaloes we encounter in current simulations is due to numerical effects. The survival of Galactic subhaloes is particularly important for our purposes. In fact, it is well known that DM substructure plays an important role in DM searches, mainly for two reasons. On one hand, both dwarf galaxies and dark satellites are excellent targets by themselves, since some of them are expected to give large DM annihilation fluxes at Earth \citep{2015PhRvL.115w1301A, 2016JCAP...05..028S, 2019JCAP...07..020C}. 
On the other, the clumpy distribution of subhaloes inside larger haloes will boost the DM annihilation flux of the host haloes significantly, since this flux is proportional to the DM density squared~\citep{mascprada14, moline, Ando:2019xlm}.

 The main goal of this work is to address and to quantify the relevance of low-mass subhaloes for gamma-ray DM searches, by computing and comparing their DM-induced signals to the ones expected from high-mass Galactic subhaloes, i.e. those hosting dwarf satellite galaxies, which are perceived as the golden targets by the community.
 We want to do so because less massive subhaloes are known to be more concentrated than more massive ones. Thus, since the DM annihilation flux is proportional to the third power of the concentration, some of these low-mass subhaloes may still yield significantly large annihilation fluxes at Earth. 
 We will first characterize in detail the subhalo population in a high-resolution $N$-body cosmological simulation, namely Via Lactea II \citep[VL-II,][]{Diemand2008}. We will then repopulate the simulation with subhaloes below its formal mass resolution limit, by extrapolating down to low masses the relevant subhalo properties (abundance, spatial distribution, inner structure) as they were found in the original simulation. 
 This procedure will be repeated so as to obtain many repopulations of VL-II, which would allow us to extract more meaningful results from the statistical point of view. 
 As it will be shown, despite having already nearly 15 years, the VL-II simulation still represents the state-of-the-art of a MW-size halo simulation and provides a high-resolution, unprecedented view of its subhalo population, critical for the purposes of this work. Working with VL-II simulation data is not exempt of potential issues though, that will also be discussed in detail. 

We note that an older, preliminary version of our repopulation machinery has already been employed in several works. In a first work, it allowed us to predict DM annihilation fluxes for the repopulated, small subhaloes and, later on, to set DM constraints by comparing simulation predictions to the number of \textit{Fermi}-LAT unidentified gamma-ray sources compatible with a DM signal~\citep{2019JCAP...07..020C}. 
In a follow-up work the latter sample was reduced and, thus, the DM constraints improved, by performing a dedicated spectral and spatial \textit{Fermi}-LAT data analysis, in which the expected spatial extension of the repopulated VL-II subhaloes was used as an additional filter~\citep{2019JCAP...11..045C}. 
A more comprehensive study of the sensitivity of the \textit{Fermi}-LAT to extended yet bright subhaloes was done in~\citet{2022PhRvD.105h3006C}, using again results from our repopulation exercise.  
Repopulation results were also adopted to investigate the potential detection of Galactic dark satellites by the next generation of Imaging Atmospheric Cherenkov Telescopes, i.e., the Cherenkov Telescope Array~\citep{2021PDU....3200845C}. In all these works, though, our repopulation machinery was only briefly introduced, as it never represented the main goal of the papers. In this work, we provide all details of the repopulation algorithm, apply it to reach even lower subhalo masses than those previously presented, and build the repopulated VL-II not only using subhalo masses, as done for our previous works, but also using subhalo velocities. As we will show, the latter allows to overcome some of the issues associated to the definition of masses in the case of subhaloes.

The paper is organized as follows. 
In Section~\ref{sec:charvl} we first describe the VL-II simulation and motivate this simulation choice. In the same section, we characterize the subhalo population of VL-II in great detail, paying particular attention to those ingredients that will be needed for the repopulation. This characterization is performed in terms of both subhalo masses and velocities.
In Section~\ref{sec:repopu} we describe the repopulation procedure and present its results. 
Section~\ref{sec:jfactorVL} is devoted to obtaining subhalo DM annihilation fluxes and apparent angular sizes of the DM emission. We compare these quantities when using either subhalo masses or velocities.
Finally, we conclude in Section~\ref{sec:conclu}, where we also discuss on potential applications and future work.

\section{Characterization of the VL-II subhalo population}\label{sec:charvl}

\subsection{Description and choice of the simulation data}
We use public data from VL-II\footnote{The simulation team made public the data at redshift 0, which can be downloaded from \url{http://www.ucolick.org/~diemand/vl}}. 
VL-II is a one-billion N-body cosmological simulation that tracks the formation and evolution of a MW-size host halo ($M_{200} = 1.9 \times 10^{12} \mathrm{M_\odot}$) from redshift $\sim$100 until now in a $\Lambda$CDM Universe, described with the WMAP3 cosmological parameters\footnote{For comparison, current Planck \citep{2020A&A...641A...6P} values are $\Omega_\mathrm{m} = 0.315, h = 0.674$ and $\sigma_8 = 0.811$. For the purposes of this work, we note that the VL-II parameters can be seen as conservative, as current ones might imply greater subhalo abundances and thus more optimistic results in Sections \ref{sec:repopu} and \ref{sec:jfactorVL} \citep[see e.g.][]{2014ApJ...786...50D}. Nonetheless, the impact of updating them will be probably negligible \citep[as suggested in][]{2021MNRAS.507.3412C}.} \citep{2007ApJS..170..377S}, 
$\Omega_\mathrm{m} = 0.238, h = 0.73$ and $\sigma_8 = 0.74$. 
The mass of a DM high-resolution particle is $M_{\mathrm{hires}} = 4.1 \times 10^3 \mathrm{M_\odot}$, which allows to resolve twenty thousand haloes and subhaloes at $z = 0$ with masses larger than $10^4 \mathrm{M_\odot}$ in a box of 4 Mpc side.

There have been other recent works that used VL-II in a context not far from ours. For instance, in \citet{2016JCAP...09..047H}, authors repopulate the host halo using the CLUMPY software~\citep{2012CoPhC.183..656C}, which among others have implemented recipes based on VL-II results. 
Other authors, such as \citet{2017PhRvD..96f3009C} and \citet{2019Galax...7...90C} have used Aquarius and/or Phat-ELVIS data to perform their repopulations instead. All in all, most works 
face issues related to the concentration of subhaloes, as they do not use a model %
of the concentration for subhaloes but rather assume similar properties for them than field haloes of the same mass \citep[see discussion in][]{moline}.
Either way, the use of VL-II data in recent works reaffirms the fact that this simulation is still perceived by the community as state-of-the-art when it comes to DM-only simulation results.

One of the main goals in our work is to compute subhalo annihilation fluxes down to masses below the resolution of current simulations. In order to do so, first it will be necessary to characterize the fraction of the subhalo population that is well resolved in VL-II. We will derive mass and velocity functions (i.e., how many subhaloes in a certain range of mass/velocity we have), radial distributions within the host halo (how the subhaloes are spatially distributed), and calculate their concentration (the precise distribution of DM inside subhaloes) by adopting the model presented in \citet{moline}, that was based on VL-II subhalo data. 
Later, in Section~\ref{sec:repopu}, we will use this detailed characterization of VL-II to repopulate the parent simulation with low-mass/velocity subhaloes below the original resolution limit. 
The main reason of dealing with both subhalo masses and velocities is that virial masses are not well defined for subhaloes: tidal stripping causes a truncation of the density profile in the subhalo outskirts, making it impossible to properly define a virial radius as done for field haloes. Instead, we can work with tidal masses, which are nevertheless still a less reliable parameter than the maximum circular velocity of particles in the subhalo when it comes to describing subhalo structural properties~\citep{moline}. Below, we will investigate how our results change by adopting either subhalo tidal masses or subhalo maximum circular velocities for the repopulation.

VL-II is not the only so-called zoom-in high-resolution numerical simulation in the market as of today. Salient examples of both DM-only and hydrodynamical simulations are  Aquarius \citep{aquarius}, Elvis \citep{Elvis}, GHALO \citep{GHALO}, Caterpillar \citep{2016ApJ...818...10G}, Apostle \citep{2016MNRAS.457..844F}, Auriga \citep{2017MNRAS.467..179G}, COCO \citep{2016MNRAS.457.3492H} and Symphony \citep{2022arXiv220902675N}. %
Yet, most of them lack the mass resolution we need for our work. 
The exceptions are Aquarius Aq-A-1 ($1.7 \times 10^3 \mathrm{M_\odot}$) and GHALO$_2$ ($10^3 \mathrm{M_\odot}$). Indeed, having a mass resolution as good  
as possible is particularly important for our purposes, as it allows us to reach lower subhalo masses directly from simulation data, thus making 
our low-mass subhalo repopulations less uncertain. 
Beyond mass particle resolution, the Aquarius simulation set has five realizations, hence possessing better statistics than VL-II. However, we note that neither VL-II nor Aquarius have Planck cosmology, and   
VL-II adopted a lower $\sigma_8$ value compared to Aquarius  \citep[$\sigma_\mathrm{8,Aq} = 0.9$, consistent with WMAP5,][]{2009ApJS..180..330K}.

Moreover, \citet{aquarius} reported that the fraction in resolved substructures among their different realizations ``varies around 11\%'' and ``is larger than the 5.3\% inside $r_{50}$ reported by \citet{Diemand:2006ik} for a Milky Way-sized halo'', where their $r_{50}$ corresponds to our $R_\mathrm{vir}$. %
The greater substructure abundance in Aquarius and higher $\sigma_8$ could lead to larger DM-induced signals from VL-II subhaloes. Thus, we will stick to VL-II data, this way making our predictions conservative. 
All in all, we decide to work with VL-II data because it is publicly available and it still possesses one of the best particle resolutions in the market as of today.
We are aware though that baryons could significantly alter the structural properties of subhaloes and their abundance, e.g.~\citet{Kelley2019,2023MNRAS.518...93A}. Their impact on DM-induced gamma-ray signals from subhaloes will be addressed in further work by means of hydrodynamical simulations.

\subsection{Characterization of subhalo properties using tidal masses}\label{sec:vlmass}

As already mentioned, subhaloes lose mass due to tidal stripping inside the host halo: the outermost material is removed, yet the inner cusp structure remains nearly intact~\citep[e.g.][]{2018MNRAS.475.4066V, 2023MNRAS.518...93A}. 
Because of that, instead of virial mass we use the so-called tidal mass of subhaloes, i.e. the mass within the tidal radius.\footnote{Both the tidal mass and radius are provided for each subhalo in the original VL-II data files.}  
The latter is the radius of the subhalo after its interaction with the tidal forces induced by the host, and can be well approximated by the \citet{1962AJ.....67..471K} radius: %
$$R_\mathrm{t} = D_\mathrm{GC} \left( \frac{M_\mathrm{sub}}{3 M (<D_\mathrm{GC})} \right)^{1/3}$$

where $M_\mathrm{sub}$ is the subhalo mass, $D_\mathrm{GC}$ is its distance to the Galactic centre (GC), and $M (<D_\mathrm{GC})$ is the host mass contained in the sphere of radius $D_\mathrm{GC}$.  \citet{1998MNRAS.299..728T} also offers a definition of the tidal radius well approximated for non-circular orbits and checked against simulations. In the following, we will be actually meaning tidal mass every time we refer to subhalo mass.

\subsubsection{Subhalo mass function}\label{sec:vlshmf}

The abundance of DM (sub)haloes as a function of their mass, i.e. the (sub)halo mass function (S/HMF), %
plays an important role in cosmology due to its sensitivity to several important parameters including the matter density of the Universe $\Omega_\mathrm{m}$ and the Hubble parameter $h$ \citep{2013MNRAS.433.1230W}. %
Since it is not easy to nail it down accurately enough with current observations, cosmological simulations have been traditionally used to study it in detail \citep{jurg1, 2019Galax...7...81Z}.

The cumulative %
SHMF at redshift 0 within VL-II can be well approximated by a power law~\citep{jurg1}:

$$N (> M_\mathrm{sub}) = c\, \left( \frac{M_\mathrm{sub}}{M_{200}} \right)^{-\alpha} $$
where $M_{200} = 1.9 \times 10^{12} \mathrm{M_\odot}$ is the mass of the host halo and $M_\mathrm{sub}$ is the (tidal) mass of the subhalo.

In practice, the cumulative number of subhaloes is not perfectly fitted by a power law, since it declines rapidly at the largest masses in the simulation, due to gravitational interactions with the host 
-- no substructures with masses larger than $\sim 10\%$ the mass of the whole halo are typically found \citep{2008MNRAS.386.2135G, RP:2016} %
-- and it decreases also at low masses, mainly due to the limited numerical resolution.  
Thus, the best-fit slope depends on the mass range and the fitting procedure.

\begin{figure}%[h]
\centering
\includegraphics[width=\columnwidth]{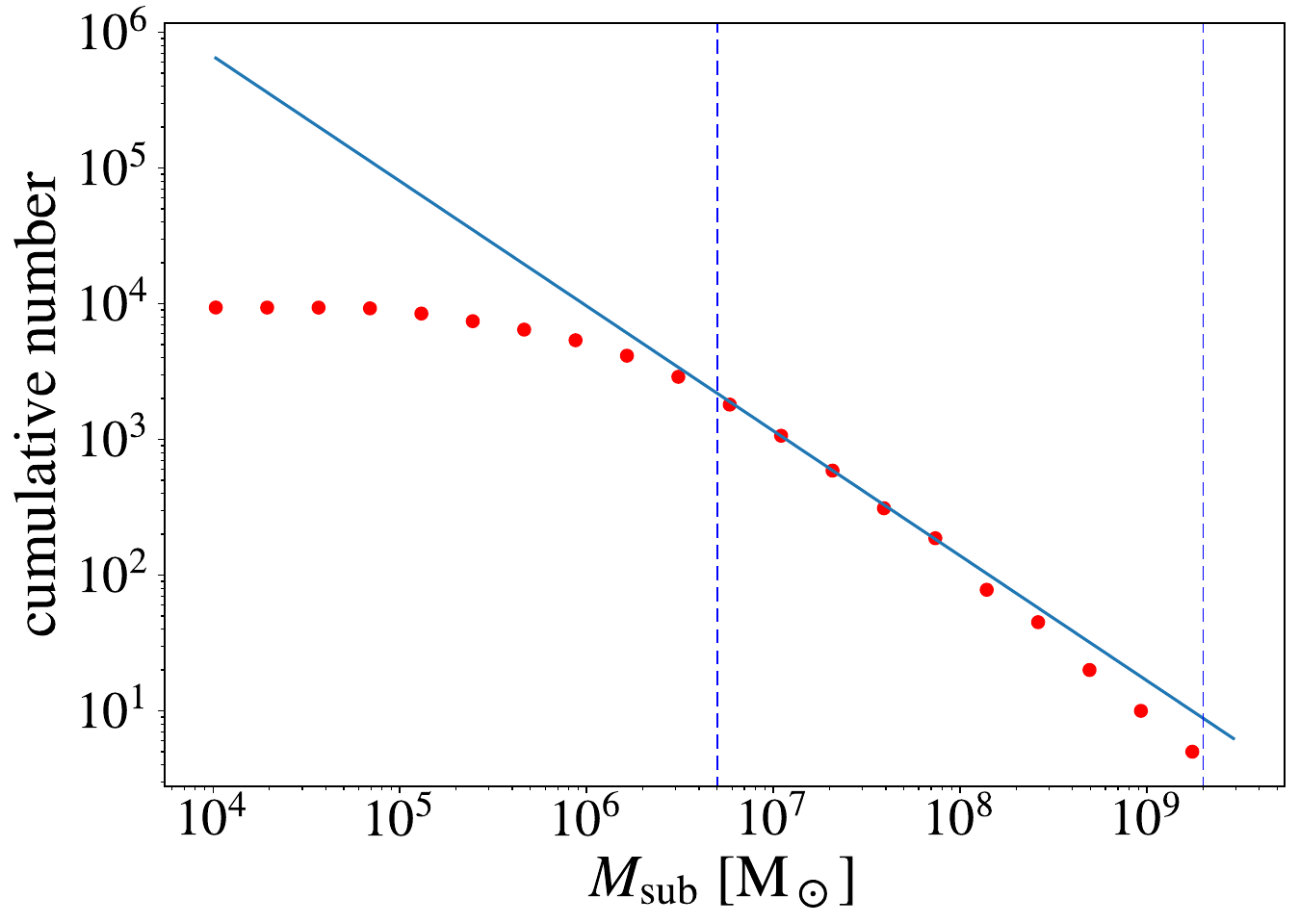} %
\caption{Cumulative SHMF of the whole VL-II simulation (dots are the data). Our fit (solid line) has been performed in the range where the SHMF roughly behaves as a power law, i.e. $\{5 \times 10^6  \mathrm{M_\odot}, 2 \times 10^9  \mathrm{M_\odot} \}$. %
The two dashed lines indicate the mentioned range, the leftmost one corresponding to $\mathrm{M_{cut}}$. 
}
\label{fig:shmf1}
\end{figure}

We perform our SHMF fit inside the mass range where the simulation is `complete', i.e. where the SHMF behaves as a power law, $\{5 \times 10^6  \mathrm{M_\odot}, 2 \times 10^9  \mathrm{M_\odot} \}$, as shown in Fig.~\ref{fig:shmf1}, and apply a bootstrapping technique to obtain more meaningful errors.\footnote{Since we only have one simulation, our fit results vary when using different intervals within the selected mass range and the errors obtained in each fit are too small. Thus, we calculate average values of the fits (and their corresponding errors) of different random samples of this population using different mass intervals.
}. From now on, this lower limit will be called $\mathrm{M_{cut}}$. We obtain the following parameters: %c
$$ \begin{matrix}
\alpha = 0.92 \pm 0.03 \\ %
 c = 0.016 \pm 0.008
 \end{matrix} $$ %
 
We note that these results are slightly different to the ones in \citet{jurg1}, where c = 0.0064, $\alpha \simeq 1$ (indeed, they get %
$\alpha = 0.97 \pm 0.03$ for $M_\mathrm{sub} > 200 M_{\mathrm{hires}}$), %
yet both sets of results are compatible with theoretical expectations in the Press-Schechter theory for structure formation, see e.g. \citet{shmf2} and \citet{shmf3}.

\subsubsection{Subhalo radial distribution}\label{sec:srdm}

\begin{figure}
\centering
\includegraphics[width=\columnwidth]{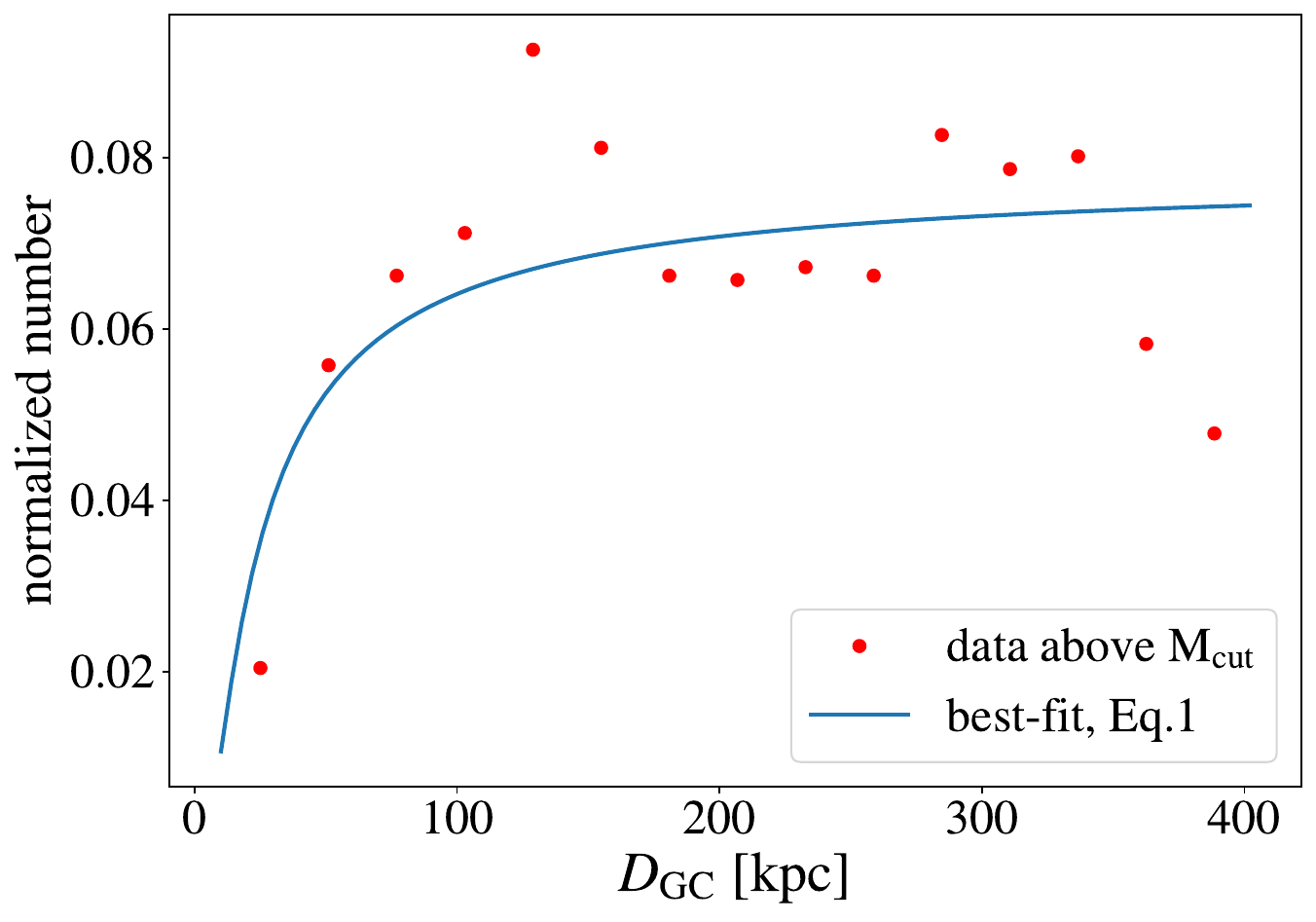} %
\caption{SRD of the VL-II simulation above $\mathrm{M_{cut}}=5\times10^6 \mathrm{M_\odot}$, i.e., number of subhaloes with respect to the distance to the GC. More precisely, the $y$ axis shows the quotient between the number of subhaloes in each bin and the total number of subhaloes above %
$\mathrm{M_{cut}}$. The solid line is the proposed fit given by Equation~\ref{eq:srdmass}. 
}
\label{fig:srdmass}
\end{figure}

We have distributed all our subhaloes in 20 radial bins to study how they are located inside the host. We have also divided our sample in mass bins to check whether the subhalo radial distribution (SRD) is mass-dependent or, on the contrary, it is universal, as commonly stated \citep{Han:2015pua}. Actually, we found that the VL-II SRD below the aforementioned $\mathrm{M_{cut}}$ exhibits a drastic change, showing in comparison significantly more substructure in the internal regions. %
Since using two completely different SRDs would have a large impact on our repopulation results of the next sections, and the subhalo population below $\mathrm{M_{cut}}$ may be already subject to numerical resolution issues \citep[see e.g.][]{2018MNRAS.474.3043V}, we decide to conservatively build our VL-II SRD using only subhalo data above $\mathrm{M_{cut}}$ for this work. The behaviour of the SRD at lower masses will be explored elsewhere. %
More details about the SRD universality in VL-II are given in Appendix \ref{sec:srdcav}.
We propose the following fitting function:

 \begin{equation}\label{eq:srdmass}
 N (D_\mathrm{GC}) = b \, e^{a/D_\mathrm{GC}} 
 \end{equation}

\noindent with best-fit parameters: 
$$ \begin{matrix} 
   a = -20 \pm 8 \\ %a =  0.8 \pm 0.2          \\
         b = 0.059 \pm 0.004 \\
         \end{matrix}
 $$
%\

 Both the SRD given by the data and our best fit are shown in Fig.~\ref{fig:srdmass}. %

We note that, typically, either the so-called `anti-biased' NFW \citep{Diemand2008} or the Einasto \citep{aquarius} fitting functions were adopted in the past to represent the subhalo mass density within the host \citep{pieri}.  
Yet, these distributions do not illustrate properly the behaviour near the GC, where no subhaloes are actually found in the simulation.\footnote{In fact, there exists some controversy about whether the lack of inner subhaloes is a result of the limited resolution of the simulations~\citep[e.g.][]{2018MNRAS.475.4066V, 2023MNRAS.518...93A}, 
however we will not take part in this debate here.
} %  
This fact will be especially relevant in our case: the use of either anti-biased NFW or Einasto SRDs would imply a larger number of subhaloes closer to Earth compared to our SRD in Eq.~\ref{eq:srdmass}, this way providing us with brighter subhaloes in terms of their annihilation luminosities. In this sense, we prefer to stay conservative in our predictions and adopt the above SRD. Note, also, that our SRD refers to the subhalo number %
instead of subhalo mass density.%

\subsubsection{Subhalo concentrations}\label{sec:concvl}

The concentration of a halo is formally defined as $c_\Delta = \frac{R_\mathrm{vir}}{R_\mathrm{s}}$, where $R_\mathrm{vir}$ %
is the virial radius of the halo, defined (at redshift $z$)
as the radius that encloses a halo mean density $\Delta$ times the
critical (or mean, depending on the chosen convention) %
density of the Universe, and $R_\mathrm{s}$ is the so-called scale radius; that is, the radius at which the logarithmic slope of the DM density profile is $-2$.
 This standard definition of halo concentration, while very useful for the study of the internal structure of well-resolved haloes, is not suitable for subhaloes, mostly because the virial radius of subhaloes is not well defined as it may not even exist: tidal stripping removes mass from the outer parts of subhaloes and, as a result, subhaloes are
truncated at smaller radii compared to field haloes of the same mass \citep[see e.g.][]{Ghigna:1998vn, Diemand:2006ik, jurg1}.

In this work, we use the subhalo concentration model of \citet{moline} to model the structural properties of subhaloes in VL-II\footnote{We note that this model is older than the one presented in \citet{2023MNRAS.518..157M}. However, we prefer to use \citet{moline} as it was obtained using the very same VL-II data.}:

\begin{multline}\label{concmoline}
c_{200} (M_\mathrm{sub}, x_\mathrm{sub}) = c_0 \left[ 1 + \sum_{i=1}^3 \left[ a_i \log_{10} \left( \frac{M_\mathrm{sub}}{10^8 h^{-1} \mathrm{M_\odot}} \right) \right]^i \right] \\ \times [1 + b \log_{10} (x_\mathrm{sub})]\end{multline}

with $M_\mathrm{sub}$ the tidal mass of the subhalo, $x_\mathrm{sub}$ its fraction distance with respect to the GC compared to the virial radius, $c_0 = 19.9,\ a_i = \{-0.195, 0.089, 0.089\}$ and $b = -0.54$. 
 This model has been mainly built using subhalo data from VL-II and ELVIS \citep{Elvis}, 
 and is also in agreement with other ones existing in the literature \citep{Bartels:2015uba, Zavala:2015ura}. 
Note that the model implies that a subhalo near the GC is significantly more concentrated than another one with the same mass but located far away. 
Also, notice that for the same subhalo mass, \citet{moline} gives a factor $\sim$1.5-2 larger concentrations than \citet{mascprada14}. This will be particularly important for the calculation of the DM annihilation signals, which are proportional to the cube of the concentration, as it will be shown in Section~\ref{sec:jfactorVL}.

We also include the scatter in subhalo concentration values that is inherent to $\Lambda$CDM. We follow \citet{Bullock:1999he, Wechsler:2001cs, pieri, moline}, where they used:
\begin{equation}\label{pconc}
P(c_{200}) = \frac{1}{c_{200} \ln 10 \sqrt{2\pi} \sigma_{\log_{10}\, c_{200}}} e^{-\frac{1}{2} \left(\frac{\log_{10}\, c_{200} - \log_{10}\, c_{200,0}}{\sigma_{\log_{10}\, c_{200}}}\right)^2},
\end{equation}

where $\sigma_{\log_{10}\, c_{200}} = 0.14$ is the scatter and $\log_{10}\, c_{200,0}$ is the median obtained with expression (\ref{concmoline}). % 

\subsection{Characterization of subhalo properties using $V_\mathrm{max}$}\label{sec:vlvmax}

Up to this point, we have used tidal masses to describe the structural
properties of subhaloes, i.e., to assign concentrations. However, as mentioned above, the very definition of masses in subhaloes is not a trivial task. Instead, it would be highly desirable to work with a subhalo concentration independent of the adopted density profile and of the particular definition used for the virial radius. Fortunately, the peak circular velocity at redshift 0, $V_\mathrm{max}$, is less prone to tidal forces and was identified as an ideal, alternative parameter for subhaloes \citep{moline, jurg1}.

In the following, we will perform a similar exercise to that in Section~\ref{sec:vlmass} but using $V_\mathrm{max}$ instead: we will obtain the subhalo velocity function (SHVF), as well as the corresponding SRD and concentration parameter.

\subsubsection{Subhalo velocity function}\label{sec:vlshvf}

The cumulative SHVF at redshift 0
within VL-II is well approximated by a power law, too \citep{Diemand:2009bm}:

$$N (> V_\mathrm{max}) = c\, \left( \frac{V_\mathrm{max}}{V_\mathrm{max,host}} \right)^{-\alpha}, $$
where $V_\mathrm{max,host} = 201$ km/s is the maximum circular velocity of VL-II.

Here, the cumulative number of subhaloes is not perfectly fitted by a power law either, due to the same reasons as with the SHMF: since subhaloes are inside a halo, there are typically no subhaloes with $V_\mathrm{max} > 0.1\, V_\mathrm{max,host}$. Besides, due to the lack of resolution, we cannot resolve subhaloes with a very small $V_\mathrm{max}$. Thus, and once again, the best-fit parameters depend on the used velocity range and the fitting procedure.

\begin{figure}%[H]
\centering
\includegraphics[width=\columnwidth]{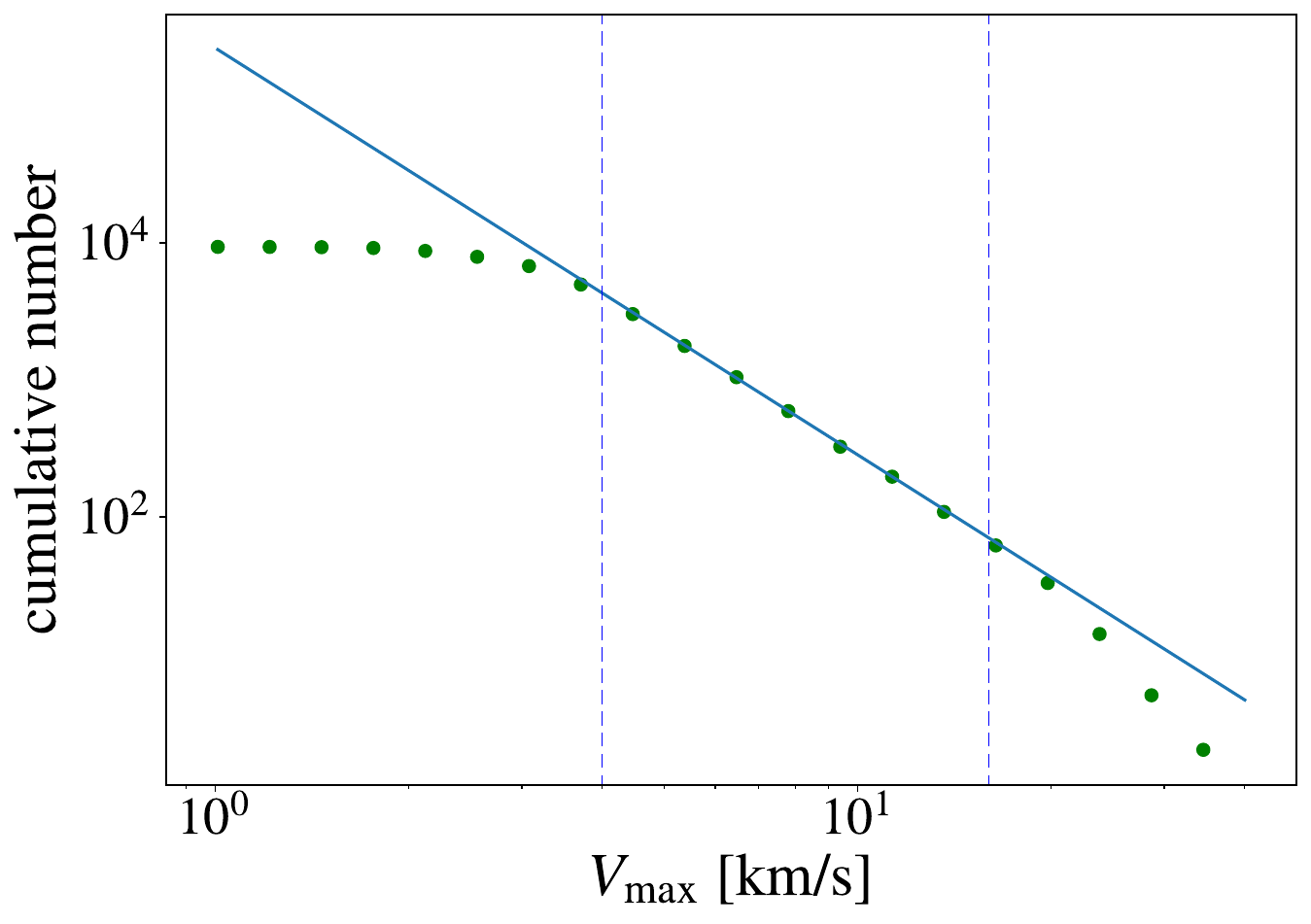} %
\caption{%
Cumulative SHVF for VL-II. Dots are the data while the solid line is the power-law fit performed in the range $4 \leq V_\mathrm{max} < 16$ km/s, i.e., where the SHVF exhibits a power-law behaviour. 
The two dashed lines indicate the mentioned range. The lower limit is our $\mathrm{V_{cut}}$; see text for details.}
\label{fig:shvf1}
\end{figure}

We fit the data within the range of completeness of the simulation\footnote{This is the range where the SHVF behaves as a power law.}, i.e., $4 \text{ km/s} < V_\mathrm{max} < 16 \text{ km/s} $ as shown in  Fig.~\ref{fig:shvf1}%
, and apply the bootstrapping technique described above in Section~\ref{sec:vlshmf} 
for the SHMF case. By doing so, we obtain the following parameters: %
$$ \begin{matrix} \alpha = 2.97 \pm 0.08 \\ %$$ 
%$$
c = 0.038 \pm 0.006  \end{matrix} $$ %
These results are in agreement with 
the ones in \citet{Diemand:2009bm}, i.e., $c = 0.036,\ \alpha \simeq 3$. %

\subsubsection{Subhalo radial distribution}\label{sec:srdv}

In this case, we use a different fitting function with respect to the one used for the case of using subhalo masses. We still distribute the subhaloes in 20 radial bins but, for the case of using $V_\mathrm{max} > 4$ km/s, we find comparatively less subhaloes in the outermost part of the host compared to the mass case (Fig.~\ref{fig:srdmass}). We observed the SRD below this $\mathrm{V_{cut}}$ to exhibit significantly more substructure in the inner regions. Yet, as done for the case of the SRD built from subhalo masses in Section~\ref{sec:srdm}, in this work we decide to conservatively stick to data above $\mathrm{V_{cut}}$ in order to build the SRD from subhalo velocities. We propose the following fitting function:

\begin{equation}\label{eq:cosmic}
N (D_\mathrm{GC}) = \left( \frac{D_\mathrm{GC}}{R_0}\right)^a \exp \left( -b\, \frac{D_\mathrm{GC} - R_0}{R_0} \right),  \end{equation}

with best-fit parameters: 
  $$ %\left \{
      \begin{matrix} 
         a =  0.8 \pm 0.1          \\
         b =  8.4 \pm 0.4 \\
         R_0 = (1040 \pm 90) \mathrm{\ kpc}
      \end{matrix}
   %\right.
   $$ 

\begin{figure}%[H]
\centering
\includegraphics[width=\columnwidth]{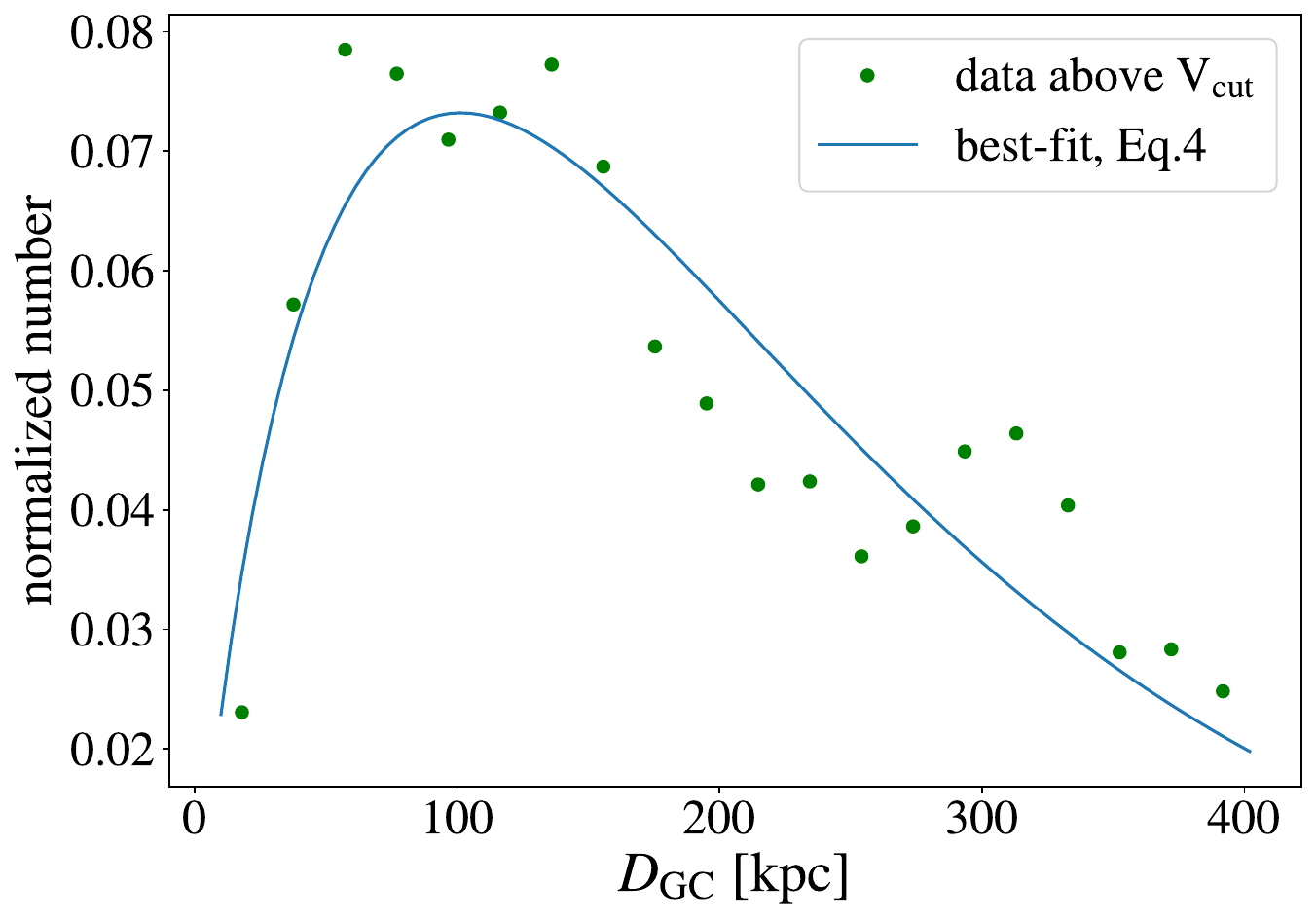}
\caption{SRD of the VL-II simulation above $\mathrm{V_{cut}}=4~$km/s, i.e., number of subhaloes with respect to the distance to the GC. More precisely, the $y$ axis shows the quotient between the number of subhaloes in each bin and the total number of subhaloes above %
$\mathrm{V_{cut}}$. The solid line is the proposed fit given by Equation~\ref{eq:cosmic}.
}
\label{fig:srdvmax}
\end{figure}

Fig.~\ref{fig:srdvmax} shows both the data and our best fit. We remind the reader that we do not compute the mass density of subhaloes nor the number density, but the number instead.

\subsubsection{Subhalo concentrations}\label{sec:cvmax}

In this case, we adopt the following definition for the concentration parameter based on subhalo velocities~\citep{Diemand2008, moline}: 
\begin{equation}
    c_\mathrm{V} = 2 \left(\frac{V_\mathrm{max}}{H_0 R_\mathrm{max}} \right)^2,
\end{equation}
where $H_0$ is the Hubble parameter. $V_\mathrm{max}$ is the maximum circular velocity of the particles inside the subhalo, and $R_\mathrm{max}$ is the radius at which this happens. Note that, in this way, $c_\mathrm{V}$ can be directly obtained independently of the assumed form for the subhalo DM density profile. At the same time, $c_\mathrm{V}$ still fully encodes the essential meaning attached to the traditional concentration parameter.

 This concentration is well-defined for subhaloes and takes implicitly into account the effect of tidal mass loss. For instance, for a fixed $V_\mathrm{max}$, the obtained $R_\mathrm{max}$ values are on average $\sim 60\%$ of those of haloes in the Aquarius simulation \citep{aquarius, 1997ApJ...490..493N, 2001ApJ...554..114E, Bullock:1999he, 2007MNRAS.381.1450N, 2007MNRAS.377L...5G}. 
A study of the relation between $V_\mathrm{max}$ and $R_\mathrm{max}$ for VL-II is provided in Appendix \ref{ap:vmaxrmax}.

To compute the concentrations from subhalo velocity data, we use the model by \citet{moline}, that uses $V_{\mathrm{max}}$ instead of $M_\mathrm{sub}$: 

\begin{multline}\label{concmolinevmax}
c_{\mathrm{V}} (V_{\mathrm{max}}, x_{\mathrm{sub}}) = c_0 \left[ 1 + \sum_{i=1}^3 \left[ a_i \log_{10} \left( \frac{V_{\mathrm{max}}}{10 \mathrm{\ km/s}} \right) \right]^i \right] \\ \times [1 + b \log_{10} (x_{\mathrm{sub}})]\end{multline}

with $c_0 = 3.5 \times 10^4,\ a_i = \{-1.38, 0.83, -0.49\}$ and $b = -2.5$. This model was also built mainly using subhalo data from VL-II and ELVIS \citep{Elvis}. Just like it happened for the case of using subhalo masses, for a given velocity, a subhalo near the GC is significantly more concentrated than one located farther away within the host.

%%%%%%%%%%%%%%%%%%%%%%%%%%%%%%%%%%%%%%%%%%

\section{VL-II repopulation %
}\label{sec:repopu}

In this Section, we aim at creating a new set of simulations that will include subhaloes with masses (velocities) well below the resolution limit we found for VL-II, i.e., $\mathrm{M_{cut}} = 5 \times 10^6 \mathrm{M_\odot}$ ($\mathrm{V_{cut}} = 4\, \mathrm{km/s}$). Indeed, as stated in Section~\ref{sec:vlshmf}, the parent simulation is complete only above $\mathrm{M_{cut}}$ ($\mathrm{V_{cut}}$) (see Fig.~\ref{fig:shmf1} (\ref{fig:shvf1})) and it is completely devoid of subhaloes below $\sim10^3 \mathrm{M_\odot}$ ($\sim1\, \mathrm{km/s})$. %
The general procedure will be to make use of that learnt in the previous sections with the help of actual VL-II data, and to extrapolate the relevant quantities down to the lower subhalo masses in a well-motivated way.

Some of the motivations of {\it repopulating} the original simulation with low-mass subhaloes are a) the opportunity to have better subhalo statistics; b) to enlarge the mass range of study; c) to solve numerical resolution issues; d) to have the freedom to vary mass and/or radial distribution functions. As stated in the Introduction, all of these motivations become particularly relevant for e.g. structure formation, Galactic archaelogy and indirect DM searches. 
Because of this, the topic has already been explored in previous works \citep{2012MNRAS.421.3343G, 2016JCAP...09..047H, 2019Galax...7...90C}, using diverse methodologies that relied on results derived from N-body cosmological simulations above their resolution limits. In this work, we will go a step forward by including some important novelties in the methodology (e.g. the derivation of more sophisticated SRDs; the use of not only $M_\mathrm{sub}$ but also $V_\mathrm{max}$ as proxies for the repopulations; J-factor calculations derived for both quantities; repopulations down to much lower subhalo masses/velocities; public release of the data...), described in detail below. 
Besides, our repopulation algorithms were already successfully applied and used in several of our previous published works \citep{2019JCAP...07..020C, 2019JCAP...11..045C, 2021PDU....3200845C, 2022PhRvD.105h3006C}, yet the full pipeline was never exhibited and dissected in detail. 

In a general perspective, our VL-II repopulation will consist of the next steps:
\begin{enumerate}
\item We compute the number of subhaloes in a certain mass ($V_\mathrm{max}$) range. 
\item We assign a mass ($V_\mathrm{max}$) to each subhalo, according to the subhalo mass (velocity) function we found for VL-II in Section~\ref{sec:charvl}, and place them inside the host halo at a distance according to the VL-II SRD. We then generate two spherical angles randomly and uniformly, $\zeta$ and $\phi$, to populate the whole halo sphere. 
\item Once all subhalo masses (velocities) and distances are settled in the desired mass (velocity) range, we assign a concentration to each subhalo using Equation~\ref{concmoline} (\ref{concmolinevmax}) in Section~\ref{sec:concvl} (\ref{sec:cvmax}). 
\item As we are also interested in obtaining the astrophysical factor of the subhalo DM annihilation flux as well as subhalo angular sizes (see later in Section~\ref{sec:jfactorVL}), we do so by placing the Earth anywhere in the repopulated Milky Way at 8.5 kpc from the GC \citep[we take 8.5 kpc as the Sun's Galactocentric distance,][]{1986MNRAS.221.1023K}. 
\end{enumerate}

We have generated two different sets of simulations for the case of using subhalo masses. The first one, $[\mathcal{A}_M]$, goes down to $10^3 \mathrm{M_\odot}$ and it repopulates the whole VL-II halo up to its virial radius. The other one, $[\mathcal{B}_M]$, reaches subhalo masses as small as $0.1 \mathrm{M_\odot}$, but it only repopulates a relatively thin spherical shell around the GC centered at the Earth's Galactocentric distance. 
It is done this way, first to avoid generating billions of low-mass subhaloes within the volume of the entire halo, which would be computationally very expensive; and second because we are particularly interested in those subhaloes exhibiting the highest DM fluxes, and thus it would be useless to repopulate regions located far from the Earth with low-mass (i.e., faint) subhaloes. Yet, as it will be shown later below, low-mass subhaloes can be potentially very relevant if they lie close enough to Earth, competing in terms of their DM annihilation fluxes with more massive, more distant subhaloes. Thus the need to repopulate the simulation with them in a volume around the Earth. A 2D representation of repopulation $[\mathcal{B}_M]$ is shown in Fig.~\ref{fig:andreashell}.

The size of the spherical shell adopted for set $[\mathcal{B}_M]$ is calculated in the following way. First, a mass range is chosen for the repopulation. Then, using the maximum mass value of that interval, $M_\mathrm{sub, max}$, the radius of the shell is calculated: 

\begin{figure}%[H]
\centering
\includegraphics[width=.8\columnwidth]{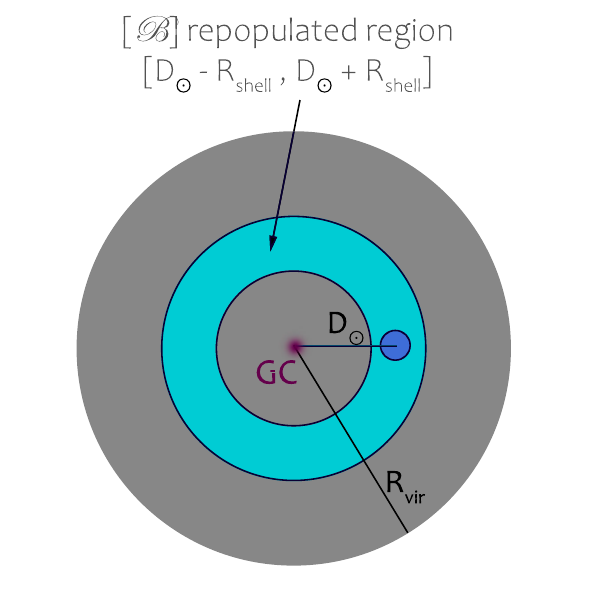} 
\caption{2D representation of a spherical shell used to generate subhalo set $[\mathcal{B}]$ (in turquoise), around the solar Galactocentric distance, $D_\odot = 8.5$ kpc. Set $[\mathcal{A}]$ is generated within the whole sphere (gray + turquoise regions). 
}
\label{fig:andreashell}
\end{figure}

\begin{equation}
   R_\mathrm{shell} = \left(\frac{f^2(c_\mathrm{D})\, M_\mathrm{sub, max}\,  D_\mathrm{D}^2\,  c_\mathrm{sub, max}^3}{0.1\,  f^2(c_\mathrm{sub, max})\,  M_\mathrm{D}\, c_\mathrm{D}^3}\right)^{1/2},
   \label{eq:rshell}
\end{equation}

where $c_\mathrm{sub, max} = c_{200}(M_\mathrm{sub, max}, x_\mathrm{sub})$ is the concentration of a subhalo with mass $M_\mathrm{sub, max}$ as given by Eq.~(\ref{concmoline})\footnote{In order not to miss any potentially bright subhaloes, we have conservatively adopted $x_\mathrm{sub}=0.01$, so that the concentration used to calculate $R_\mathrm{shell}$ is always large enough.}; $D_\mathrm{D}$ and  $M_\mathrm{D}$ are, respectively, the distance of the Draco dwarf spheroidal to the GC and its mass \citep{ 2011JCAP...12..011S}; $c_\mathrm{D}=19$ its concentration \citep{2002MNRAS.333..697L}; 
and $f(c) = \ln(1+c) - c/(1+c)$. 
We calculate $R_\mathrm{shell}$ this way as a good compromise in terms of computational time considering the purpose of our work, i.e. accounting for the small subhaloes with J-factors large enough to be among the 10000 most brilliant ones, while not generating lots of meaningless subhaloes. Indeed, by doing so we only generate subhaloes with astrophysical annihilation factors typically 
larger than a 10\% of Draco's (thus the 0.1 factor in Equation~\ref{eq:rshell}). We adopt Draco as our reference here since this object has been identified recurrently as one of the best targets for indirect DM searches in the literature, e.g. \citet{2004PhRvD..69l3501E, 2011JCAP...12..011S, 2015PhRvL.115w1301A, 2015MNRAS.453..849B, 2019MNRAS.482.3480P, 2020JCAP...10..041A}.

The volume around the Earth considered for set $[\mathcal{B}_M]$ is thus the one enclosed by the spherical shell $8.5 \text{ kpc} - R_\mathrm{shell} < D_\mathrm{GC} < 8.5 \text{ kpc} + R_\mathrm{shell}$\,.
With this approach, we have been able to generate 1000 repopulations of set $[\mathcal{B}_M]$. From now on, we will make use of set $[\mathcal{A}_M]$ alone for checks related to the whole subhalo population of the Milky Way, and a combination of both sets to analyze the viability of low-mass subhaloes for DM searches (see next section).

In addition to these $[\mathcal{A}_M]$ and $[\mathcal{B}_M]$ repopulation sets built from that found in Section~\ref{sec:vlmass} using VL-II subhalo masses, we also generate repopulation sets adopting $V_\mathrm{max}$-based quantities instead (Section~\ref{sec:vlvmax}). More precisely, simulation set labeled $[\mathcal{A}_V]$, the analogous to $[\mathcal{A}_M]$ but for subhalo velocities, includes subhaloes down to %
0.5 km/s, while $[\mathcal{B}_V]$, the analogous to $[\mathcal{B}_M]$, populates the corresponding spherical shell with subhaloes down to 0.05 km/s. A summary of the main characteristics of these data sets is given in Table \ref{tab:repsets}. %

\begin{table}
	\centering
	\caption{Main characteristics of the different repopulation sets created for this work; see Section~\ref{sec:repopu}. Each set has been generated 1000 times and the 10000 brightest subhaloes were selected and saved in each case. $N_\mathrm{sub}$ is the approximate number of subhaloes generated in a single run. For comparison, VL-II has around $10^4$ subhaloes inside the virial radius. 
    All this data is publicly available and can be found at \\ \url{https://projects.ift.uam-csic.es/damasco/?page_id=831}. %
	}
	\label{tab:repsets}
	\begin{tabular}{ccccc} 
		\hline
		 Name & Parameter  & Min & Max & $N_\mathrm{sub}$ \\
   \hline
		 $[\mathcal{A}_M]$ & $M_\mathrm{sub}$ & $10^{3} \mathrm{M_\odot}$  & $3 \times 10^{9} \mathrm{M_\odot}$ & $\sim 5.5 \times 10^6$  \\
		 $[\mathcal{B}_M]$ & $M_\mathrm{sub}$ & $0.1 \mathrm{M_\odot}$  & $10^{3} \mathrm{M_\odot}$ & $\sim 2.0 \times 10^6$  \\
          $[\mathcal{A}_V]$ & $V_\mathrm{max}$ & 0.5 km/s  & 40.4 km/s & $\sim 2.1 \times 10^6$ \\
          $[\mathcal{B}_V]$ & $V_\mathrm{max}$ & 0.05 km/s  & 0.5 km/s & $\sim 4.7 \times 10^5$  \\
		\hline
	\end{tabular}
\end{table}

\begin{figure}%[htbp]
\centering
\includegraphics[width=.99\columnwidth]{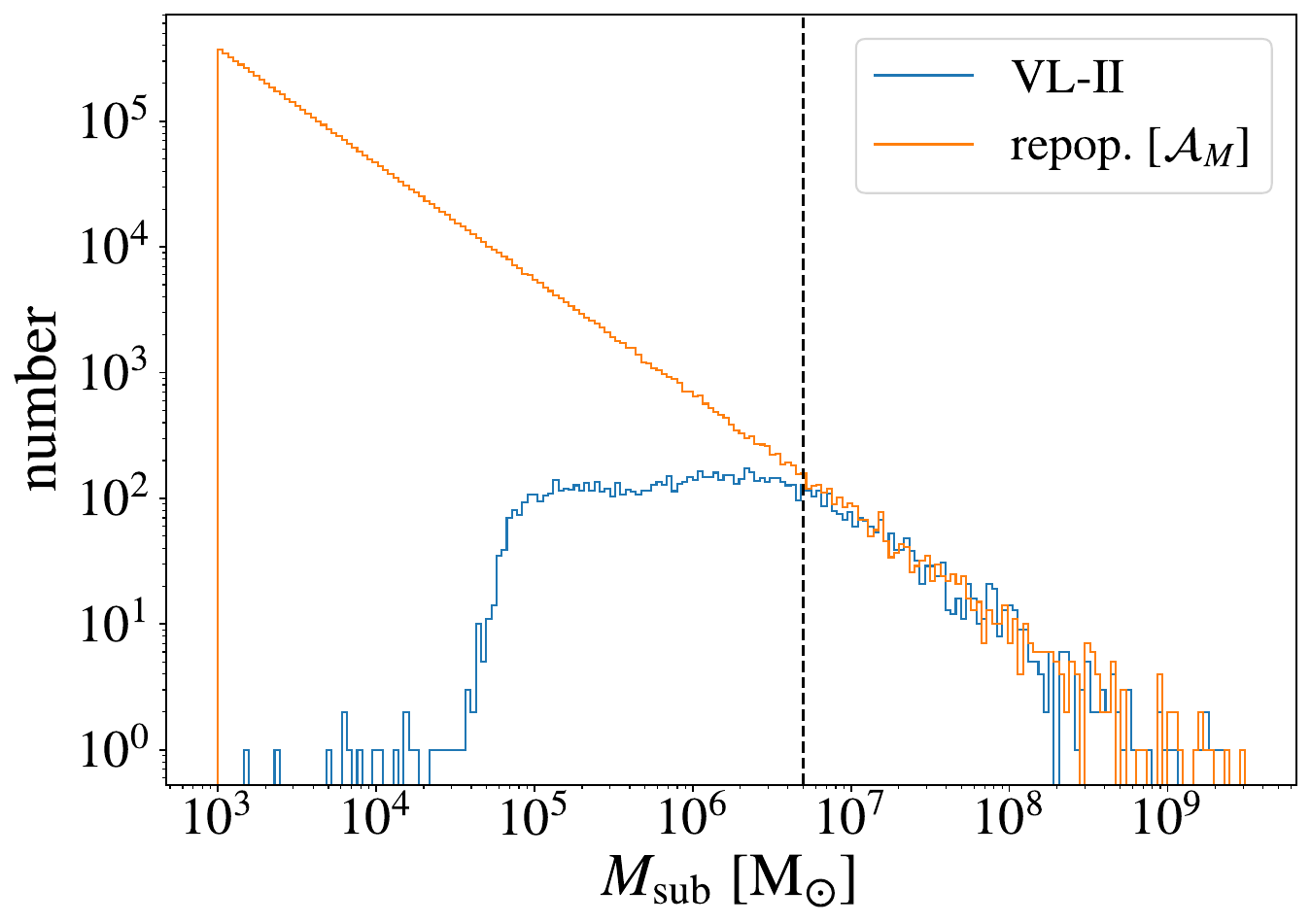} %
\includegraphics[width=.99\columnwidth]{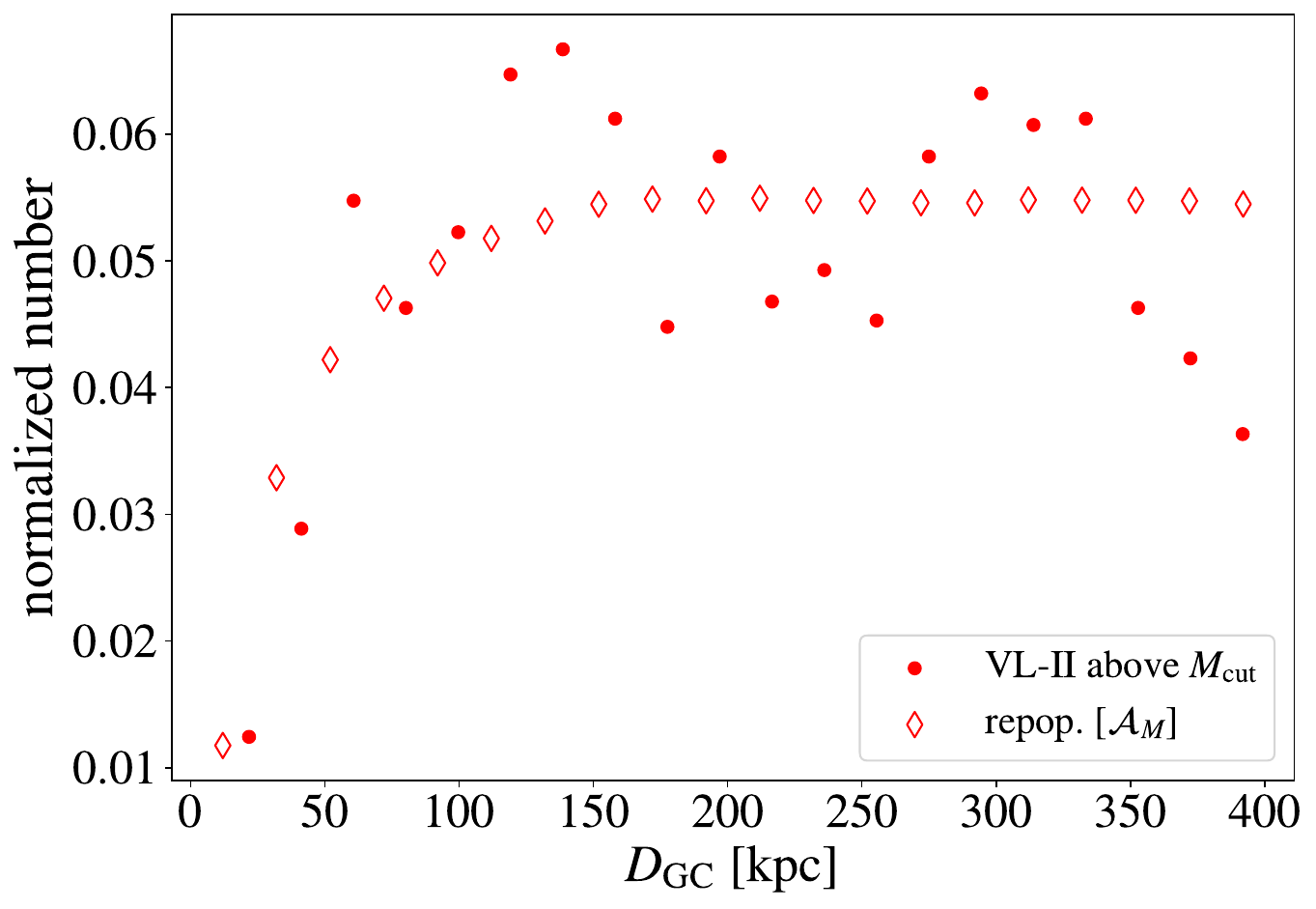}
\includegraphics[width=.99\columnwidth]{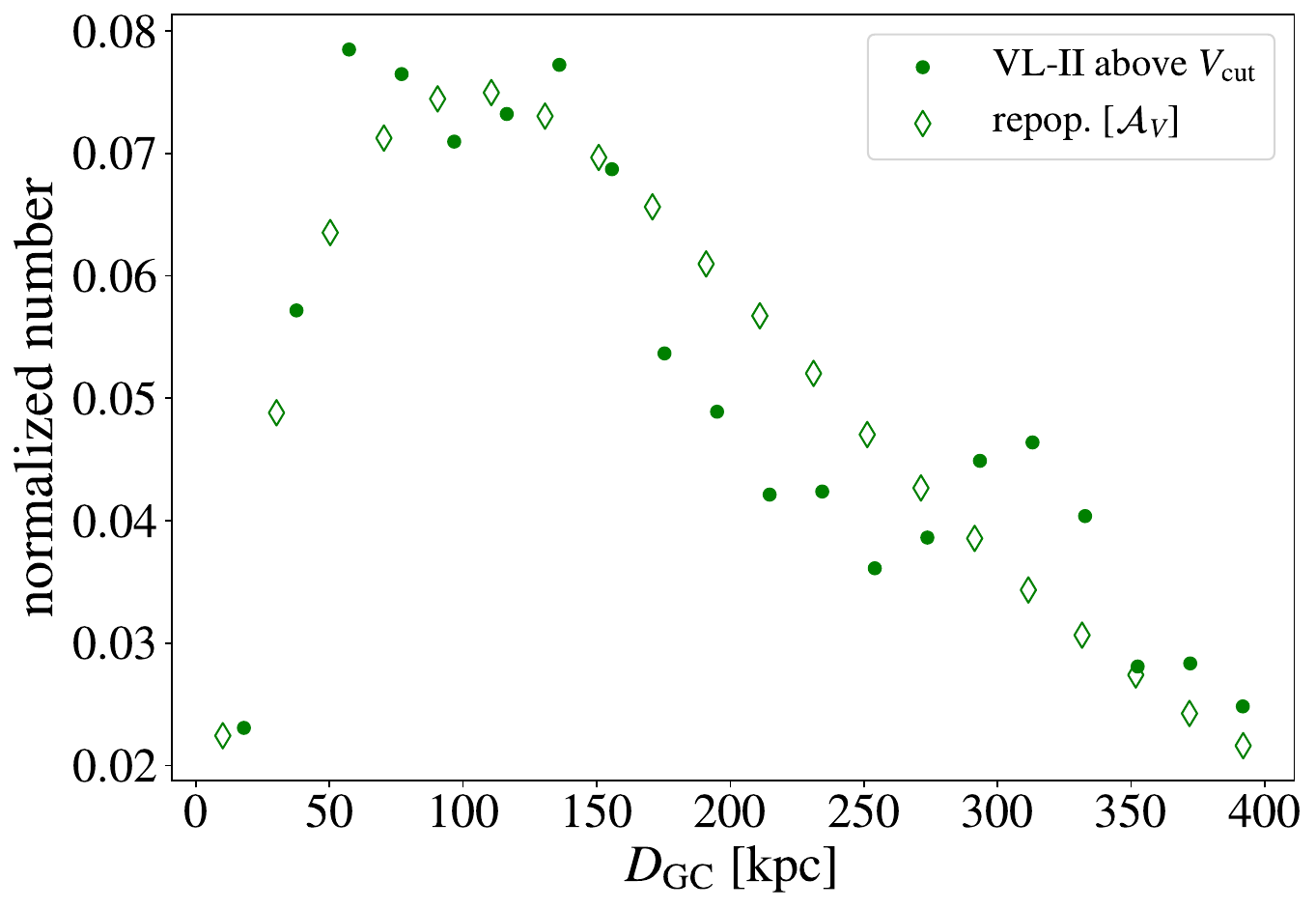}
\caption{Above: SHMF for set $[\mathcal{A}_M]$ compared to the original simulation. Both the original VL-II data and the repopulation agree down to $\mathrm{M_{cut}}$ (dashed line). Below that, VL-II shows a departure from the power-law due to resolution effects, while set $[\mathcal{A}_M]$ continues as expected down to $10^3 \mathrm{M_\odot}$. 
Middle: 
SRD for set $[\mathcal{A}_M]$ %
compared to VL-II for subhaloes with $M_\mathrm{sub}$ above $\mathrm{M_{cut}}$. %
Bottom: SRD for set $[\mathcal{A}_V]$ %
compared to VL-II for subhaloes with $V_\mathrm{max}$ above $\mathrm{V_{cut}}$. 
Recall that the SRD only gives the radial distance of each subhalo; two random angles $\zeta$ and $\phi$ are also created in order to assign a position. 
}
\label{fig:repmshmf}
\end{figure}

As an example, the upper panel of Fig.~\ref{fig:repmshmf} shows how the repopulation mimics the SHMF of the original simulation down to $M_{\mathrm{cut}} = 5 \times 10^6 \mathrm{M_\odot}$, and generates lots of lower mass subhaloes via the used power-law extrapolation. The repopulated SRDs both for the case of using subhalo masses or velocities are also shown in the middle and lower panels %
and agree with expectations.

We also apply the Roche criterium \citep{Binney2008} %
in our repopulated mass simulations in order to get rid of any subhalo that might have been included but may have been destroyed by tidal forces within the host.
It consists on removing subhaloes whose scale radii are larger than their tidal radii, i.e. 
$R_\mathrm{t} \leq R_\mathrm{s} $.\footnote{$R_\mathrm{t}$ has already been defined in Section \ref{sec:vlmass}, %
while $R_\mathrm{s}$ is obtained from the $R_\mathrm{max}-R_\mathrm{s}$ relation found for VL-II. %
} %
Yet, we obtain the reduction in the number of subhaloes due to the Roche criterium to be almost negligible.\footnote{In particular we find that, by doing so, most times the population remains unaltered, while in only a few repopulations %
a subhalo within the innermost 10 kpc is removed.% 
}  Indeed, this is an expected result: the SRD adopted in our mass-based realizations comes from a fit to VL-II data, and these data already account for tidal disruption in a natural way.
This also means that our proposed SRD fit in Equation~\ref{eq:srdmass}, that hardly provides subhaloes within the inner $\sim 15-20$ kpc of the Galaxy is, indeed, a good representation of the actual VL-II SRD.\footnote{Note, however, that subhalo survival below 20 kpc is still an open question in the field, e.g.~\citet{10.1093/mnras/stx1710, 2019MNRAS.488.4585G, vdb1, vandenBosch:2016hjf, 2018MNRAS.474.3043V, 2018MNRAS.475.4066V, 2023MNRAS.518...93A, 2022arXiv220700604S}.
% :)
}

\section{Subhalo J-factors %
and angular sizes} \label{sec:jfactorVL}

The expression used to calculate the DM annihilation flux reaching the Earth from a DM source is composed by two main and differentiated ingredients~\citep{BERGSTROM1998137, PhysRevD.69.123501, jfac1}: a particle physics factor and an astrophysics factor, also called J-factor. 
In this work, we mainly focus on the precise computation of the latter for DM subhaloes in our Galaxy.

We assume a single DM candidate $\chi$ that does not belong to the Standard Model and cannot decay directly into photons. Instead, it annihilates producing Standard Model particles which can eventually generate photons. The annihilation flux is then given by:

\begin{multline}\label{annflux}
\phi (\Delta \Omega, E_\mathrm{min}, E_\mathrm{max}) = \boxandcomment[fill=yellow!15]{X}{\small \textcolor{orange!10!black}{particle physics factor}}{\frac{1}{4\pi} \dfrac{\expval{\sigma v}_{\chi\chi}}{2 m^2_\mathrm{\chi}} \int_{E_\mathrm{min}}^{E_\mathrm{max}} \dfrac{dN_\mathrm{\gamma}}{dE_\mathrm{\gamma}} dE_\mathrm{\gamma}} \\ \times \boxandcomment[fill=blue!10]{X}{\small \textcolor{blue!10!black}{astrophysical J-factor}}{\int_{\Delta \Omega} \int_{l.o.s.} \rho_\mathrm{DM}^{2}(r) \, dl\, d\Omega}  \end{multline}

Here, $\expval{\sigma v}_{\chi\chi}$ is the thermally-averaged cross section of the DM particle, $m_\mathrm{\chi}$ its mass, $\rho_\mathrm{DM}$ is the DM density profile of the object under consideration, $E_\mathrm{min}$ and $E_\mathrm{max}$ are the minimum and maximum energies considered, $\Delta \Omega$ is the solid angle of the region, $\dfrac{dN_\mathrm{\gamma}}{dE_\mathrm{\gamma}}$ is the differential annihilation flux and $l.o.s.$ stands for `line of sight'.

In this section, we will compute J-factors starting both from the mass and maximum circular velocity of subhaloes, and will compare them. 
Also, note that DM subhaloes may have an angular extension on the sky as seen from Earth, that is, might not be point-like sources (even for gamma-ray telescopes, whose angular resolution is typically around 0.1 degrees or worse). Thus, such observables may be relevant for future subhalo search strategies in gamma rays. For that purpose, in the following we will also investigate their typical angular sizes.

\subsection{J-factors based on $M_\mathrm{sub}$}

 The total, integrated J-factor for a given subhalo can be expressed in terms of its concentration and mass in the following way \citep{moline}: 
\begin{multline} \label{eqJ} %
J_{T}=\frac{1}{D_\mathrm{Earth}^{2}}\int_{V}\rho_\mathrm{DM}^{2}(r)dV=\\ \frac{1}{D_\mathrm{Earth}^{2}}\frac{M_\mathrm{sub} \,c^3_{200}}{[f(c_{200})]^{2}}\frac{200\,\rho_\mathrm{crit}}{9}\left(1-\frac{1}{(1+(\frac{R_\mathrm{t}}{R_\mathrm{s}})^3)}\right),  \end{multline} %%

where $D_\mathrm{Earth}$ is the distance from the Earth to the centre of the subhalo, $f(x)=\ln(1+x)-x/(1+x)$, $r$ is the Galactocentric distance inside it, $c_{200} = c_{200}(M_\mathrm{sub}, x_\mathrm{sub})$ is the concentration model, for which we will keep using the one in \citet{moline} for subhaloes as stated in Section~\ref{sec:concvl}, $R_\mathrm{t}$ is the subhalo tidal radius, $R_\mathrm{s}$ is the subhalo scale radius and $\rho_\mathrm{crit}=275.027\,h^{2}\,\mathrm{M_\odot}/\text{kpc}^{3}$ is the critical density of the Universe. 

Equation~\ref{eqJ} implicitly assumes NFW density profiles for the DM distribution inside the subhalo. Yet, it is well known that subhaloes exhibit truncated NFW profiles instead due to tidal stripping \citep{2018MNRAS.474.3043V, 2020MNRAS.491.4591E, 2023MNRAS.518...93A, 2022arXiv220700604S}. 
Note that this is solved in Equation~\ref{eqJ} by integrating the J-factor only up to $R_\mathrm{t}$.  % 

\begin{figure*}%[H]
\centering
\includegraphics[width=\columnwidth]{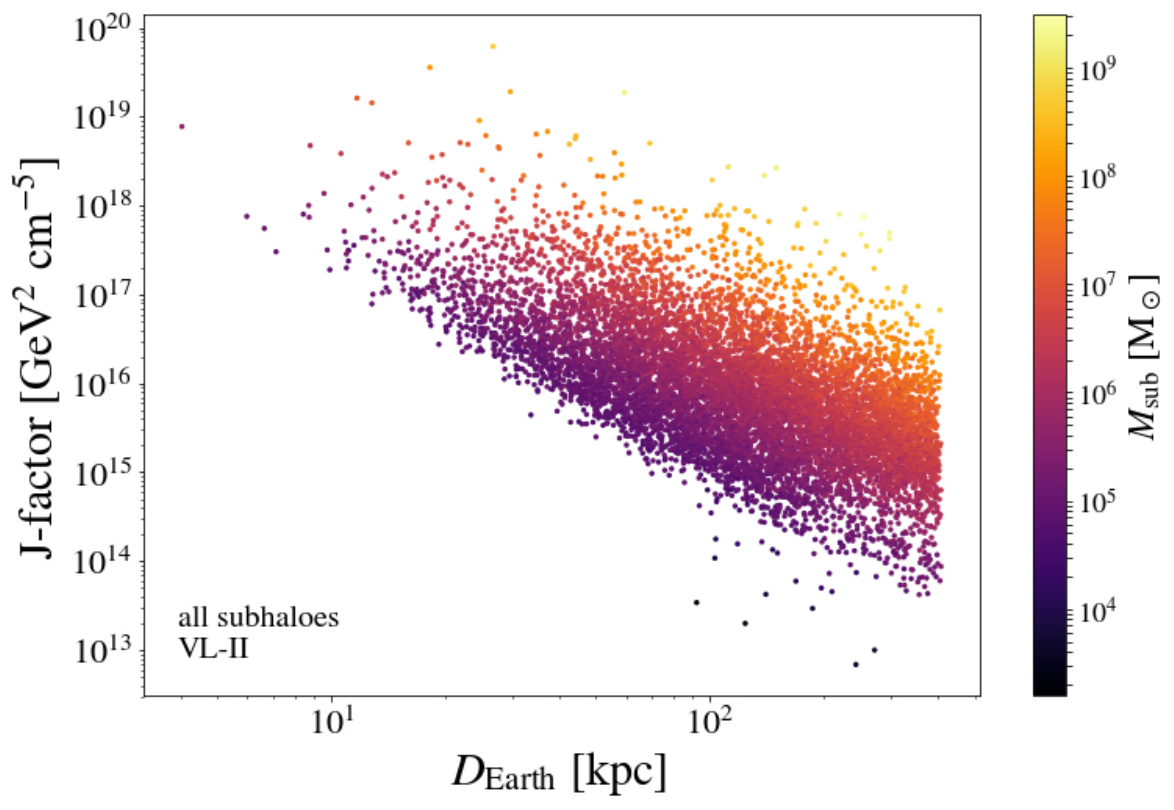}
\includegraphics[width=\columnwidth]{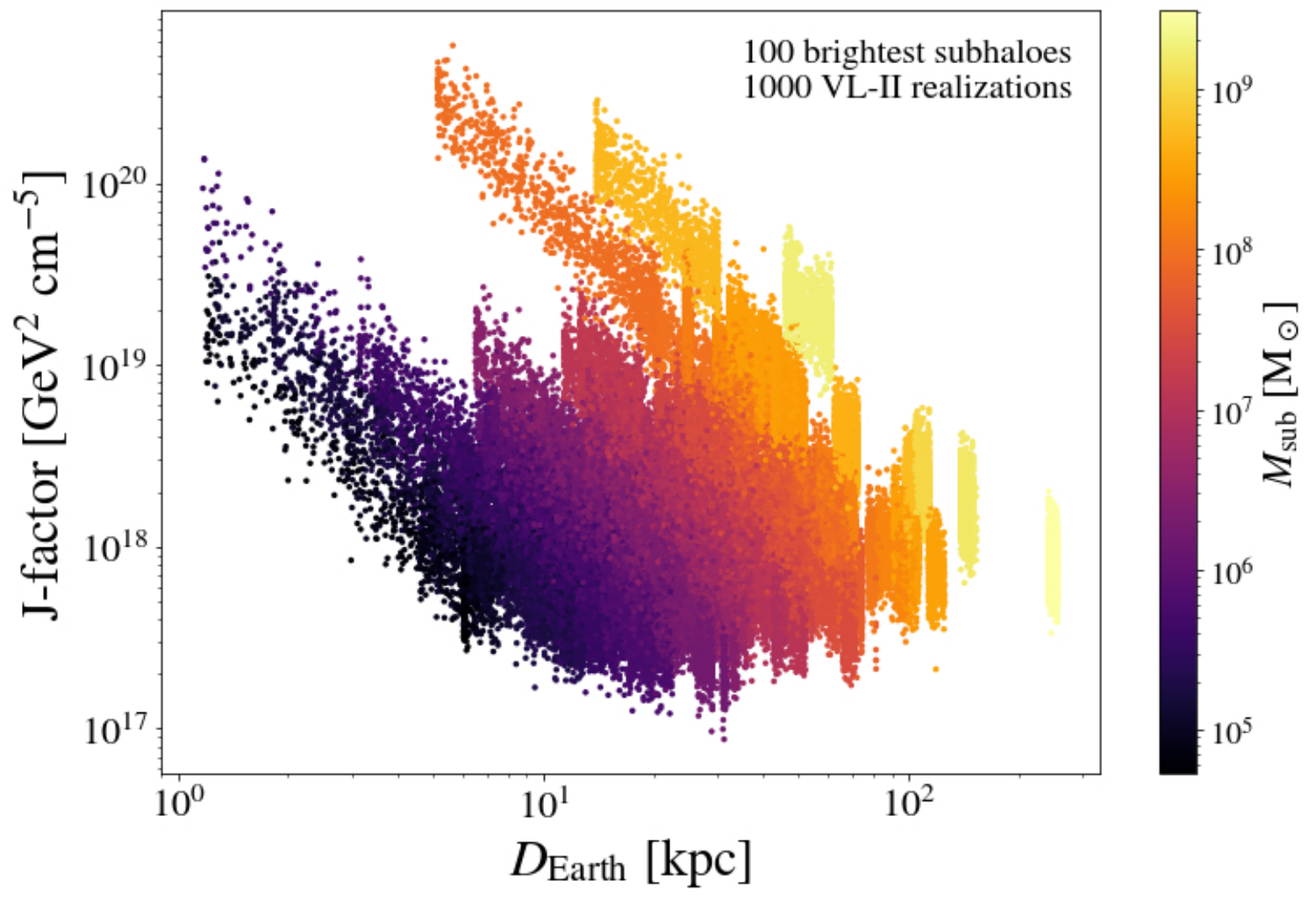} %
\includegraphics[width=\columnwidth]{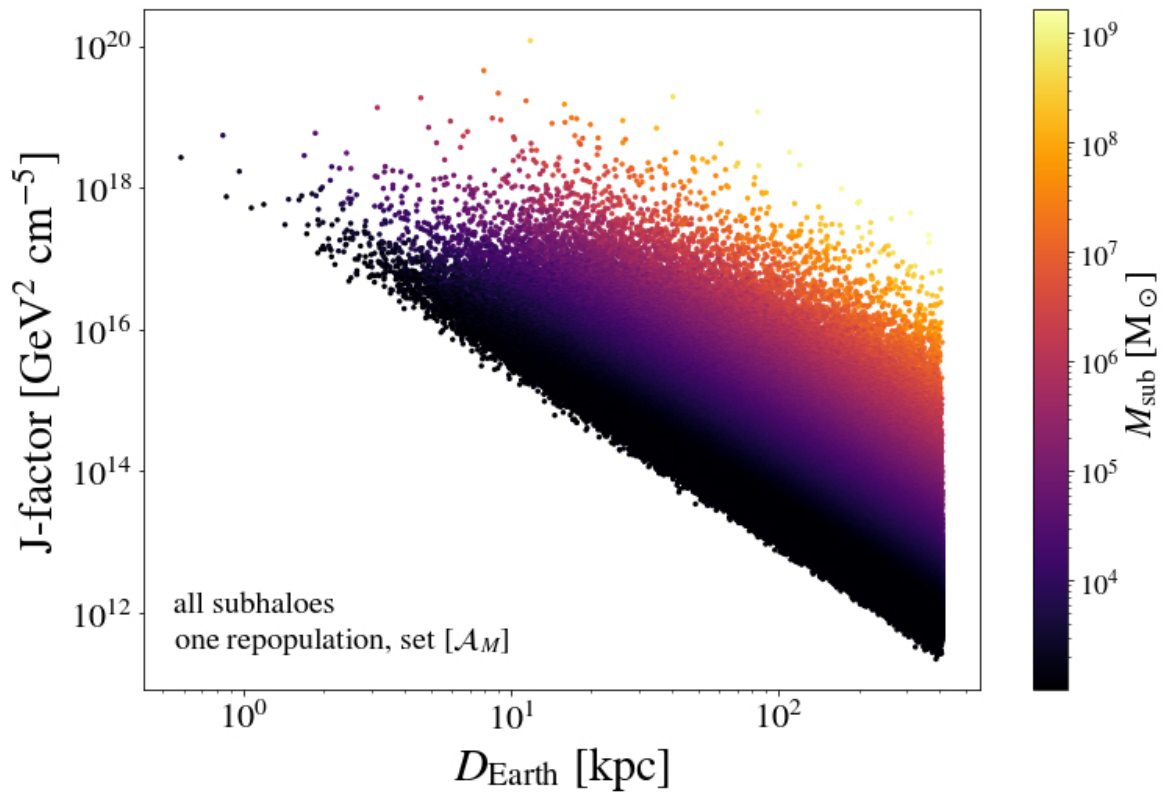}
\includegraphics[width=\columnwidth]{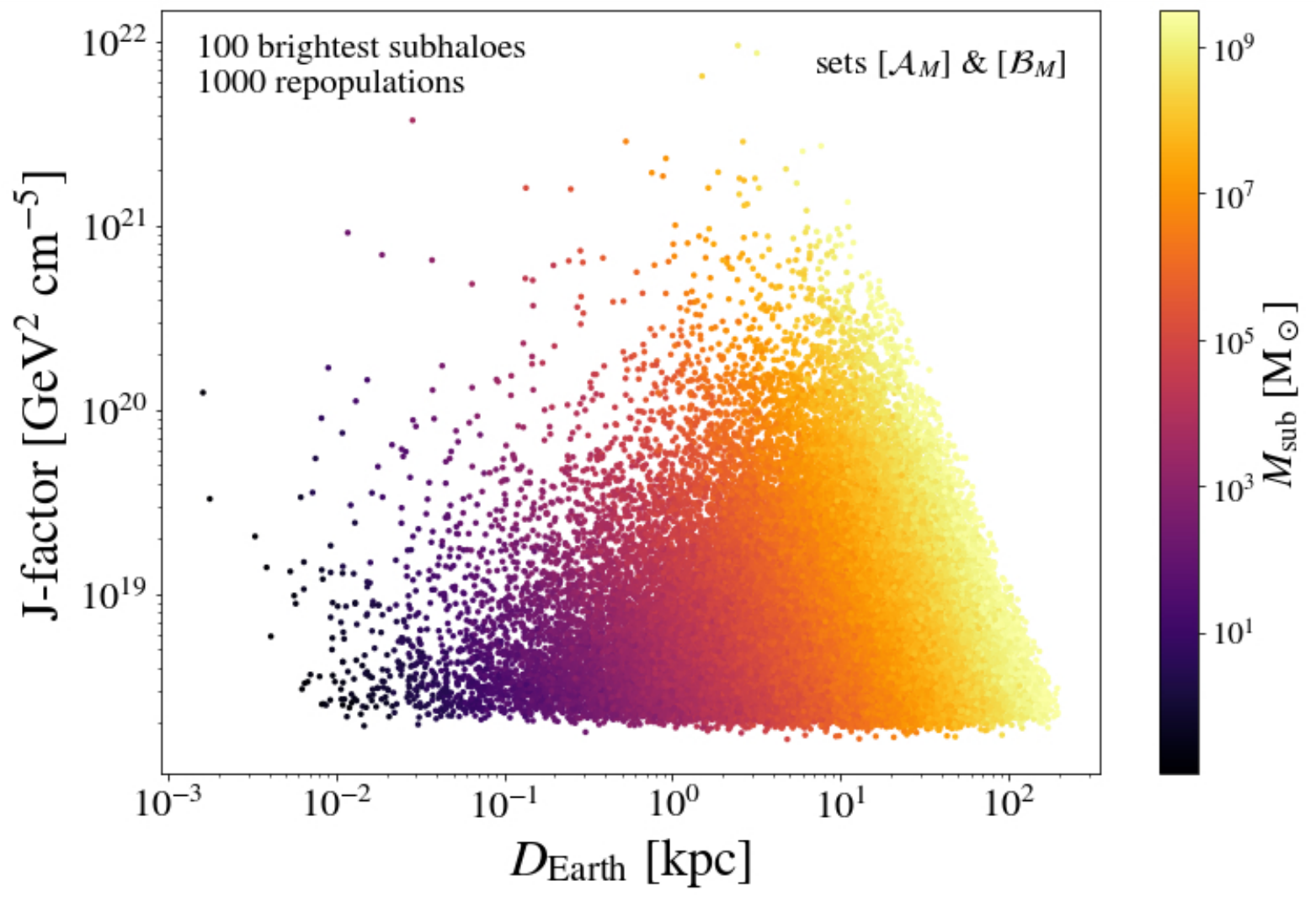}%
\caption{J-factors calculated from subhalo masses by means of Equation~\ref{eqJ}, in both VL-II (top panels) and our repopulations (bottom panels). The color represents the subhalo mass in all cases. Top left: for all subhaloes in VL-II in a single realization. We place the Earth in a random position at 8.5 kpc from the GC in the simulation. Bottom left:  for all subhaloes in a single repopulation, down to $10^3 \mathrm{M_\odot}$ (set $[\mathcal{A}_M]$).  
Top right: for 1000 realizations and the 100 brightest subhaloes in VL-II. The distance to the GC is fixed for a subhalo in the original simulation, yet the distance to the Earth varies as a result of locating it at different places in the original simulation, always at a Galactocentric distance of 8.5 kpc. Bottom right: for the 100 brightest subhaloes in 1000 repopulations of the simulation  down to $10^{-1} \mathrm{M_\odot}$ (sets $[\mathcal{A}_M]$ and $[\mathcal{B}_M]$). 
}
\label{fig:jfac1}
\end{figure*}

Given that the J-factor depends on the distance to the Earth, but VL-II does not place our planet in any specific position, we can locate the observer wherever we want in the simulation. This allows us to perform many realizations by placing the Earth in different positions -- just keeping the distance to the GC constant and equal to 8.5 kpc. We do so and obtain the J-factor of the subhaloes in each realization using Equation~\ref{eqJ}.   %
Subhalo J-factors computed this way are shown in Fig.~\ref{fig:jfac1}, both for the original VL-II, in the top panels, and for the repopulations down to lower subhalo masses, in the bottom ones. Left and right panels show, respectively, the J-factors of all subhaloes in a single realization, and the J-factors of the 100 brightest subhaloes in 1000 repopulations. 
Note that the repopulation exercise provides, statistically, around an order of magnitude larger J-factors compared to the original VL-II simulation. Also, many among the brightest subhaloes in the repopulations are still light subhaloes located at small distances to Earth (these are not present in the original VL-II as they have masses below its mass resolution limit). 

\begin{figure*}%[H]
\centering
\includegraphics[width=\columnwidth]{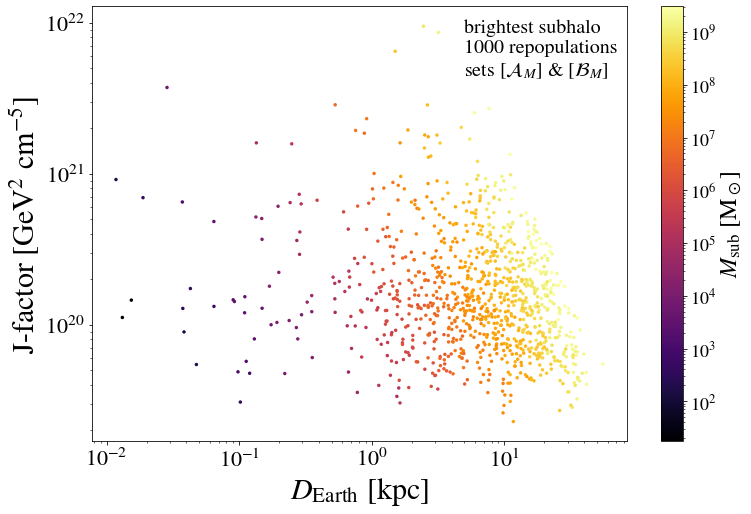} %
\includegraphics[width=\columnwidth]{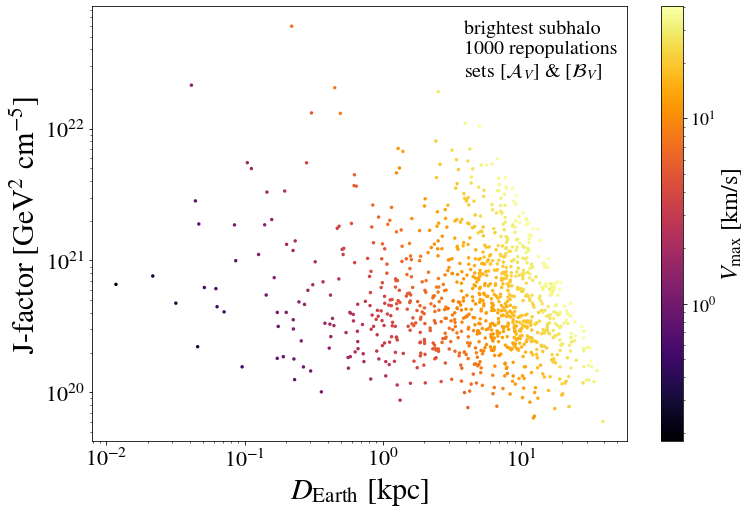} %
\caption{Brightest subhalo in each of the 1000 repopulations of VL-II combining sets $[\mathcal{A}_M]$ and $[\mathcal{B}_M]$ (left) and combining sets $[\mathcal{A}_V]$ and $[\mathcal{B}_V]$ (right). 
 Colours represent the subhalo mass (left) or its $V_\mathrm{max}$ (right).}
\label{fig:jfacmassbri}
\end{figure*}

 The left panel of Fig.~\ref{fig:jfacmassbri} shows the J-factor of the brightest subhalo in each of the 1000 repopulations combining sets $[\mathcal{A}_M]$ and $[\mathcal{B}_M]$. %
 As it can be seen, in several cases the brightest subhalo has a very low mass, well below the original VL-II resolution limit, and is located at just a few pc from Earth.

\subsection{J-factors based on $V_\mathrm{max}$}

The total, integrated J-factor can also be obtained in terms of subhalo concentration and velocity with the following expression~\citep{moline}:

\begin{multline} \label{eqJv} J_{T}=\frac{1}{D_\mathrm{Earth}^{2}}\int_{V}\rho_\mathrm{DM}^{2}(r)dV=\\ \frac{1}{D_\mathrm{Earth}^2} \left(\frac{2.163}{f(2.163)}\right)^2\frac{2.163 H_0}{12 \pi G^2}\sqrt{\frac{c_\mathrm{V} (V_\mathrm{max})}{2}} V^3_\mathrm{max},  \end{multline}

where $c_\mathrm{V} = c_\mathrm{V}(V_\mathrm{max}, x_\mathrm{sub})$ is the velocity-based concentration model, for which we adopt the one in \citet{moline} for subhaloes as stated in Section~\ref{sec:cvmax}, and $\rho_\mathrm{crit}$ %
is the critical density of the Universe. Again, we note that we implicitly assume an NFW profile when using this expression.

\begin{figure*}%[H]
\centering
\includegraphics[width=\columnwidth]{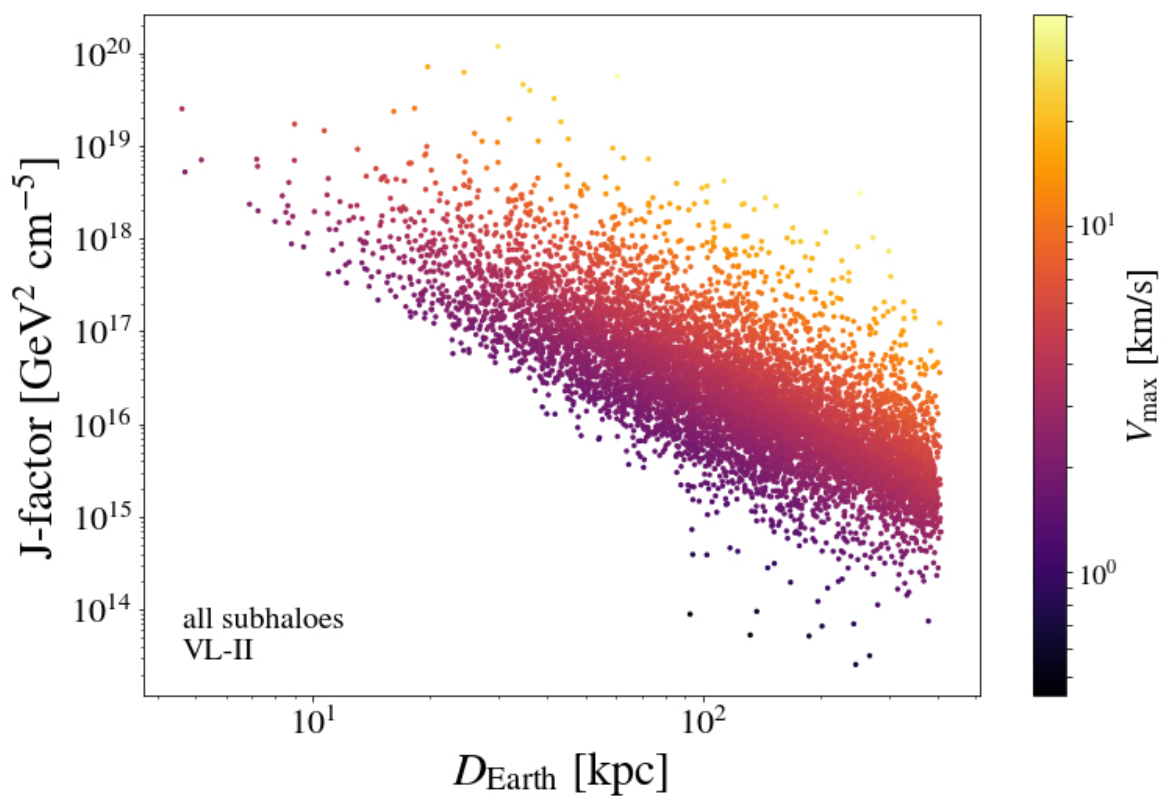}
\includegraphics[width=\columnwidth]{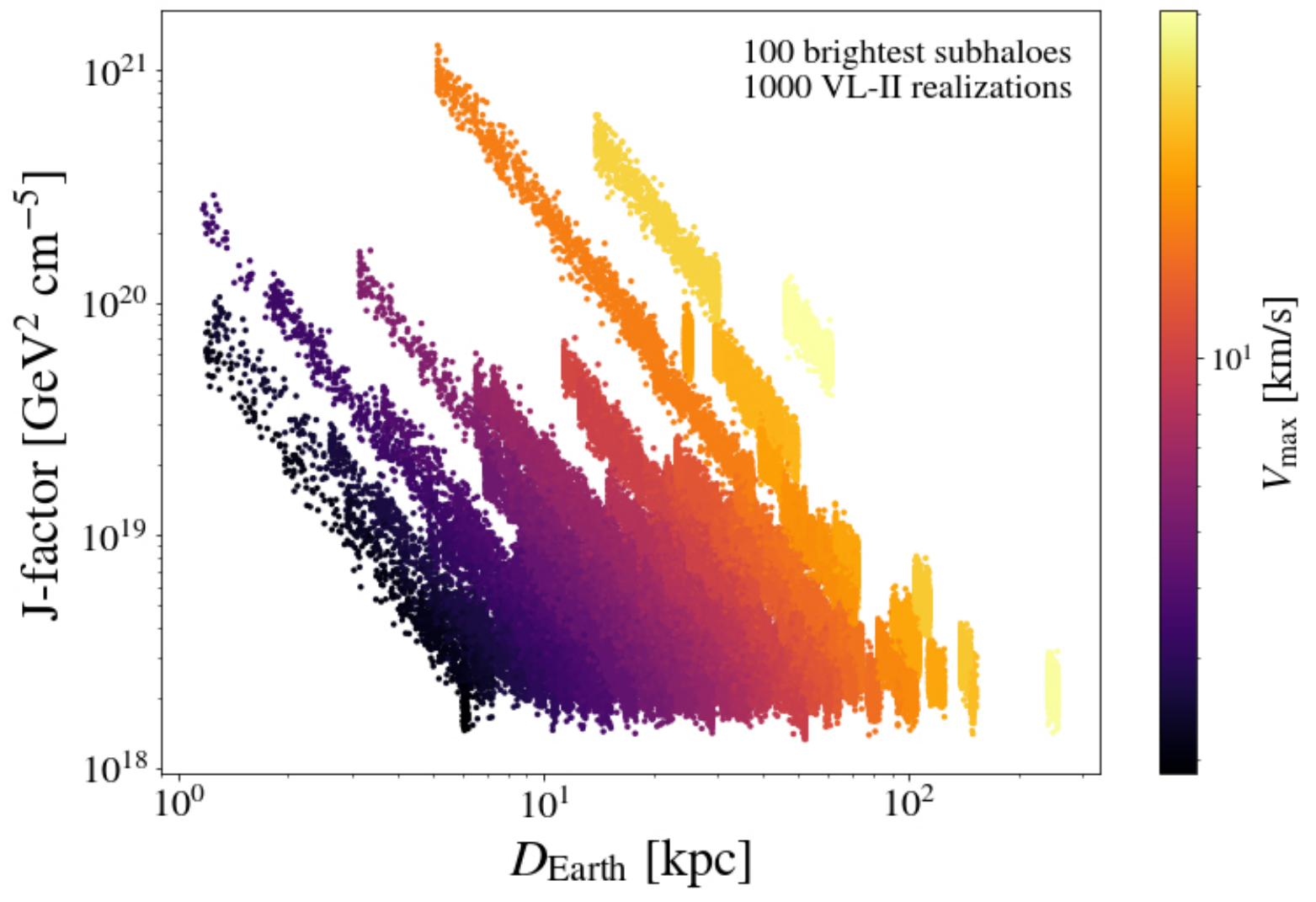} %
\includegraphics[width=\columnwidth]{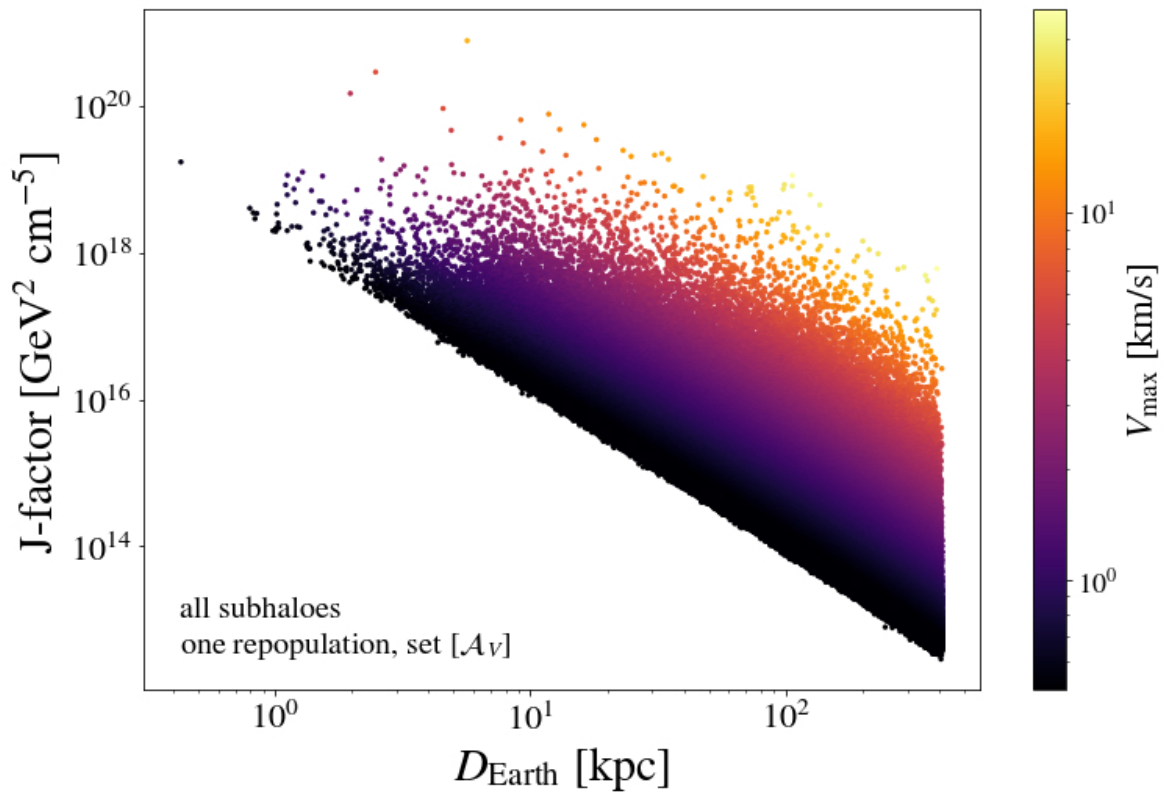}
\includegraphics[width=\columnwidth]{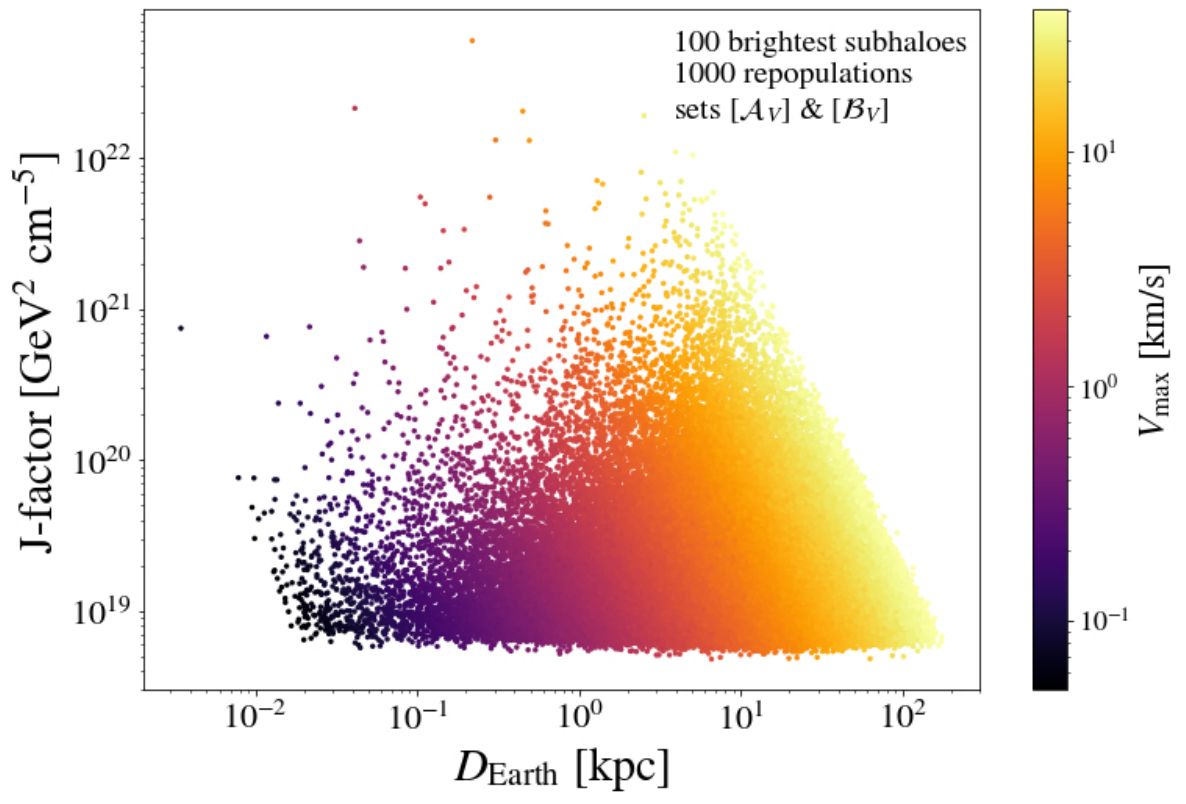} 
\caption{J-factors calculated from subhalo masses by means of Equation~\ref{eqJv}, in both VL-II (top panels) and our repopulations (bottom panels). The color represents the subhalo maximum circular velocity in all cases. Top left: for all subhaloes in VL-II in a single realization. We place the Earth in a random position at 8.5 kpc from the GC in the simulation. Bottom left:  for all subhaloes in a single repopulation, down to $0.5$ km/s (set $[\mathcal{A}_V]$). 
Top right: for 1000 realizations and the 100 brightest subhaloes in VL-II. The distance to the GC is fixed for a subhalo in the original simulation, yet the distance to the Earth varies as a result of locating it at different places in the original simulation, always at a Galactocentric distance of 8.5 kpc. Bottom right: for the 100 brightest subhaloes in 1000 repopulations of the simulation down to $0.05$ km/s (sets $[\mathcal{A}_V]$ and $[\mathcal{B}_V]$). 
}
\label{fig:jfac3}
\end{figure*}

As in the mass case, we place the Earth in the simulation at a random position 8.5 kpc away from the GC in order to derive subhalo J-factors. 
The results are shown in the four panels of Fig.~\ref{fig:jfac3} organized in the same way they were shown for the case of mass-based J-factors. 
Again, we obtain an increase of around one order of magnitude in the case of the repopulation to lower velocities compared to the J-factors in the original simulation. Also, many low-velocity subhaloes are among the brightest ones, reaffirming once more their potential relevance for gamma-ray DM searches. 
The brightest subhalo in each repopulation, combining sets $[\mathcal{A}_V]$ and $[\mathcal{B}_V]$, %
is shown in the right panel of Fig.~\ref{fig:jfacmassbri}. This figure shows, once again, that it is likely to have as the brightest subhalo in the Galaxy one with a very small $V_\mathrm{max}$ located at just a few pc from Earth.

\subsection{Subhalo angular sizes %
}
\label{sec:angsizeVL}

As said, it can be particularly useful to investigate the angular extension subtended by subhaloes in the sky as seen from Earth, as this may have important implications for designing current or future DM search analysis strategies. 
This is the solid angle $\Omega$ subtended by the subhalo, which is the two-dimensional angle in three-dimensional space that an object subtends at a point and measures how large the object appears to the observer who is looking from that point. 
In practice, we will work with the projection $\theta$ of this angle, knowing that: 
\begin{equation}\label{angsize}
\ \ \ \ \ \ \theta = \atan \dfrac{R_\mathrm{s}(M_\mathrm{sub}) }{ D_\mathrm{Earth}}; \ \ \ \ \  
\Omega = 2 \pi (1- \cos \theta). \end{equation}
Here $R_\mathrm{s}$ is the scale radius of the subhalo. We recall that for NFW haloes $90\%$ of the total J-factor is originated inside $R_\mathrm{s}$. However, in the case of subhaloes, since the subhalo profile is a truncated NFW, more than the mentioned $90\%$ of the annihilation flux will be actually originated within this $R_\mathrm{s}$ \citep{2011JCAP...12..011S}.  
This means that $R_\mathrm{s}$ is a good estimate of the angular size as it would be seen in gamma rays.\footnote{Note that choosing $R_\mathrm{max}$ instead of $R_\mathrm{s}$ would lead to larger angular sizes, since for an NFW profile $R_\mathrm{max} = 2.163 R_\mathrm{s}$. %
} 

\begin{figure}%[!ht]
\centering
\includegraphics[width=\columnwidth]{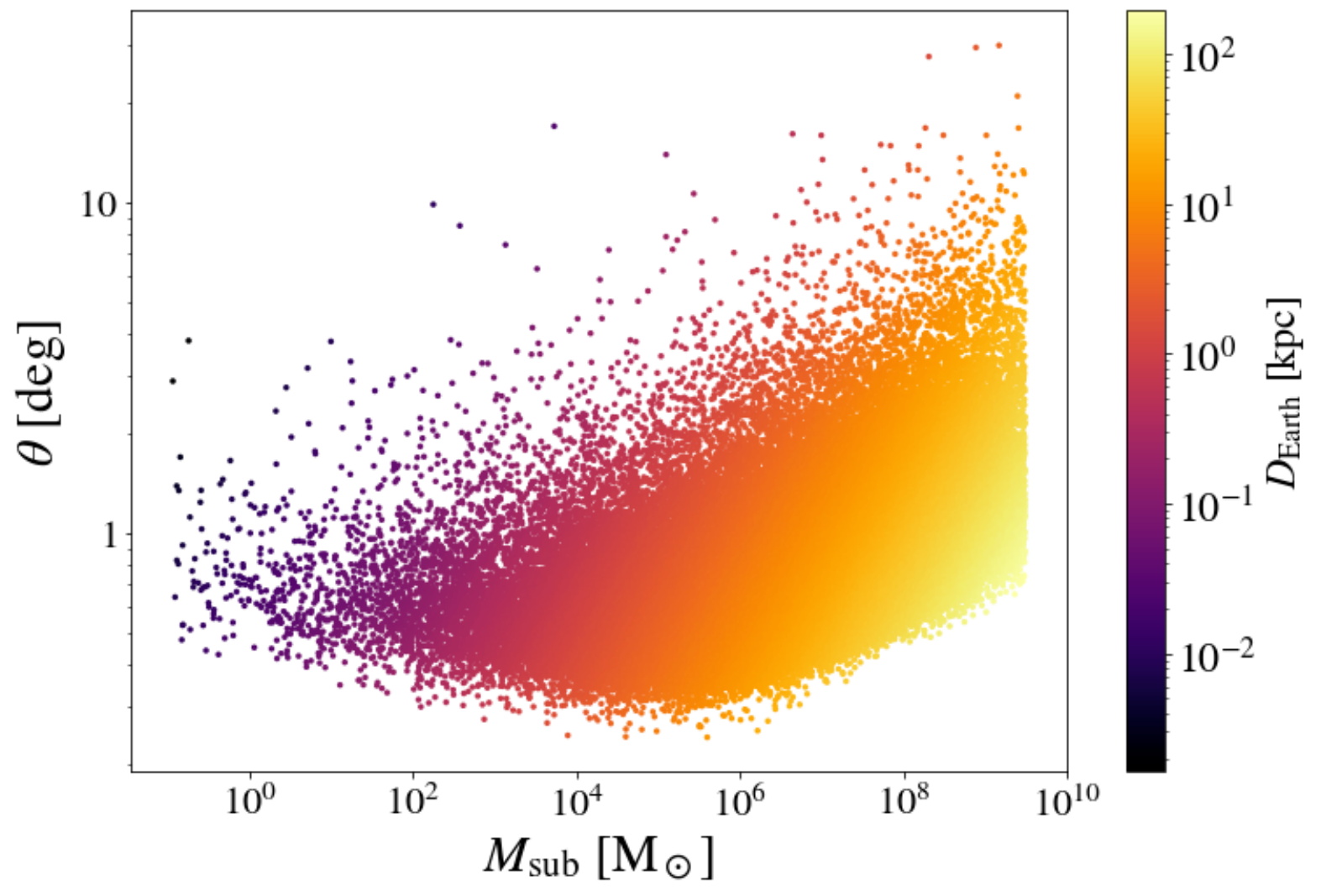} %
\caption{Angular extension, in degrees, of subhaloes as seen from Earth versus their mass. %
The colour refers to the distance to the Earth. %
Shown are the 100 brightest subhaloes in each of the 1000 repopulations of the simulation combining sets $[\mathcal{A}_M]$ and $[\mathcal{B}_M]$. %
}
\label{fig:ang_extension}
\end{figure}

Figure~\ref{fig:ang_extension} depicts the angular size of the 100 brightest subhaloes in 1000 repopulations, combining sets $[\mathcal{A}_M]$ and $[\mathcal{B}_M]$. 
We observe that the annihilation signal from most of the brightest subhaloes is expected to be spatially extended, with typical angular sizes of a few tenths to few degrees, and even $\mathcal{O}$(10 deg) in some cases. This might have very important implications for gamma-ray DM searches since, for instance, none of the \textit{Fermi}-LAT unidentified source analyses has shown any preference for
a spatially-extended signal over a point-like model~\citep{2019JCAP...11..045C}. Nevertheless, we note that gamma-ray telescopes might still observe these subhaloes as point-like sources due to instrumental limitations (sensitivity, angular resolution...); see e.g. ~\citet{2020PhRvD.102j3010D, 2022PhRvD.105h3006C}. Further work and detailed analyses may be needed for each particular instrument to clarify this matter, that should include proper, realistic spatial templates of the subhalo annihilation emission following our findings.

\subsection{Comparison between J-factors}\label{sec:jfaccomp}

\begin{figure}%[H]
\centering
\includegraphics[width=\columnwidth]{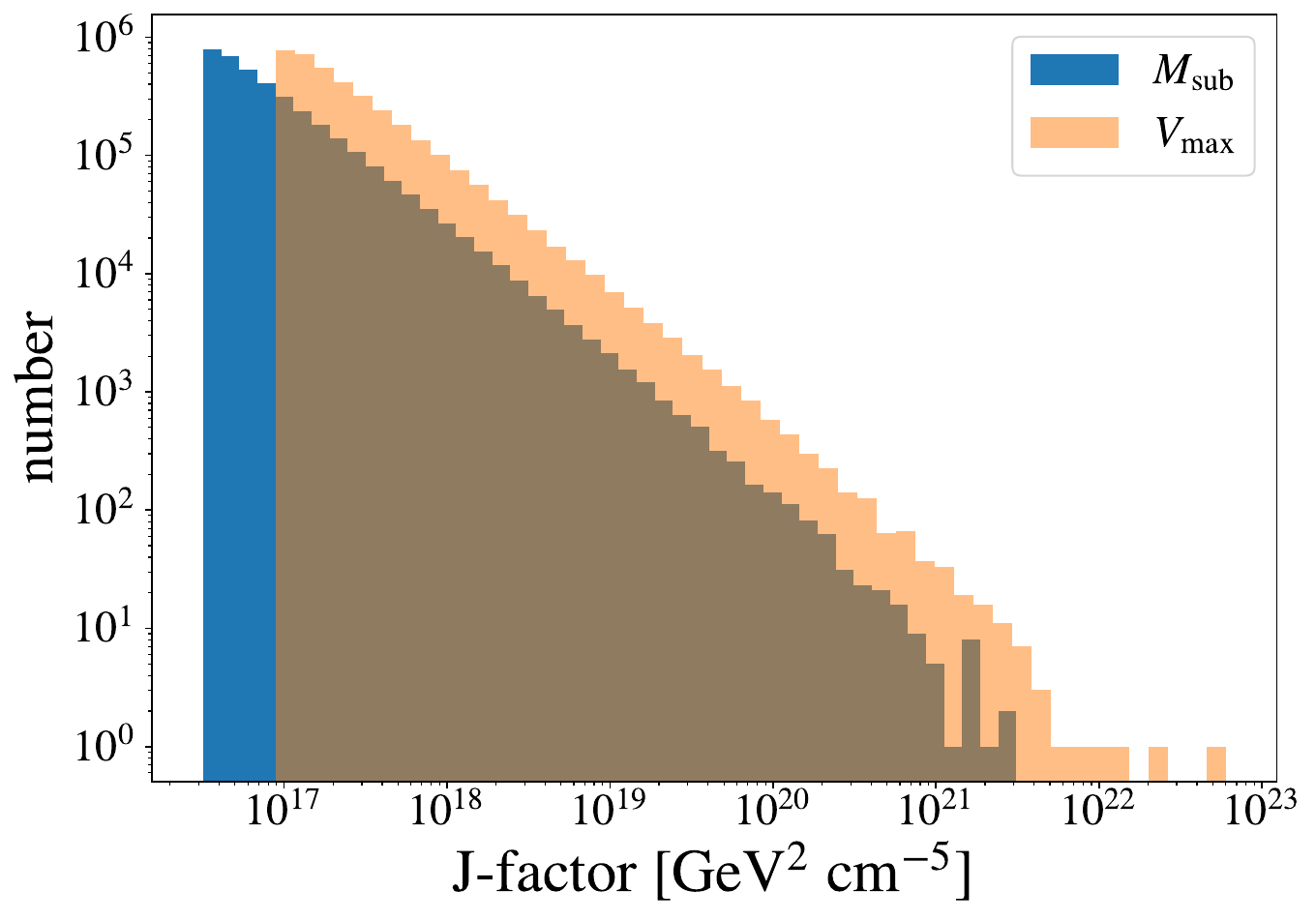}
\caption{%
Comparison of J-factors obtained by either combining sets $[\mathcal{A}_M]$ and $[\mathcal{B}_M]$ %$M_\mathrm{sub}$ 
(Equation \ref{eqJ}), in blue, and combining sets $[\mathcal{A}_V]$ and $[\mathcal{B}_V]$ %$V_\mathrm{max}$ 
(Equation \ref{eqJv}), in orange. The data correspond to the 10000 brightest subhaloes of 1000 repopulations. %
}
\label{fig:compah}
\end{figure}

It is now time to compare the J-factors obtained by means of subhalo masses (Equation~\ref{eqJ} and Figure~\ref{fig:jfac1}) with those derived from subhalo maximum circular velocities (Equation~\ref{eqJv} and Figure~\ref{fig:jfac3}).
This is shown in the form of histograms in Fig.~\ref{fig:compah}. As it can be seen, we obtain  brighter subhaloes when repopulating using $V_\mathrm{max}$, the brightest ones in the simulation reaching up to one order of magnitude larger values compared to the brightest subhaloes whose J-factors were derived from masses.

Reasons for this apparent discrepancy are multiple. Most notably, in each case we adopt a different SRD (Equations~\ref{eq:srdmass} and~\ref{eq:cosmic}), and the integration of the J-factor for the case of using either mass or velocity is done in different ways. On one hand, the J-factor obtained using $M_\mathrm{sub}$ relies on $R_\mathrm{t}$, which shrinks due to tidal stripping, while the one calculated with $V_\mathrm{max}$ makes use of $R_\mathrm{max}$ instead. We recall that the latter is about twice $R_\mathrm{s}$ for NFW profiles, which could generate a noticeable difference in the direction of that seen in Fig.~\ref{fig:compah}. %
Integrating up to the same radius in both cases would probably lead to a fairer comparison, however we prefer not to mix mass and velocity variables together here. We show the outcome of adopting the same angular radius for both cases in Appendix~\ref{sec:intjfac}.

\section{Conclusions and outlook}\label{sec:conclu}

$N$-body cosmological simulations are computationally expensive and they are prone to both mass and spatial resolution limits. Conversely, multiple realizations of those same simulations are much cheaper to perform. Even more, a full characterization of the original simulations may enable to extrapolate their results beyond the original resolution limits, this way reaching much lower (sub)halo masses. The latter can be particularly useful, not only because lower (sub)halo masses than those resolved in current simulations are indeed expected in $\Lambda$CDM cosmology, but also because it allows for studies for which low-mass structures can be especially relevant.

With this in mind, in this work we have characterized the subhalo population in the Via Lactea II (VL-II) N-body cosmological simulation \citep{Diemand2008}. Despite the years, VL-II still represents a unique, state-of-the-art N-body simulation of a MW-size halo. 
Indeed, VL-II possesses a superb resolution compared to other, more recent simulations in the market, i.e. a particle mass of $\sim 4 \times 10^3 \mathrm{M_\odot}$, and about $10^9$ particles. Mass resolution is critical for our purposes, since in this work we were mainly interested in reaching extremely low subhalo masses and, thus, any extrapolations from VL-II data to lower masses are expected to be more robust compared to other simulations. Although the VL-II cosmological parameters differ significantly from current ones (WMAP3 vs Planck), their impact is expected to be not significant for our subhalo repopulation purposes.\footnote{Indeed, it is very likely that the most recent cosmological parameters would lead to an increase of the subhalo abundance, thus our study can be considered conservative in that sense.}

The first step was to use VL-II data to study in detail the subhalo mass/velocity functions (SHMF/SHVF; sections \ref{sec:vlshmf} and \ref{sec:vlshvf}) and radial distribution (SRD; sections \ref{sec:srdm} and \ref{sec:srdv}) in a galaxy like our own. We also calculated subhalo concentrations applying the recipes given in \citet{moline}. %
Once the subhalo population in the original simulations was fully characterized via the mentioned quantities, in section \ref{sec:repopu} we repopulated VL-II with low-mass (low-velocity) subhaloes, indeed including subhaloes with masses (velocities) $\sim$7 ($\sim$2) orders of magnitude below the original VL-II resolution mass (velocity) limit. We did so by assuming reasonable extrapolations of the SHMF/SHVF, SRD and concentrations as expected in $\Lambda$CDM.
We then analyzed, still in section~\ref{sec:repopu}, the potential role of these tiny subhaloes for indirect DM searches, by comparing them to more massive, well resolved subhaloes in the parent simulation. In order to do so, we calculated the astrophysical J-factors -- a measure of the DM annihilation flux -- of all subhaloes in the Galaxy as seen from Earth. In order to obtain statistically meaningful results, we performed 1000 repopulations of this kind and compared them to 1000 realizations of the original simulation (in this case simply varying the %
Earth's position but keeping its Galactocentric distance fixed to 8.5 kpc).%}

All this repopulation work showed the viability of light yet close subhaloes as excellent DM targets. More precisely, some of the main conclusions of this work for DM search related studies are: 
i) low-mass/velocity subhaloes below $\mathrm{M_{cut}} = 5 \times 10^6 \mathrm{M_\odot} $ ($\mathrm{V_{cut}} = 4 $ km/s) %}
and close to Earth can be as bright as more massive subhaloes above this $\mathrm{M_{cut}} (\mathrm{V_{cut}}$) located farther away (Figs.~\ref{fig:jfac1} and \ref{fig:jfac3}); ii) very low-mass/velocity ($\sim 10^{-1} \mathrm{M_\odot} / \sim 0.05$ km/s) subhaloes are still able to be competitive in terms of their J-factors; iii) in a few ($\sim 1\%$) repopulations 
the brightest subhalo is indeed a tiny subhalo with mass (velocity) below $\sim 10^{3} \mathrm{M_\odot}$ ($\sim 0.5$ km/s)  (Fig.~\ref{fig:jfacmassbri}). %

We notice that by means of the repopulation procedure we obtain J-factors that are around one order of magnitude larger with respect to those given by the original simulation. This increase occurs at all masses/velocities. This is mainly due to the radial repopulation. %
For example, in our repopulations we may end up placing some $\mathcal{O}(10^{8} \mathrm{M_\odot})$ subhaloes closer than in the original simulation -- because the adopted SRD allows this to happen -- which would in turn produce a very bright subhalo. Although this effect only occurs for a very small fraction of the repopulated subhaloes in each realization, it has an important impact on J-factor final results, since we particularly focus on the brightest subhaloes in this work. Nevertheless, the comparison of a single realization of VL-II versus a single repopulation shows that the general behaviour above $\mathrm{M_{cut}}$/$\mathrm{V_{cut}}$ is similar. 
Besides, the J-factors calculated with $V_\mathrm{max}$ are up to about one order of magnitude larger than those obtained using $M_\mathrm{sub}$ instead (Fig.~\ref{fig:compah}). We recall that this comparison is not one-to-one though, as it implicitly implies the use of different SRDs and integration angles for the computation of the J-factors in either case. 

Another important result, obtained in section~\ref{sec:angsizeVL}, is that the brightest subhaloes in our simulations are always expected to be extended sources ($\mathcal{O} (1$ deg); see Fig.~\ref{fig:ang_extension}) for current gamma-ray telescopes. This may have important implications for planning the most optimal gamma-ray DM search and/or data analysis strategy~\citep{2014PhRvD..89a6014B}, as well as to differentiate between DM and other astrophysical sources. We note, however, that gamma-ray telescopes might still observe extended subhaloes as point-like, depending of their precise instrumental performances; see e.g. ~\citet{2020PhRvD.102j3010D, 2022PhRvD.105h3006C} for the particular case of Fermi LAT. In any case, these results claim for the use of realistic spatial templates for the subhalo emission in this type of DM analyses.

We recall that results from an earlier version of the repopulation algorithm here presented has already been to set DM constraints, by comparing our simulation predictions to the number of unidentified gamma-ray sources in the \textit{Fermi}-LAT catalogs in~\citet{2019JCAP...07..020C}. A similar work was also done for the Cherenkov Telescope Array in~\citet{2021PDU....3200845C}. Our study of the sensitivity of the \textit{Fermi}-LAT to extended subhaloes~in~\citet{2019JCAP...11..045C, 2022PhRvD.105h3006C} also made use of repopulation predictions. Other possible applications of our low-mass subhalo repopulation machinery include, for instance, its use for studies aimed at setting more robust constraints on DM particle properties; %
a more precise computation of the so-called subhalo annihilation boost factor \citep[see, e.g.,][]{moline, Ando:2019xlm}; and a further optimization of both observation and data analysis strategies for the search of DM subhalo sources with current or future gamma-ray telescopes.

In the future, we will repopulate newer simulations based on the most updated cosmological parameters, both for the cases of DM-only and hydrodynamical simulations, such as Caterpillar \citep{2016ApJ...818...10G}, Phat-Elvis \citep{2016arXiv160706479F}, Apostle \citep{2017MNRAS.467..179G}, Auriga \citep{2019MNRAS.486.4790B} or Uchuu \citep{Ishiyama2021}. This should provide a more accurate and nearly definitive answer to the role of low-mass subhaloes for DM searches. 
On the other hand, this same methodology could also be applied to studies of field haloes in large-scale structure simulations. 
In all these cases, we will pay special attention to the role of the smallest structures for indirect DM searches and their viability as competitive DM targets. Yet, we note that, although very powerful and cheap, all this repopulation work also shows the need for even higher-resolution numerical simulations to avoid the need of extrapolations of the relevant quantities down to low subhalo masses.

%%%%%%%%%%%%%%%%%%%%%%%%%%%%%%%

\section*{Acknowledgements}

% The Acknowledgements section is not numbered. Here you can thank helpful
% colleagues, acknowledge funding agencies, telescopes and facilities used etc.
% Try to keep it short.

AAS and MASC would like to thank Javier Coronado-Bl\'azquez and \'Angeles Molin\'e for useful discussions and comments, as well as David Gordo and Roi Naveiro Flores for their help with the bootstrapping technique. We would also like to thank Andrea Klein for developing an earlier version of the code used for repopulating. Finally, the authors acknowledge the anonymous referee for their insightful feedback which helped to improve the article.

The work of AAS and MASC %
was supported by the grants PID2021-125331NB-I00 and CEX2020-001007-S, both funded by MCIN/AEI/10.13039/501100011033 and by ``ERDF A way of making Europe''. The work of AAS was also supported by the Spanish Ministry of Science and Innovation through the grant FPI-UAM 2018.

Furthermore, initial numerical computations were made using the Hydra cluster at the Instituto de F\'isica Te\'orica (Universidad Aut\'onoma de Madrid) and the computational resources of the High Energy Physics Group at Universidad Complutense de Madrid.

This research made use of Python, along with community-developed or maintained software packages, including IPython \citep{Ipython_paper}, Matplotlib \citep{Matplotlib_paper}, NumPy \citep{Numpy_paper} and SciPy \citep{scipy_paper}. This work made use of NASA’s Astrophysics Data System for bibliographic information.

%%%%%%%%%%%%%%%%%%%%%%%%%%%%
\section*{Data availability}
The data underlying this article are publicly available in the website of the DAMASCO group at \url{https://projects.ift.uam-csic.es/damasco/?page_id=831}.

%%%%%%%%%%%%%%%%%%%%%%%%%%%%%%%%%%%%%%%%%%%%%%%%%%

%%%%%%%%%%%%%%%%%%%% REFERENCES %%%%%%%%%%%%%%%%%%

\bibliographystyle{mnras}
\bibliography{References}

\begin{thebibliography}{}
\makeatletter
\relax
\def\mn@urlcharsother{\let\do\@makeother \do\$\do\&\do\#\do\^\do\_\do\%\do\~}
\def\mn@doi{\begingroup\mn@urlcharsother \@ifnextchar [ {\mn@doi@}
  {\mn@doi@[]}}
\def\mn@doi@[#1]#2{\def\@tempa{#1}\ifx\@tempa\@empty \href
  {http://dx.doi.org/#2} {doi:#2}\else \href {http://dx.doi.org/#2} {#1}\fi
  \endgroup}
\def\mn@eprint#1#2{\mn@eprint@#1:#2::\@nil}
\def\mn@eprint@arXiv#1{\href {http://arxiv.org/abs/#1} {{\tt arXiv:#1}}}
\def\mn@eprint@dblp#1{\href {http://dblp.uni-trier.de/rec/bibtex/#1.xml}
  {dblp:#1}}
\def\mn@eprint@#1:#2:#3:#4\@nil{\def\@tempa {#1}\def\@tempb {#2}\def\@tempc
  {#3}\ifx \@tempc \@empty \let \@tempc \@tempb \let \@tempb \@tempa \fi \ifx
  \@tempb \@empty \def\@tempb {arXiv}\fi \@ifundefined
  {mn@eprint@\@tempb}{\@tempb:\@tempc}{\expandafter \expandafter \csname
  mn@eprint@\@tempb\endcsname \expandafter{\@tempc}}}

\bibitem[\protect\citeauthoryear{{Ackermann} et~al.,}{{Ackermann}
  et~al.}{2015}]{2015PhRvL.115w1301A}
{Ackermann} M.,  et~al., 2015, \mn@doi [\prl] {10.1103/PhysRevLett.115.231301},
  \href {https://ui.adsabs.harvard.edu/abs/2015PhRvL.115w1301A} {115, 231301}

\bibitem[\protect\citeauthoryear{{Aghanim} et~al.,}{{Aghanim}
  et~al.}{2020}]{2020A&A...641A...6P}
{Aghanim} N.,  et~al., 2020, \mn@doi [\aap] {10.1051/0004-6361/201833910},
  \href {https://ui.adsabs.harvard.edu/abs/2020A&A...641A...6P} {641, A6}

\bibitem[\protect\citeauthoryear{{Aguirre-Santaella}, {Gammaldi},
  {S{\'a}nchez-Conde}  \& {Nieto}}{{Aguirre-Santaella}
  et~al.}{2020}]{2020JCAP...10..041A}
{Aguirre-Santaella} A.,  {Gammaldi} V.,  {S{\'a}nchez-Conde} M.~A.,   {Nieto}
  D.,  2020, \mn@doi [\jcap] {10.1088/1475-7516/2020/10/041}, \href
  {https://ui.adsabs.harvard.edu/abs/2020JCAP...10..041A} {2020, 041}

\bibitem[\protect\citeauthoryear{{Aguirre-Santaella}, {S{\'a}nchez-Conde},
  {Ogiya}, {St{\"u}cker}  \& {Angulo}}{{Aguirre-Santaella}
  et~al.}{2023}]{2023MNRAS.518...93A}
{Aguirre-Santaella} A.,  {S{\'a}nchez-Conde} M.~A.,  {Ogiya} G.,  {St{\"u}cker}
  J.,   {Angulo} R.~E.,  2023, \mn@doi [\ MNRAS] {10.1093/mnras/stac2921},
  \href {https://ui.adsabs.harvard.edu/abs/2023MNRAS.518...93A} {518, 93}

\bibitem[\protect\citeauthoryear{Albert et~al.}{Albert et~al.}{2017}]{jfac1}
Albert A.,  et~al., 2017, \mn@doi [Astrophys. J.]
  {10.3847/1538-4357/834/2/110}, 834, 110

\bibitem[\protect\citeauthoryear{{Amorisco}}{{Amorisco}}{2021}]{2021arXiv211101148A}
{Amorisco} N.~C.,  2021, arXiv e-prints, \href
  {https://ui.adsabs.harvard.edu/abs/2021arXiv211101148A} {p. arXiv:2111.01148}

\bibitem[\protect\citeauthoryear{Ando, Ishiyama  \& Hiroshima}{Ando
  et~al.}{2019}]{Ando:2019xlm}
Ando S.,  Ishiyama T.,   Hiroshima N.,  2019, \mn@doi [Galaxies]
  {10.3390/galaxies7030068}, 7, 68

\bibitem[\protect\citeauthoryear{{Angulo} \& {Hahn}}{{Angulo} \&
  {Hahn}}{2022}]{2022LRCA....8....1A}
{Angulo} R.~E.,  {Hahn} O.,  2022, \mn@doi [Living Reviews in Computational
  Astrophysics] {10.1007/s41115-021-00013-z}, \href
  {https://ui.adsabs.harvard.edu/abs/2022LRCA....8....1A} {8, 1}

\bibitem[\protect\citeauthoryear{{Angulo}, {Baugh}, {Frenk}  \&
  {Lacey}}{{Angulo} et~al.}{2014}]{2014MNRAS.442.3256A}
{Angulo} R.~E.,  {Baugh} C.~M.,  {Frenk} C.~S.,   {Lacey} C.~G.,  2014, \mn@doi
  [\mnras] {10.1093/mnras/stu1084}, \href
  {https://ui.adsabs.harvard.edu/abs/2014MNRAS.442.3256A} {442, 3256}

\bibitem[\protect\citeauthoryear{Bartels \& Ando}{Bartels \&
  Ando}{2015}]{Bartels:2015uba}
Bartels R.,  Ando S.,  2015, \mn@doi [Phys. Rev.] {10.1103/PhysRevD.92.123508},
  D92, 123508

\bibitem[\protect\citeauthoryear{Bergstr{\"o}m, Ullio  \&
  Buckley}{Bergstr{\"o}m et~al.}{1998}]{BERGSTROM1998137}
Bergstr{\"o}m L.,  Ullio P.,   Buckley J.~H.,  1998, \mn@doi [Astroparticle
  Physics] {https://doi.org/10.1016/S0927-6505(98)00015-2}, 9, 137

\bibitem[\protect\citeauthoryear{{Berlin} \& {Hooper}}{{Berlin} \&
  {Hooper}}{2014}]{2014PhRvD..89a6014B}
{Berlin} A.,  {Hooper} D.,  2014, \mn@doi [\prd] {10.1103/PhysRevD.89.016014},
  \href {https://ui.adsabs.harvard.edu/abs/2014PhRvD..89a6014B} {89, 016014}

\bibitem[\protect\citeauthoryear{Bertone}{Bertone}{2010}]{Bertone10}
Bertone G.,  2010, \mn@doi [Nature 468] {10.1038/nature09509}

\bibitem[\protect\citeauthoryear{{Bertone} \& {Hooper}}{{Bertone} \&
  {Hooper}}{2018}]{2018RvMP...90d5002B}
{Bertone} G.,  {Hooper} D.,  2018, \mn@doi [Rev. Mod. Phys.]
  {10.1103/RevModPhys.90.045002}, \href
  {https://ui.adsabs.harvard.edu/abs/2018RvMP...90d5002B} {90, 045002}

\bibitem[\protect\citeauthoryear{{Bertone} \& {Merritt}}{{Bertone} \&
  {Merritt}}{2005}]{2005MPLA...20.1021B}
{Bertone} G.,  {Merritt} D.,  2005, \mn@doi [Modern Physics Letters A]
  {10.1142/S0217732305017391}, \href
  {https://ui.adsabs.harvard.edu/abs/2005MPLA...20.1021B} {20, 1021}

\bibitem[\protect\citeauthoryear{Bertone, Hooper  \& Silk}{Bertone
  et~al.}{2005}]{Bertone+05}
Bertone G.,  Hooper D.,   Silk J.,  2005, \mn@doi [Physics Reports]
  {10.1016/j.physrep.2004.08.031}, 405, 279

\bibitem[\protect\citeauthoryear{Binney \& Tremaine}{Binney \&
  Tremaine}{2008}]{Binney2008}
Binney J.,  Tremaine S.,  2008, Galactic Dynamics: Second Edition (Princeton
  Series in Astrophysics).
Princeton University Press

\bibitem[\protect\citeauthoryear{Blanchet \& Lavalle}{Blanchet \&
  Lavalle}{2012}]{shmf3}
Blanchet S.,  Lavalle J.,  2012, \mn@doi [JCAP]
  {10.1088/1475-7516/2012/11/021}, 2012, 021

\bibitem[\protect\citeauthoryear{{Bonnivard} et~al.,}{{Bonnivard}
  et~al.}{2015}]{2015MNRAS.453..849B}
{Bonnivard} V.,  et~al., 2015, \mn@doi [\mnras] {10.1093/mnras/stv1601}, \href
  {https://ui.adsabs.harvard.edu/abs/2015MNRAS.453..849B} {453, 849}

\bibitem[\protect\citeauthoryear{{Bose} et~al.,}{{Bose}
  et~al.}{2019}]{2019MNRAS.486.4790B}
{Bose} S.,  et~al., 2019, \mn@doi [\mnras] {10.1093/mnras/stz1168}, \href
  {https://ui.adsabs.harvard.edu/abs/2019MNRAS.486.4790B} {486, 4790}

\bibitem[\protect\citeauthoryear{{Boveia} \& {Doglioni}}{{Boveia} \&
  {Doglioni}}{2018}]{2018ARNPS..68..429B}
{Boveia} A.,  {Doglioni} C.,  2018, \mn@doi [Annual Review of Nuclear and
  Particle Science] {10.1146/annurev-nucl-101917-021008}, \href
  {https://ui.adsabs.harvard.edu/abs/2018ARNPS..68..429B} {68, 429}

\bibitem[\protect\citeauthoryear{Bullock, Kolatt, Sigad, Somerville, Kravtsov,
  Klypin, Primack  \& Dekel}{Bullock et~al.}{2001}]{Bullock:1999he}
Bullock J.~S.,  Kolatt T.~S.,  Sigad Y.,  Somerville R.~S.,  Kravtsov A.~V.,
  Klypin A.~A.,  Primack J.~R.,   Dekel A.,  2001, \mn@doi [Mon. Not. Roy.
  Astron. Soc.] {10.1046/j.1365-8711.2001.04068.x}, 321, 559

\bibitem[\protect\citeauthoryear{{Calore}, {De Romeri}, {Di Mauro}, {Donato}
  \& {Marinacci}}{{Calore} et~al.}{2017}]{2017PhRvD..96f3009C}
{Calore} F.,  {De Romeri} V.,  {Di Mauro} M.,  {Donato} F.,   {Marinacci} F.,
  2017, \mn@doi [\prd] {10.1103/PhysRevD.96.063009}, \href
  {https://ui.adsabs.harvard.edu/abs/2017PhRvD..96f3009C} {96, 063009}

\bibitem[\protect\citeauthoryear{{Calore}, {H{\"u}tten}  \& {Stref}}{{Calore}
  et~al.}{2019}]{2019Galax...7...90C}
{Calore} F.,  {H{\"u}tten} M.,   {Stref} M.,  2019, \mn@doi [Galaxies]
  {10.3390/galaxies7040090}, \href
  {https://ui.adsabs.harvard.edu/abs/2019Galax...7...90C} {7, 90}

\bibitem[\protect\citeauthoryear{{Cerde{\~n}o} \& {Green}}{{Cerde{\~n}o} \&
  {Green}}{2010}]{2010arXiv1002.1912C}
{Cerde{\~n}o} D.~G.,  {Green} A.~M.,  2010, arXiv e-prints, \href
  {https://ui.adsabs.harvard.edu/abs/2010arXiv1002.1912C} {p. arXiv:1002.1912}

\bibitem[\protect\citeauthoryear{{Charbonnier}, {Combet}  \&
  {Maurin}}{{Charbonnier} et~al.}{2012}]{2012CoPhC.183..656C}
{Charbonnier} A.,  {Combet} C.,   {Maurin} D.,  2012, \mn@doi [Computer Physics
  Communications] {10.1016/j.cpc.2011.10.017}, \href
  {https://ui.adsabs.harvard.edu/abs/2012CoPhC.183..656C} {183, 656}

\bibitem[\protect\citeauthoryear{{Contreras}, {Chaves-Montero}, {Zennaro}  \&
  {Angulo}}{{Contreras} et~al.}{2021}]{2021MNRAS.507.3412C}
{Contreras} S.,  {Chaves-Montero} J.,  {Zennaro} M.,   {Angulo} R.~E.,  2021,
  \mn@doi [\mnras] {10.1093/mnras/stab2367}, \href
  {https://ui.adsabs.harvard.edu/abs/2021MNRAS.507.3412C} {507, 3412}

\bibitem[\protect\citeauthoryear{{Coronado-Bl{\'a}zquez}, {S{\'a}nchez-Conde},
  {Dom{\'\i}nguez}, {Aguirre-Santaella}, {Di Mauro}, {Mirabal}, {Nieto}  \&
  {Charles}}{{Coronado-Bl{\'a}zquez} et~al.}{2019a}]{2019JCAP...07..020C}
{Coronado-Bl{\'a}zquez} J.,  {S{\'a}nchez-Conde} M.~A.,  {Dom{\'\i}nguez} A.,
  {Aguirre-Santaella} A.,  {Di Mauro} M.,  {Mirabal} N.,  {Nieto} D.,
  {Charles} E.,  2019a, \mn@doi [\jcap] {10.1088/1475-7516/2019/07/020}, \href
  {https://ui.adsabs.harvard.edu/abs/2019JCAP...07..020C} {2019, 020}

\bibitem[\protect\citeauthoryear{{Coronado-Bl{\'a}zquez}, {S{\'a}nchez-Conde},
  {Di Mauro}, {Aguirre-Santaella}, {Ciuc{\u{a}}}, {Dom{\'\i}nguez}, {Kawata}
  \& {Mirabal}}{{Coronado-Bl{\'a}zquez} et~al.}{2019b}]{2019JCAP...11..045C}
{Coronado-Bl{\'a}zquez} J.,  {S{\'a}nchez-Conde} M.~A.,  {Di Mauro} M.,
  {Aguirre-Santaella} A.,  {Ciuc{\u{a}}} I.,  {Dom{\'\i}nguez} A.,  {Kawata}
  D.,   {Mirabal} N.,  2019b, \mn@doi [\jcap] {10.1088/1475-7516/2019/11/045},
  \href {https://ui.adsabs.harvard.edu/abs/2019JCAP...11..045C} {2019, 045}

\bibitem[\protect\citeauthoryear{{Coronado-Bl{\'a}zquez}, {Doro},
  {S{\'a}nchez-Conde}  \& {Aguirre-Santaella}}{{Coronado-Bl{\'a}zquez}
  et~al.}{2021}]{2021PDU....3200845C}
{Coronado-Bl{\'a}zquez} J.,  {Doro} M.,  {S{\'a}nchez-Conde} M.~A.,
  {Aguirre-Santaella} A.,  2021, \mn@doi [Physics of the Dark Universe]
  {10.1016/j.dark.2021.100845}, \href
  {https://ui.adsabs.harvard.edu/abs/2021PDU....3200845C} {32, 100845}

\bibitem[\protect\citeauthoryear{{Coronado-Bl{\'a}zquez}, {S{\'a}nchez-Conde},
  {P{\'e}rez-Romero}, {Aguirre-Santaella}  \& {Fermi-LAT
  Collaboration}}{{Coronado-Bl{\'a}zquez} et~al.}{2022}]{2022PhRvD.105h3006C}
{Coronado-Bl{\'a}zquez} J.,  {S{\'a}nchez-Conde} M.~A.,  {P{\'e}rez-Romero} J.,
   {Aguirre-Santaella} A.,   {Fermi-LAT Collaboration} 2022, \mn@doi [\prd]
  {10.1103/PhysRevD.105.083006}, \href
  {https://ui.adsabs.harvard.edu/abs/2022PhRvD.105h3006C} {105, 083006}

\bibitem[\protect\citeauthoryear{{De Lucia}, {Kauffmann}, {Springel}, {White},
  {Lanzoni}, {Stoehr}, {Tormen}  \& {Yoshida}}{{De Lucia}
  et~al.}{2004}]{2004MNRAS.348..333D}
{De Lucia} G.,  {Kauffmann} G.,  {Springel} V.,  {White} S.~D.~M.,  {Lanzoni}
  B.,  {Stoehr} F.,  {Tormen} G.,   {Yoshida} N.,  2004, \mn@doi [\mnras]
  {10.1111/j.1365-2966.2004.07372.x}, \href
  {https://ui.adsabs.harvard.edu/abs/2004MNRAS.348..333D} {348, 333}

\bibitem[\protect\citeauthoryear{{Despali} \& {Vegetti}}{{Despali} \&
  {Vegetti}}{2017}]{2017MNRAS.469.1997D}
{Despali} G.,  {Vegetti} S.,  2017, \mn@doi [\mnras] {10.1093/mnras/stx966},
  \href {https://ui.adsabs.harvard.edu/abs/2017MNRAS.469.1997D} {469, 1997}

\bibitem[\protect\citeauthoryear{{Di Mauro}, {Stref}  \& {Calore}}{{Di Mauro}
  et~al.}{2020}]{2020PhRvD.102j3010D}
{Di Mauro} M.,  {Stref} M.,   {Calore} F.,  2020, \mn@doi [\prd]
  {10.1103/PhysRevD.102.103010}, \href
  {https://ui.adsabs.harvard.edu/abs/2020PhRvD.102j3010D} {102, 103010}

\bibitem[\protect\citeauthoryear{Diemand \& Moore}{Diemand \&
  Moore}{2011}]{Diemand:2009bm}
Diemand J.,  Moore B.,  2011, \mn@doi [Adv. Sci. Lett.]
  {10.1166/asl.2011.1211}, 4, 297

\bibitem[\protect\citeauthoryear{Diemand, Kuhlen  \& Madau}{Diemand
  et~al.}{2007a}]{Diemand:2006ik}
Diemand J.,  Kuhlen M.,   Madau P.,  2007a, \mn@doi [Astrophys. J.]
  {10.1086/510736}, 657, 262

\bibitem[\protect\citeauthoryear{Diemand, Kuhlen  \& Madau}{Diemand
  et~al.}{2007b}]{jurg1}
Diemand J.,  Kuhlen M.,   Madau P.,  2007b, \mn@doi [Astrophys. J.]
  {10.1086/520573}, 667, 859

\bibitem[\protect\citeauthoryear{{Diemand}, {Kuhlen}, {Madau}, {Zemp}, {Moore},
  {Potter}  \& {Stadel}}{{Diemand} et~al.}{2008}]{Diemand2008}
{Diemand} J.,  {Kuhlen} M.,  {Madau} P.,  {Zemp} M.,  {Moore} B.,  {Potter} D.,
    {Stadel} J.,  2008, \mn@doi [Nature] {10.1038/nature07153}, \href
  {https://ui.adsabs.harvard.edu/abs/2008Natur.454..735D} {454, 735}

\bibitem[\protect\citeauthoryear{{Dooley}, {Griffen}, {Zukin}, {Ji},
  {Vogelsberger}, {Hernquist}  \& {Frebel}}{{Dooley}
  et~al.}{2014}]{2014ApJ...786...50D}
{Dooley} G.~A.,  {Griffen} B.~F.,  {Zukin} P.,  {Ji} A.~P.,  {Vogelsberger} M.,
   {Hernquist} L.~E.,   {Frebel} A.,  2014, \mn@doi [\apj]
  {10.1088/0004-637X/786/1/50}, \href
  {https://ui.adsabs.harvard.edu/abs/2014ApJ...786...50D} {786, 50}

\bibitem[\protect\citeauthoryear{{Eke}, {Navarro}  \& {Steinmetz}}{{Eke}
  et~al.}{2001}]{2001ApJ...554..114E}
{Eke} V.~R.,  {Navarro} J.~F.,   {Steinmetz} M.,  2001, \mn@doi [\apj]
  {10.1086/321345}, \href
  {https://ui.adsabs.harvard.edu/abs/2001ApJ...554..114E} {554, 114}

\bibitem[\protect\citeauthoryear{{Errani} \& {Pe{\~n}arrubia}}{{Errani} \&
  {Pe{\~n}arrubia}}{2020}]{2020MNRAS.491.4591E}
{Errani} R.,  {Pe{\~n}arrubia} J.,  2020, \mn@doi [MNRAS]
  {10.1093/mnras/stz3349}, \href
  {https://ui.adsabs.harvard.edu/abs/2020MNRAS.491.4591E} {491, 4591}

\bibitem[\protect\citeauthoryear{Evans, Ferrer  \& Sarkar}{Evans
  et~al.}{2004a}]{PhysRevD.69.123501}
Evans N.~W.,  Ferrer F.,   Sarkar S.,  2004a, \mn@doi [Phys. Rev. D]
  {10.1103/PhysRevD.69.123501}, 69, 123501

\bibitem[\protect\citeauthoryear{{Evans}, {Ferrer}  \& {Sarkar}}{{Evans}
  et~al.}{2004b}]{2004PhRvD..69l3501E}
{Evans} N.~W.,  {Ferrer} F.,   {Sarkar} S.,  2004b, \mn@doi [\prd]
  {10.1103/PhysRevD.69.123501}, \href
  {https://ui.adsabs.harvard.edu/abs/2004PhRvD..69l3501E} {69, 123501}

\bibitem[\protect\citeauthoryear{{Fattahi}, {Navarro}, {Sawala}, {Frenk},
  {Sales}, {Oman}, {Schaller}  \& {Wang}}{{Fattahi}
  et~al.}{2016a}]{2016arXiv160706479F}
{Fattahi} A.,  {Navarro} J.~F.,  {Sawala} T.,  {Frenk} C.~S.,  {Sales} L.~V.,
  {Oman} K.,  {Schaller} M.,   {Wang} J.,  2016a, arXiv e-prints, \href
  {https://ui.adsabs.harvard.edu/abs/2016arXiv160706479F} {p. arXiv:1607.06479}

\bibitem[\protect\citeauthoryear{{Fattahi} et~al.,}{{Fattahi}
  et~al.}{2016b}]{2016MNRAS.457..844F}
{Fattahi} A.,  et~al., 2016b, \mn@doi [\mnras] {10.1093/mnras/stv2970}, \href
  {https://ui.adsabs.harvard.edu/abs/2016MNRAS.457..844F} {457, 844}

\bibitem[\protect\citeauthoryear{{Frenk} \& {White}}{{Frenk} \&
  {White}}{2012}]{2012AnP...524..507F}
{Frenk} C.~S.,  {White} S.~D.~M.,  2012, \mn@doi [Annalen der Physik]
  {10.1002/andp.201200212}, \href
  {https://ui.adsabs.harvard.edu/abs/2012AnP...524..507F} {524, 507}

\bibitem[\protect\citeauthoryear{{Gao} \& {White}}{{Gao} \&
  {White}}{2007}]{2007MNRAS.377L...5G}
{Gao} L.,  {White} S. D.~M.,  2007, \mn@doi [\mnras]
  {10.1111/j.1745-3933.2007.00292.x}, \href
  {https://ui.adsabs.harvard.edu/abs/2007MNRAS.377L...5G} {377, L5}

\bibitem[\protect\citeauthoryear{Garrett \& Duda}{Garrett \&
  Duda}{2011}]{Garrett:2010hd}
Garrett K.,  Duda G.,  2011, \mn@doi [Adv. Astron.] {10.1155/2011/968283},
  2011, 968283

\bibitem[\protect\citeauthoryear{Garrison-Kimmel, Boylan-Kolchin, Bullock  \&
  Lee}{Garrison-Kimmel et~al.}{2014}]{Elvis}
Garrison-Kimmel S.,  Boylan-Kolchin M.,  Bullock J.~S.,   Lee K.,  2014,
  \mn@doi [MNRAS] {10.1093/mnras/stt2377}, 438, 2578

\bibitem[\protect\citeauthoryear{{Garrison-Kimmel} et~al.,}{{Garrison-Kimmel}
  et~al.}{2017a}]{2017MNRAS.471.1709G}
{Garrison-Kimmel} S.,  et~al., 2017a, \mn@doi [MNRAS] {10.1093/mnras/stx1710},
  \href {https://ui.adsabs.harvard.edu/abs/2017MNRAS.471.1709G} {471, 1709}

\bibitem[\protect\citeauthoryear{{Garrison-Kimmel} et~al.,}{{Garrison-Kimmel}
  et~al.}{2017b}]{10.1093/mnras/stx1710}
{Garrison-Kimmel} S.,  et~al., 2017b, \mn@doi [\mnras] {10.1093/mnras/stx1710},
  \href {https://ui.adsabs.harvard.edu/abs/2017MNRAS.471.1709G} {471, 1709}

\bibitem[\protect\citeauthoryear{{Gehrels} \& {Michelson}}{{Gehrels} \&
  {Michelson}}{1999}]{fermiglast}
{Gehrels} N.,  {Michelson} P.,  1999, \mn@doi [Astroparticle Physics]
  {10.1016/S0927-6505(99)00066-3}, \href
  {http://adsabs.harvard.edu/abs/1999APh....11..277G} {11, 277}

\bibitem[\protect\citeauthoryear{Ghigna, Moore, Governato, Lake, Quinn  \&
  Stadel}{Ghigna et~al.}{1998}]{Ghigna:1998vn}
Ghigna S.,  Moore B.,  Governato F.,  Lake G.,  Quinn T.~R.,   Stadel J.,
  1998, \mn@doi [Mon. Not. Roy. Astron. Soc.]
  {10.1046/j.1365-8711.1998.01918.x}, 300, 146

\bibitem[\protect\citeauthoryear{{Giocoli}, {Tormen}  \& {van den
  Bosch}}{{Giocoli} et~al.}{2008a}]{2008MNRAS.386.2135G}
{Giocoli} C.,  {Tormen} G.,   {van den Bosch} F.~C.,  2008a, \mn@doi [\mnras]
  {10.1111/j.1365-2966.2008.13182.x}, \href
  {https://ui.adsabs.harvard.edu/abs/2008MNRAS.386.2135G} {386, 2135}

\bibitem[\protect\citeauthoryear{Giocoli, Pieri  \& Tormen}{Giocoli
  et~al.}{2008b}]{shmf2}
Giocoli C.,  Pieri L.,   Tormen G.,  2008b, \mn@doi [MNRAS]
  {10.1111/j.1365-2966.2008.13283.x}, 387, 689

\bibitem[\protect\citeauthoryear{{Giocoli}, {Tormen}, {Sheth}  \& {van den
  Bosch}}{{Giocoli} et~al.}{2010}]{2010MNRAS.404..502G}
{Giocoli} C.,  {Tormen} G.,  {Sheth} R.~K.,   {van den Bosch} F.~C.,  2010,
  \mn@doi [\mnras] {10.1111/j.1365-2966.2010.16311.x}, \href
  {https://ui.adsabs.harvard.edu/abs/2010MNRAS.404..502G} {404, 502}

\bibitem[\protect\citeauthoryear{{Giocoli}, {Meneghetti}, {Bartelmann},
  {Moscardini}  \& {Boldrin}}{{Giocoli} et~al.}{2012}]{2012MNRAS.421.3343G}
{Giocoli} C.,  {Meneghetti} M.,  {Bartelmann} M.,  {Moscardini} L.,   {Boldrin}
  M.,  2012, \mn@doi [\mnras] {10.1111/j.1365-2966.2012.20558.x}, \href
  {https://ui.adsabs.harvard.edu/abs/2012MNRAS.421.3343G} {421, 3343}

\bibitem[\protect\citeauthoryear{{Grand} \& {White}}{{Grand} \&
  {White}}{2021}]{2021MNRAS.501.3558G}
{Grand} R. J.~J.,  {White} S. D.~M.,  2021, \mn@doi [MNRAS]
  {10.1093/mnras/staa3993}, \href
  {https://ui.adsabs.harvard.edu/abs/2021MNRAS.501.3558G} {501, 3558}

\bibitem[\protect\citeauthoryear{{Grand} et~al.,}{{Grand}
  et~al.}{2017}]{2017MNRAS.467..179G}
{Grand} R.~J.~J.,  et~al., 2017, \mn@doi [\mnras] {10.1093/mnras/stx071}, \href
  {https://ui.adsabs.harvard.edu/abs/2017MNRAS.467..179G} {467, 179}

\bibitem[\protect\citeauthoryear{{Grand} et~al.,}{{Grand}
  et~al.}{2021}]{2021MNRAS.507.4953G}
{Grand} R. J.~J.,  et~al., 2021, \mn@doi [\mnras] {10.1093/mnras/stab2492},
  \href {https://ui.adsabs.harvard.edu/abs/2021MNRAS.507.4953G} {507, 4953}

\bibitem[\protect\citeauthoryear{{Graus}, {Bullock}, {Kelley},
  {Boylan-Kolchin}, {Garrison-Kimmel}  \& {Qi}}{{Graus}
  et~al.}{2019}]{2019MNRAS.488.4585G}
{Graus} A.~S.,  {Bullock} J.~S.,  {Kelley} T.,  {Boylan-Kolchin} M.,
  {Garrison-Kimmel} S.,   {Qi} Y.,  2019, \mn@doi [\mnras]
  {10.1093/mnras/stz1992}, \href
  {https://ui.adsabs.harvard.edu/abs/2019MNRAS.488.4585G} {488, 4585}

\bibitem[\protect\citeauthoryear{{Green} \& {van den Bosch}}{{Green} \& {van
  den Bosch}}{2019}]{2019MNRAS.490.2091G}
{Green} S.~B.,  {van den Bosch} F.~C.,  2019, \mn@doi [\mnras]
  {10.1093/mnras/stz2767}, \href
  {https://ui.adsabs.harvard.edu/abs/2019MNRAS.490.2091G} {490, 2091}

\bibitem[\protect\citeauthoryear{{Green}, {van den Bosch}  \& {Jiang}}{{Green}
  et~al.}{2022}]{2022MNRAS.509.2624G}
{Green} S.~B.,  {van den Bosch} F.~C.,   {Jiang} F.,  2022, \mn@doi [\mnras]
  {10.1093/mnras/stab3130}, \href
  {https://ui.adsabs.harvard.edu/abs/2022MNRAS.509.2624G} {509, 2624}

\bibitem[\protect\citeauthoryear{{Griffen}, {Ji}, {Dooley}, {G{\'o}mez},
  {Vogelsberger}, {O'Shea}  \& {Frebel}}{{Griffen}
  et~al.}{2016}]{2016ApJ...818...10G}
{Griffen} B.~F.,  {Ji} A.~P.,  {Dooley} G.~A.,  {G{\'o}mez} F.~A.,
  {Vogelsberger} M.,  {O'Shea} B.~W.,   {Frebel} A.,  2016, \mn@doi [\apj]
  {10.3847/0004-637X/818/1/10}, \href
  {https://ui.adsabs.harvard.edu/abs/2016ApJ...818...10G} {818, 10}

\bibitem[\protect\citeauthoryear{Han, Cole, Frenk  \& Jing}{Han
  et~al.}{2016}]{Han:2015pua}
Han J.,  Cole S.,  Frenk C.~S.,   Jing Y.,  2016, \mn@doi [Mon. Not. Roy.
  Astron. Soc.] {10.1093/mnras/stv2900}, 457, 1208

\bibitem[\protect\citeauthoryear{{Hellwing}, {Frenk}, {Cautun}, {Bose},
  {Helly}, {Jenkins}, {Sawala}  \& {Cytowski}}{{Hellwing}
  et~al.}{2016}]{2016MNRAS.457.3492H}
{Hellwing} W.~A.,  {Frenk} C.~S.,  {Cautun} M.,  {Bose} S.,  {Helly} J.,
  {Jenkins} A.,  {Sawala} T.,   {Cytowski} M.,  2016, \mn@doi [\mnras]
  {10.1093/mnras/stw214}, \href
  {https://ui.adsabs.harvard.edu/abs/2016MNRAS.457.3492H} {457, 3492}

\bibitem[\protect\citeauthoryear{Hinton}{Hinton}{2004}]{hinton2004}
Hinton J.,  2004, \mn@doi [New Astronomy Reviews]
  {10.1016/j.newar.2003.12.004}, 48, 331

\bibitem[\protect\citeauthoryear{Hunter}{Hunter}{2007}]{Matplotlib_paper}
Hunter J.~D.,  2007, \mn@doi [Comput. Sci. Eng.] {10.1109/mcse.2007.55}, 9, 90

\bibitem[\protect\citeauthoryear{{H{\"u}tten}, {Combet}, {Maier}  \&
  {Maurin}}{{H{\"u}tten} et~al.}{2016}]{2016JCAP...09..047H}
{H{\"u}tten} M.,  {Combet} C.,  {Maier} G.,   {Maurin} D.,  2016, \mn@doi
  [\jcap] {10.1088/1475-7516/2016/09/047}, \href
  {https://ui.adsabs.harvard.edu/abs/2016JCAP...09..047H} {2016, 047}

\bibitem[\protect\citeauthoryear{{Ishiyama} et~al.,}{{Ishiyama}
  et~al.}{2021}]{Ishiyama2021}
{Ishiyama} T.,  et~al., 2021, \mn@doi [\mnras] {10.1093/mnras/stab1755}, \href
  {https://ui.adsabs.harvard.edu/abs/2021MNRAS.506.4210I} {506, 4210}

\bibitem[\protect\citeauthoryear{{Kelley}, {Bullock}, {Garrison-Kimmel},
  {Boylan-Kolchin}, {Pawlowski}  \& {Graus}}{{Kelley}
  et~al.}{2019}]{Kelley2019}
{Kelley} T.,  {Bullock} J.~S.,  {Garrison-Kimmel} S.,  {Boylan-Kolchin} M.,
  {Pawlowski} M.~S.,   {Graus} A.~S.,  2019, \mn@doi [\mnras]
  {10.1093/mnras/stz1553}, \href
  {https://ui.adsabs.harvard.edu/abs/2019MNRAS.487.4409K} {487, 4409}

\bibitem[\protect\citeauthoryear{{Kerr} \& {Lynden-Bell}}{{Kerr} \&
  {Lynden-Bell}}{1986}]{1986MNRAS.221.1023K}
{Kerr} F.~J.,  {Lynden-Bell} D.,  1986, \mn@doi [\mnras]
  {10.1093/mnras/221.4.1023}, \href
  {https://ui.adsabs.harvard.edu/abs/1986MNRAS.221.1023K} {221, 1023}

\bibitem[\protect\citeauthoryear{{King}}{{King}}{1962}]{1962AJ.....67..471K}
{King} I.,  1962, \mn@doi [\aj] {10.1086/108756}, \href
  {http://adsabs.harvard.edu/abs/1962AJ.....67..471K} {67, 471}

\bibitem[\protect\citeauthoryear{{Komatsu} et~al.,}{{Komatsu}
  et~al.}{2009}]{2009ApJS..180..330K}
{Komatsu} E.,  et~al., 2009, \mn@doi [\apjs] {10.1088/0067-0049/180/2/330},
  \href {https://ui.adsabs.harvard.edu/abs/2009ApJS..180..330K} {180, 330}

\bibitem[\protect\citeauthoryear{{{\L}okas}}{{{\L}okas}}{2002}]{2002MNRAS.333..697L}
{{\L}okas} E.~L.,  2002, \mn@doi [\mnras] {10.1046/j.1365-8711.2002.05457.x},
  \href {https://ui.adsabs.harvard.edu/abs/2002MNRAS.333..697L} {333, 697}

\bibitem[\protect\citeauthoryear{{Lorenz}}{{Lorenz}}{2004}]{Lorenz:2004ah}
{Lorenz} E.,  2004, \mn@doi [\nar] {10.1016/j.newar.2003.12.059}, \href
  {https://ui.adsabs.harvard.edu/abs/2004NewAR..48..339L} {48, 339}

\bibitem[\protect\citeauthoryear{{Molin{\'e}}, {S{\'a}nchez-Conde},
  {Palomares-Ruiz}  \& {Prada}}{{Molin{\'e}} et~al.}{2017}]{moline}
{Molin{\'e}} {\'A}.,  {S{\'a}nchez-Conde} M.~A.,  {Palomares-Ruiz} S.,
  {Prada} F.,  2017, \mn@doi [\mnras] {10.1093/mnras/stx026}, \href
  {https://ui.adsabs.harvard.edu/abs/2017MNRAS.466.4974M} {466, 4974}

\bibitem[\protect\citeauthoryear{{Molin{\'e}}, {S{\'a}nchez-Conde},
  {Aguirre-Santaella}  et~al.}{{Molin{\'e}} et~al.}{2023}]{2023MNRAS.518..157M}
{Molin{\'e}} {\'A}.,  {S{\'a}nchez-Conde} M.~A.,  {Aguirre-Santaella} A.,
  et~al., 2023, \mn@doi [\ MNRAS] {10.1093/mnras/stac2930}, \href
  {https://ui.adsabs.harvard.edu/abs/2023MNRAS.518..157M} {518, 157}

\bibitem[\protect\citeauthoryear{{Nadler} et~al.,}{{Nadler}
  et~al.}{2023}]{2022arXiv220902675N}
{Nadler} E.~O.,  et~al., 2023, \mn@doi [\apj] {10.3847/1538-4357/acb68c}, \href
  {https://ui.adsabs.harvard.edu/abs/2023ApJ...945..159N} {945, 159}

\bibitem[\protect\citeauthoryear{{Navarro}, {Frenk}  \& {White}}{{Navarro}
  et~al.}{1997}]{1997ApJ...490..493N}
{Navarro} J.~F.,  {Frenk} C.~S.,   {White} S.~D.~M.,  1997, \mn@doi [\apj]
  {10.1086/304888}, \href
  {https://ui.adsabs.harvard.edu/abs/1997ApJ...490..493N} {490, 493}

\bibitem[\protect\citeauthoryear{{Neto} et~al.,}{{Neto}
  et~al.}{2007}]{2007MNRAS.381.1450N}
{Neto} A.~F.,  et~al., 2007, \mn@doi [\mnras]
  {10.1111/j.1365-2966.2007.12381.x}, \href
  {https://ui.adsabs.harvard.edu/abs/2007MNRAS.381.1450N} {381, 1450}

\bibitem[\protect\citeauthoryear{{Ogiya}, {van den Bosch}, {Hahn}, {Green},
  {Miller}  \& {Burkert}}{{Ogiya} et~al.}{2019}]{Ogiya2019}
{Ogiya} G.,  {van den Bosch} F.~C.,  {Hahn} O.,  {Green} S.~B.,  {Miller}
  T.~B.,   {Burkert} A.,  2019, \mn@doi [MNRAS] {10.1093/mnras/stz375}, \href
  {https://ui.adsabs.harvard.edu/abs/2019MNRAS.485..189O} {485, 189}

\bibitem[\protect\citeauthoryear{{Pace} \& {Strigari}}{{Pace} \&
  {Strigari}}{2019}]{2019MNRAS.482.3480P}
{Pace} A.~B.,  {Strigari} L.~E.,  2019, \mn@doi [\mnras]
  {10.1093/mnras/sty2839}, \href
  {https://ui.adsabs.harvard.edu/abs/2019MNRAS.482.3480P} {482, 3480}

\bibitem[\protect\citeauthoryear{Perez \& Granger}{Perez \&
  Granger}{2007}]{Ipython_paper}
Perez F.,  Granger B.~E.,  2007, \mn@doi [Comput. Sci. Eng.]
  {10.1109/mcse.2007.53}, 9, 21

\bibitem[\protect\citeauthoryear{Pieri, Lavalle, Bertone  \& Branchini}{Pieri
  et~al.}{2011}]{pieri}
Pieri L.,  Lavalle J.,  Bertone G.,   Branchini E.,  2011, \mn@doi [Phys. Rev.
  D] {10.1103/physrevd.83.023518}, 83

\bibitem[\protect\citeauthoryear{{Porter}, {Johnson}  \& {Graham}}{{Porter}
  et~al.}{2011}]{2011ARA&A..49..155P}
{Porter} T.~A.,  {Johnson} R.~P.,   {Graham} P.~W.,  2011, \mn@doi [\araa]
  {10.1146/annurev-astro-081710-102528}, \href
  {https://ui.adsabs.harvard.edu/abs/2011ARA&A..49..155P} {49, 155}

\bibitem[\protect\citeauthoryear{{Rodr{\'\i}guez-Puebla}, {Behroozi},
  {Primack}, {Klypin}, {Lee}  \& {Hellinger}}{{Rodr{\'\i}guez-Puebla}
  et~al.}{2016}]{RP:2016}
{Rodr{\'\i}guez-Puebla} A.,  {Behroozi} P.,  {Primack} J.,  {Klypin} A.,  {Lee}
  C.,   {Hellinger} D.,  2016, \mn@doi [\mnras] {10.1093/mnras/stw1705}, \href
  {https://ui.adsabs.harvard.edu/abs/2016MNRAS.462..893R} {462, 893}

\bibitem[\protect\citeauthoryear{{Salvador-Sol{\'e}}, {Manrique}, {Canales}  \&
  {Botella}}{{Salvador-Sol{\'e}} et~al.}{2022}]{2022MNRAS.511..641S}
{Salvador-Sol{\'e}} E.,  {Manrique} A.,  {Canales} D.,   {Botella} I.,  2022,
  \mn@doi [\mnras] {10.1093/mnras/stac067}, \href
  {https://ui.adsabs.harvard.edu/abs/2022MNRAS.511..641S} {511, 641}

\bibitem[\protect\citeauthoryear{S\'anchez-Conde \& Prada}{S\'anchez-Conde \&
  Prada}{2014}]{mascprada14}
S\'anchez-Conde M.~A.,  Prada F.,  2014, \mn@doi [Mon. Not. Roy. Astron. Soc.]
  {10.1093/mnras/stu1014}, 442, 2271

\bibitem[\protect\citeauthoryear{{S{\'a}nchez-Conde}, {Cannoni}, {Zandanel},
  {G{\'o}mez}  \& {Prada}}{{S{\'a}nchez-Conde}
  et~al.}{2011}]{2011JCAP...12..011S}
{S{\'a}nchez-Conde} M.~A.,  {Cannoni} M.,  {Zandanel} F.,  {G{\'o}mez} M.~E.,
  {Prada} F.,  2011, \mn@doi [\jcap] {10.1088/1475-7516/2011/12/011}, \href
  {https://ui.adsabs.harvard.edu/abs/2011JCAP...12..011S} {2011, 011}

\bibitem[\protect\citeauthoryear{{Sawala} et~al.,}{{Sawala}
  et~al.}{2016}]{2016MNRAS.457.1931S}
{Sawala} T.,  et~al., 2016, \mn@doi [\mnras] {10.1093/mnras/stw145}, \href
  {https://ui.adsabs.harvard.edu/abs/2016MNRAS.457.1931S} {457, 1931}

\bibitem[\protect\citeauthoryear{{Schoonenberg}, {Gaskins}, {Bertone}  \&
  {Diemand}}{{Schoonenberg} et~al.}{2016}]{2016JCAP...05..028S}
{Schoonenberg} D.,  {Gaskins} J.,  {Bertone} G.,   {Diemand} J.,  2016, \mn@doi
  [\jcap] {10.1088/1475-7516/2016/05/028}, \href
  {https://ui.adsabs.harvard.edu/abs/2016JCAP...05..028S} {2016, 028}

\bibitem[\protect\citeauthoryear{{Spergel} et~al.,}{{Spergel}
  et~al.}{2007}]{2007ApJS..170..377S}
{Spergel} D.~N.,  et~al., 2007, \mn@doi [\apjs] {10.1086/513700}, \href
  {https://ui.adsabs.harvard.edu/abs/2007ApJS..170..377S} {170, 377}

\bibitem[\protect\citeauthoryear{{Springel}, {Frenk}  \& {White}}{{Springel}
  et~al.}{2006}]{2006Natur.440.1137S}
{Springel} V.,  {Frenk} C.~S.,   {White} S. D.~M.,  2006, \mn@doi [\nat]
  {10.1038/nature04805}, \href
  {https://ui.adsabs.harvard.edu/abs/2006Natur.440.1137S} {440, 1137}

\bibitem[\protect\citeauthoryear{Springel et~al.,}{Springel
  et~al.}{2008}]{aquarius}
Springel V.,  et~al., 2008, \mn@doi [MNRAS] {10.1111/j.1365-2966.2008.14066.x}

\bibitem[\protect\citeauthoryear{Stadel, Potter, Moore, Diemand, Madau, Zemp,
  Kuhlen  \& Quilis}{Stadel et~al.}{2009}]{GHALO}
Stadel J.,  Potter D.,  Moore B.,  Diemand J.,  Madau P.,  Zemp M.,  Kuhlen M.,
    Quilis V.,  2009, \mn@doi [MNRAS] {10.1111/j.1745-3933.2009.00699.x}, 398,
  L21

\bibitem[\protect\citeauthoryear{{St{\"u}cker}, {Ogiya}, {Angulo},
  {Aguirre-Santaella}  \& {S{\'a}nchez-Conde}}{{St{\"u}cker}
  et~al.}{2023}]{2022arXiv220700604S}
{St{\"u}cker} J.,  {Ogiya} G.,  {Angulo} R.~E.,  {Aguirre-Santaella} A.,
  {S{\'a}nchez-Conde} M.~A.,  2023, \mn@doi [\mnras] {10.1093/mnras/stad844},
  \href {https://ui.adsabs.harvard.edu/abs/2023MNRAS.521.4432S} {521, 4432}

\bibitem[\protect\citeauthoryear{{Tormen}, {Diaferio}  \& {Syer}}{{Tormen}
  et~al.}{1998}]{1998MNRAS.299..728T}
{Tormen} G.,  {Diaferio} A.,   {Syer} D.,  1998, \mn@doi [\mnras]
  {10.1046/j.1365-8711.1998.01775.x}, \href
  {https://ui.adsabs.harvard.edu/abs/1998MNRAS.299..728T} {299, 728}

\bibitem[\protect\citeauthoryear{{Virtanen}, {Gommers}, {Oliphant}
  et~al.}{{Virtanen} et~al.}{2020}]{scipy_paper}
{Virtanen} P.,  {Gommers} R.,  {Oliphant} T.~E.,   et~al., 2020, \mn@doi [Nat.
  Methods] {https://doi.org/10.1038/s41592-019-0686-2}, \href
  {https://rdcu.be/b08Wh} {17, 261}

\bibitem[\protect\citeauthoryear{{Vogelsberger} et~al.,}{{Vogelsberger}
  et~al.}{2014}]{2014MNRAS.444.1518V}
{Vogelsberger} M.,  et~al., 2014, \mn@doi [\mnras] {10.1093/mnras/stu1536},
  \href {https://ui.adsabs.harvard.edu/abs/2014MNRAS.444.1518V} {444, 1518}

\bibitem[\protect\citeauthoryear{{Vogelsberger}, {Marinacci}, {Torrey}  \&
  {Puchwein}}{{Vogelsberger} et~al.}{2020}]{2020NatRP...2...42V}
{Vogelsberger} M.,  {Marinacci} F.,  {Torrey} P.,   {Puchwein} E.,  2020,
  \mn@doi [Nature Reviews Physics] {10.1038/s42254-019-0127-2}, \href
  {https://ui.adsabs.harvard.edu/abs/2020NatRP...2...42V} {2, 42}

\bibitem[\protect\citeauthoryear{{Watson}, {Iliev}, {D'Aloisio}, {Knebe},
  {Shapiro}  \& {Yepes}}{{Watson} et~al.}{2013}]{2013MNRAS.433.1230W}
{Watson} W.~A.,  {Iliev} I.~T.,  {D'Aloisio} A.,  {Knebe} A.,  {Shapiro} P.~R.,
    {Yepes} G.,  2013, \mn@doi [\mnras] {10.1093/mnras/stt791}, \href
  {https://ui.adsabs.harvard.edu/abs/2013MNRAS.433.1230W} {433, 1230}

\bibitem[\protect\citeauthoryear{Wechsler, Bullock, Primack, Kravtsov  \&
  Dekel}{Wechsler et~al.}{2002}]{Wechsler:2001cs}
Wechsler R.~H.,  Bullock J.~S.,  Primack J.~R.,  Kravtsov A.~V.,   Dekel A.,
  2002, \mn@doi [Astrophys. J.] {10.1086/338765}, 568, 52

\bibitem[\protect\citeauthoryear{Weekes et~al.}{Weekes
  et~al.}{2002}]{Weekes:2001pd}
Weekes T.~C.,  et~al., 2002, \mn@doi [Astropart. Phys.]
  {10.1016/S0927-6505(01)00152-9}, 17, 221

\bibitem[\protect\citeauthoryear{Zavala \& Afshordi}{Zavala \&
  Afshordi}{2016}]{Zavala:2015ura}
Zavala J.,  Afshordi N.,  2016, \mn@doi [Mon. Not. Roy. Astron. Soc.]
  {10.1093/mnras/stw048}, 457, 986

\bibitem[\protect\citeauthoryear{{Zavala} \& {Frenk}}{{Zavala} \&
  {Frenk}}{2019}]{2019Galax...7...81Z}
{Zavala} J.,  {Frenk} C.~S.,  2019, \mn@doi [Galaxies]
  {10.3390/galaxies7040081}, \href
  {https://ui.adsabs.harvard.edu/abs/2019Galax...7...81Z} {7, 81}

\bibitem[\protect\citeauthoryear{van~den Bosch}{van~den
  Bosch}{2017}]{vandenBosch:2016hjf}
van~den Bosch F.~C.,  2017, \mn@doi [Mon. Not. Roy. Astron. Soc.]
  {10.1093/mnras/stx520}, 468, 885

\bibitem[\protect\citeauthoryear{{van den Bosch} \& {Ogiya}}{{van den Bosch} \&
  {Ogiya}}{2018}]{2018MNRAS.475.4066V}
{van den Bosch} F.~C.,  {Ogiya} G.,  2018, \mn@doi [\mnras]
  {10.1093/mnras/sty084}, \href
  {https://ui.adsabs.harvard.edu/abs/2018MNRAS.475.4066V} {475, 4066}

\bibitem[\protect\citeauthoryear{van~den Bosch, Jiang, Campbell  \&
  Behroozi}{van~den Bosch et~al.}{2015}]{vdb1}
van~den Bosch F.,  Jiang F.,  Campbell D.,   Behroozi P.,  2015, \mn@doi
  [Monthly Notices of the Royal Astronomical Society] {10.1093/mnras/stv2338},
  455

\bibitem[\protect\citeauthoryear{{van den Bosch}, {Ogiya}, {Hahn}  \&
  {Burkert}}{{van den Bosch} et~al.}{2018}]{2018MNRAS.474.3043V}
{van den Bosch} F.~C.,  {Ogiya} G.,  {Hahn} O.,   {Burkert} A.,  2018, \mn@doi
  [\mnras] {10.1093/mnras/stx2956}, \href
  {https://ui.adsabs.harvard.edu/abs/2018MNRAS.474.3043V} {474, 3043}

\bibitem[\protect\citeauthoryear{van~der Walt, Colbert  \& Varoquaux}{van~der
  Walt et~al.}{2011}]{Numpy_paper}
van~der Walt S.,  Colbert S.~C.,   Varoquaux G.,  2011, \mn@doi [Comput. in
  Sci. Eng.] {10.1109/mcse.2011.37}, 13, 22

\makeatother
\end{thebibliography}

%%%%%%%%%%%%%%%%%%%%%%%%%%%%%%%%%%%%%%%%%%%%%%%%%%

%%%%%%%%%%%%%%%%% APPENDICES %%%%%%%%%%%%%%%%%%%%%

\appendix

%\section{Discussion}

%\sub
\section{Impact of different SRD choices on J-factor values}
\label{sec:srdcav}

\begin{figure}%[H]
\centering
\includegraphics[width=\columnwidth]{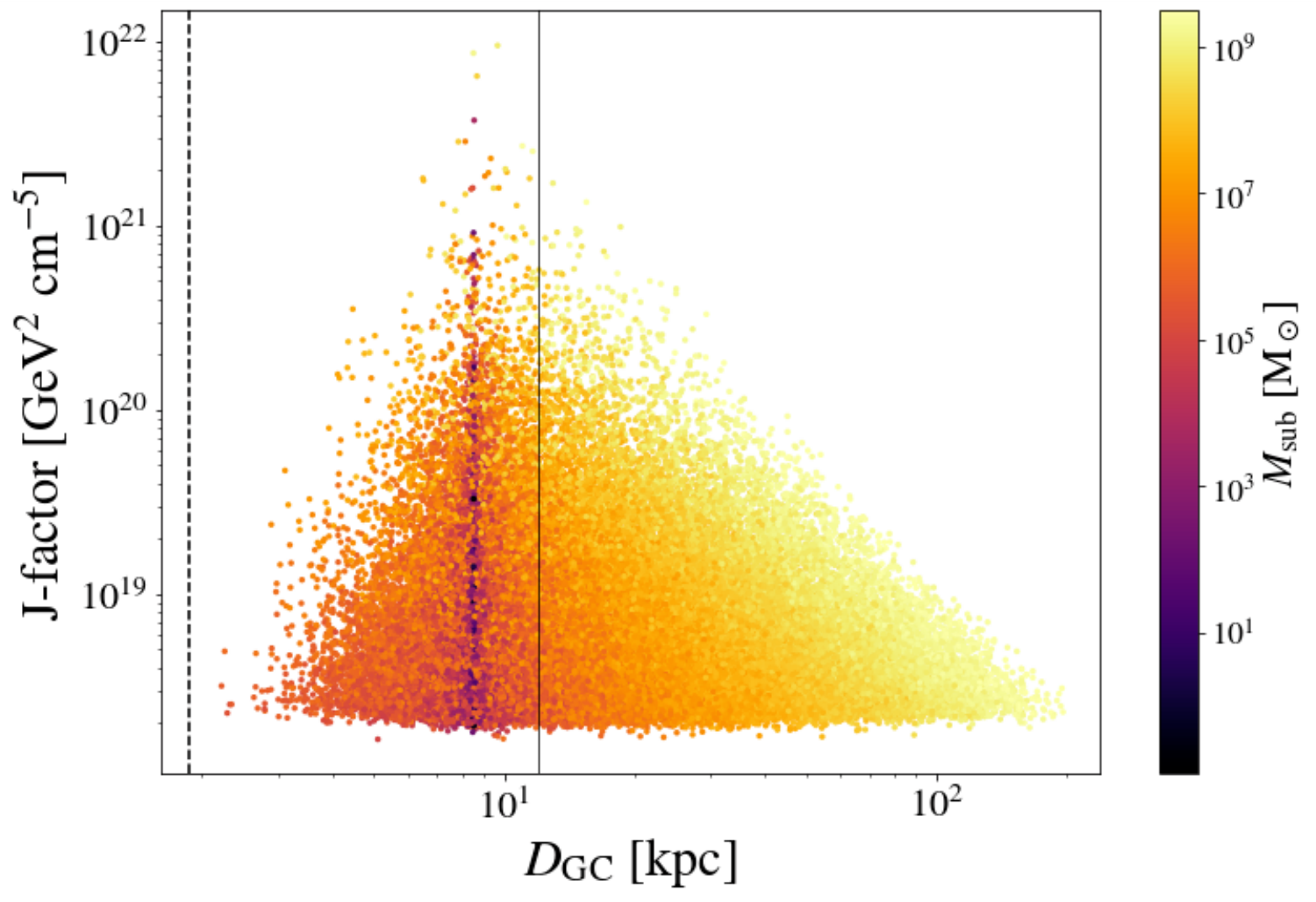} %
\caption{J-factors as a function of subhalo distance to the GC, for 1000 repopulations and the 100 brightest subhaloes in each of them, combining sets $[\mathcal{A}_M]$ and $[\mathcal{B}_M]$. The vertical dashed line, placed at 1.87 kpc, indicates the distance of the closest subhalo to the GC in the original VL-II simulation. The solid vertical line, located at 12 kpc, shows the distance of the closest VL-II subhalo above $\mathrm{M_{cut}}$. The color represents $M_\mathrm{sub}$. }
\label{fig:dgcjfac}
\end{figure}

In this section, we analyze the robustness of our results to different assumptions with respect to the SRD. The results in Section~\ref{sec:srdm} are based on an SRD that closely follows the original data in VL-II and reaches rather inner parts of the host, namely as close to the GC as 1.87 kpc (i.e., the distance of the closest VL-II subhalo to the centre of the host). %
Because of that, since there are VL-II data at such small radii and our default SRD used for repopulations (Equation~\ref{eq:srdmass}) does not exhibit any cutoff down to this point, this SRD can provide us with subhaloes in the innermost region of the host with large J-factor values, especially after so many repopulations as done in this work. This is shown in Fig.~\ref{fig:dgcjfac}, where the dashed vertical line corresponds to the distance of the closest subhalo to the GC in the original VL-II simulation, while the solid one indicates the distance of the closest one with a mass above $\mathrm{M_{cut}}$ (which, we recall, is the value below which VL-II is not complete; see Section~\ref{sec:vlshmf}).

\begin{figure}%[H]
\centering
\includegraphics[width=\columnwidth]{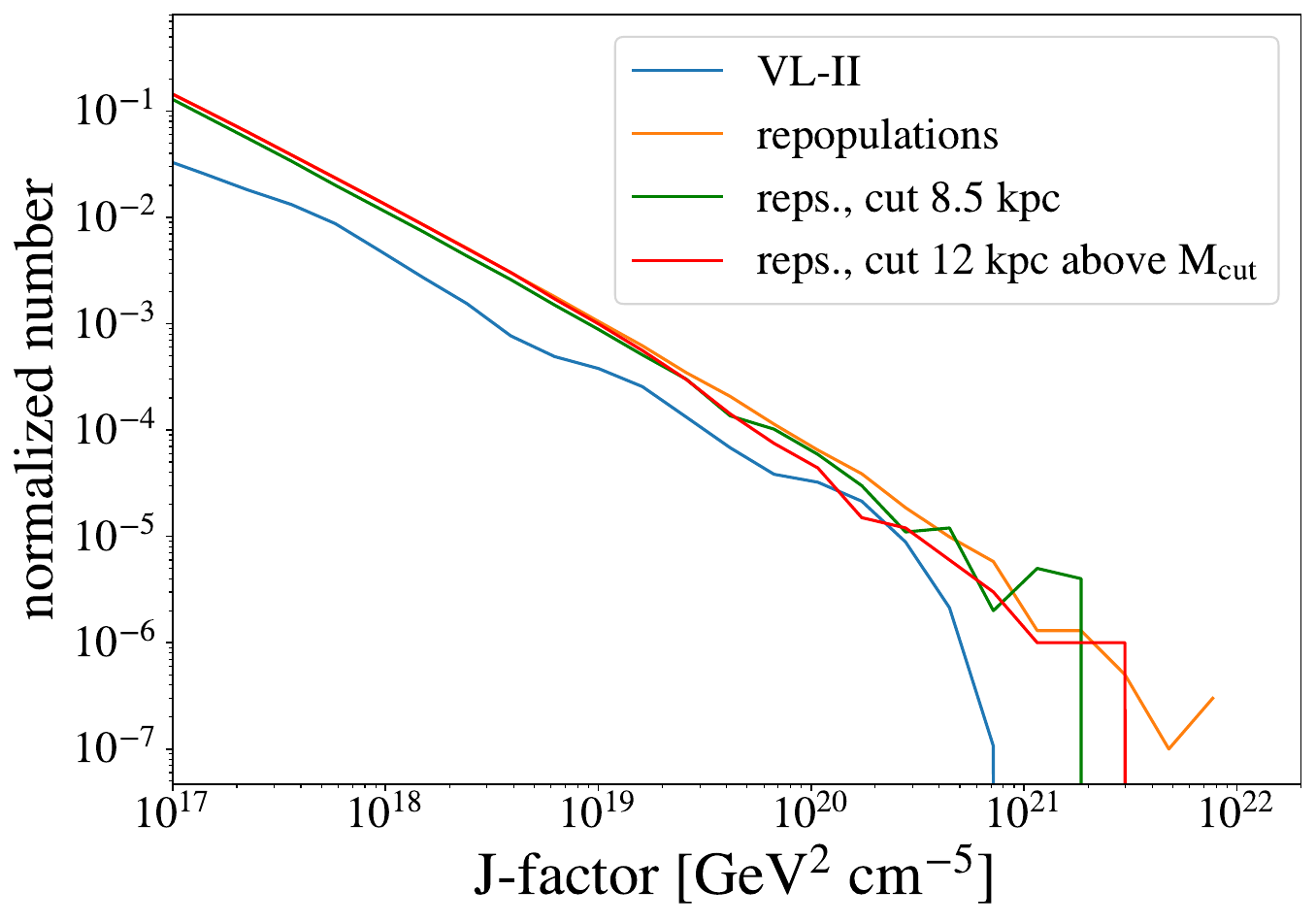} %
\caption{Histogram of J-factors normalized to the total number of subhaloes. Only those above $10^{17} \mathrm{GeV}^2 \mathrm{cm}^{-5}$ are shown. The blue line is for 1000 realizations of VL-II. The orange line corresponds to the 10000 brightest subhaloes in 1000 repopulations using set $[\mathcal{A}_M]$. %
The green line shows the 10000 brightest subhaloes in 100 repopulations where the SRD exhibits a hard cutoff at 8.5 kpc from the host halo centre. The red line is for the 10000 brightest subhaloes in 100 repopulations when the SRD does not provide subhaloes with masses above $\mathrm{M_{cut}}$ within the innermost 12 kpc instead; see text for details on these choices. All repopulations considered for this plot have been done above $10^{3} \mathrm{M_\odot}$. %
}
\label{fig:dgcjfaccomp}
\end{figure}

A more conservative SRD could have been assumed in our repopulations and derivation of J-factors, by simply including a cutoff at a particular Galactocentric distance below which no subhaloes are allowed to exist. The difference in J-factor values obtained for different cutoff choices of the SRD is shown in Fig.~\ref{fig:dgcjfaccomp}. In particular, in this figure we show results for the default SRD, for the case of applying a strict cut at 8.5 kpc, i.e., the solar Galactocentric radius, and for the case of not allowing subhaloes with masses above $\mathrm{M_{cut}}$ to be located at Galactocentric distances smaller than the closest distance of subhaloes with masses above $\mathrm{M_{cut}}$, i.e. 12 kpc.   
These checks helped us to understand the usefulness of repopulating VL-II, even when no subhaloes are included in the inner region of the host. Moreover, Fig.~\ref{fig:dgcjfaccomp} shows that adding these innermost subhaloes (as given by our default SRD) does not result in a substantially brighter population.

\begin{figure}
\centering
\includegraphics[width=\columnwidth]{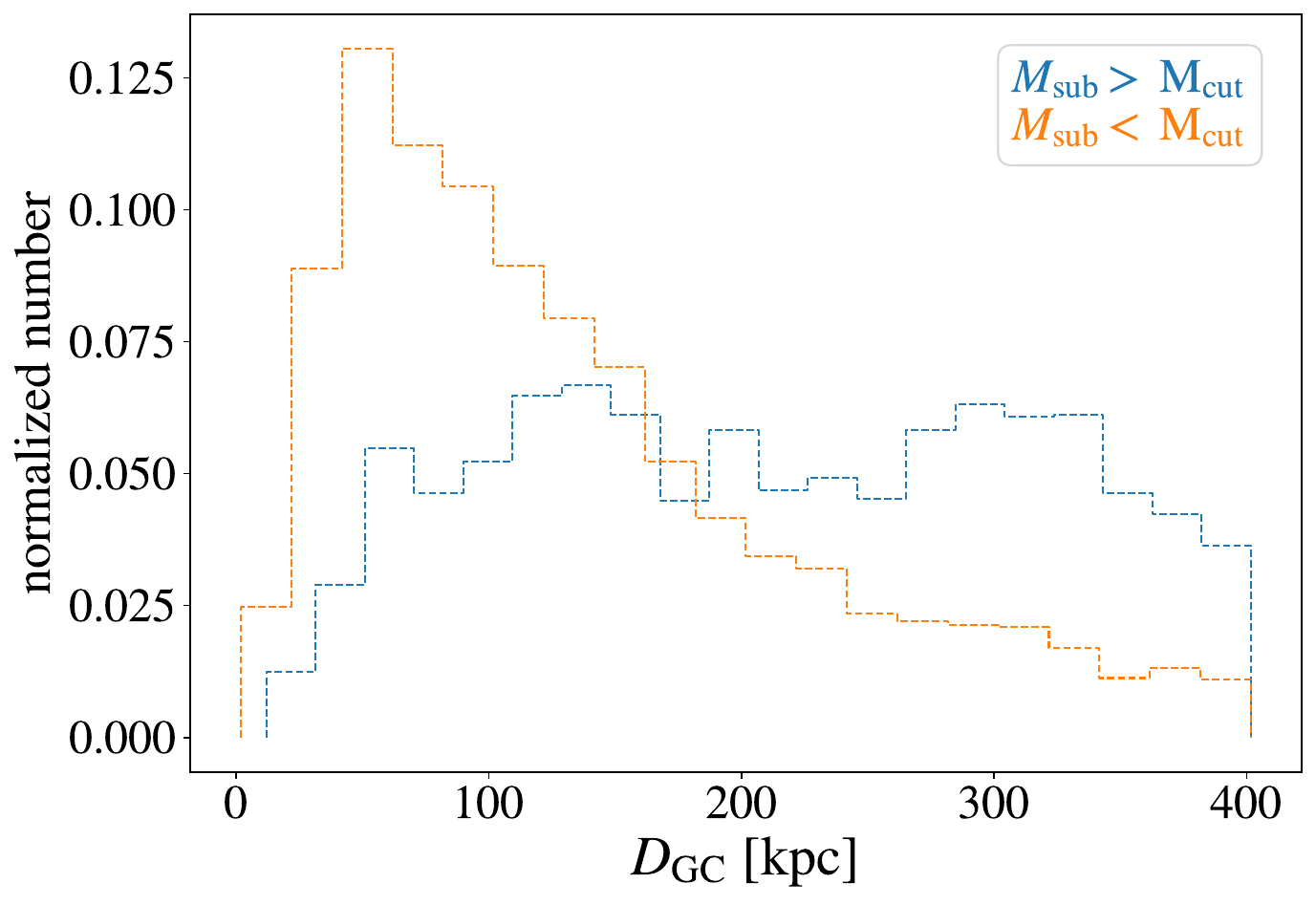} %
\caption{Previous SRD using two mass subsets, used for the repopulations in \citet{2022PhRvD.105h3006C}. 
}
\label{fig:srdmassapp}
\end{figure}

One other main concern is related to the `universality' of the SRD. %
In addition to placing our subhaloes in %
20 radial bins to obtain the radial distribution within the host, we also wanted to explore whether this distribution has a dependence with subhalo mass or not, the latter being the dominant view in the previous literature
 \citep[e.g.][]{Han:2015pua, 2022MNRAS.511..641S, 2023MNRAS.518..157M}. Thus, we divided our sample according to the subhalo mass and built the corresponding SRDs for each mass bin. Interestingly,  %
we found significant differences between the subhalo sample above $\mathrm{M_{cut}}$ and below, which seems to support a dependence on subhalo mass. In particular, as shown in  Fig.~\ref{fig:srdmassapp}, the SRD for subhaloes below $\mathrm{M_{cut}}$ is significantly different from the one above this value, indeed exhibiting a pronounced peak at smaller Galactocentric distances, that is not present above $\mathrm{M_{cut}}$. As a result, we have many more low-mass subhaloes closer to the centre of the host, while the most massive subhalo sample is distributed in a more homogeneous way. %
Despite there is no reason to believe that VL-II would resolve subhaloes more efficiently near the GC when they are small, in the end we decided to adopt a more conservative approach (from the point of view of J-factor values) in which we perform our repopulation work using the SRD obtained above $\mathrm{M_{cut}}$, irrespectively of subhalo mass. Indeed, adopting an SRD more peaked towards the center of the host would imply a general increase of the J-factor values, since a larger number of subhaloes in this region would be located closer to Earth, thus yielding larger annihilation fluxes.

\section{Impact of different integration radii on J-factor values}
\label{sec:intjfac}

\begin{figure}
    \centering
	\includegraphics[width=\columnwidth]{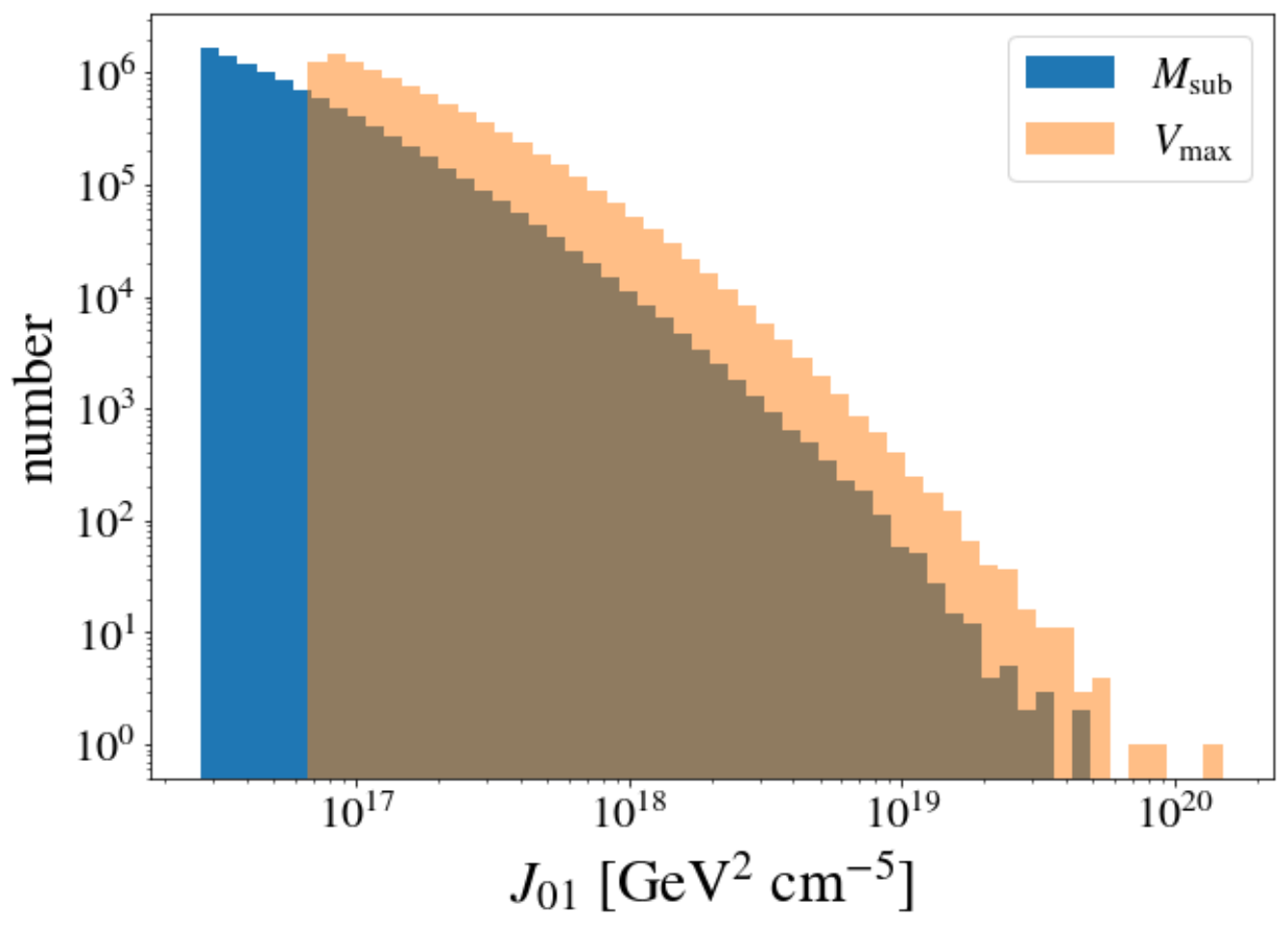}
	%\vspace{-0.55cm}
	\caption{Same as Fig.~\ref{fig:compah}, but for $J_{01}$. See text for details. }
	\label{fig:jfac01}
\end{figure}

Here we want to clarify the effect of calculating the subhalo J-factors by integrating up to the same  radius for both mass and velocity repopulations, as discussed in Section~\ref{sec:jfaccomp}. We have chosen an angular diameter of $\varphi =$ 0.1 deg, thus taking into account only the central cusp in most cases (see Fig.~\ref{fig:ang_extension}).%\masc{is this really true for all these objects? probably not, look at fig.10: in some cases around 1e4-1e5 your subhaloes are actually nearly this size. I'd just add above 'in most cases (see Fig.10).'} 
The J-factors for a fixed angular diameter $\varphi$ are obtained the following way, using $M_\mathrm{sub}$ and $V_\mathrm{max}$ respectively: %\aastomasc{now better?}
\begin{multline}
    J_\varphi = \frac{200}{9}  \frac{\rho_\mathrm{crit}}{D_\mathrm{Earth}^2}  \frac{M_\mathrm{sub} (c_{200})^3}{ (f(c_{200}))^2} \ \times  \\ \left(1 - \frac{1}{\left(1 + D_\mathrm{Earth}  \frac{\tan(\varphi/2)}{r_\mathrm{s}}\right)^3} \right)
\end{multline}
\begin{multline}
    J_\varphi = \frac{H_0}{12\pi G^2 D_\mathrm{Earth}^2} \frac{2.163^3}{f(2.163)^2}  \sqrt{\frac{c_\mathrm{V}}{2}} \ \times \\ V^3_\mathrm{max} \left(1 - \frac{1}{\left(1 + D_\mathrm{Earth}  \frac{\tan(\varphi/2)}{R_\mathrm{max} / 2.163}\right)^3}\right)
\end{multline}
%We have found that we still obtain larger J-factors when using $V_\mathrm{max}$, as shown in
The results are shown in Fig.~\ref{fig:jfac01}. 
Note that in this exercise the maximum J-factor values are obviously lower, as expected, since we are only integrating the annihilation signal produced in the inner cusp here. Also, for a small interval around $10^{18}\ \mathrm{GeV}^2 \mathrm{cm}^{-5}$ we find $5 \times 10^4$ subhaloes when using $V_\mathrm{max}$  and $10^4$ subhaloes when using $M_\mathrm{sub}$. More importantly, we still obtain larger J-factors when using $V_\mathrm{max}$, although the maximum J-factor values now differ by less than one order of magnitude.  
This can still be explained since each repopulation method relies on its own SHM/VF and SRD. On one hand, the SHVF slope found is steeper than the SHMF one, %is closer to the nearest unit, while the SHMF slope is closer to .9 \aastomasc{if you like the argument I guess it should be written in a clearer way but I am not finding the words now. Maybe just "the SHVF slope found is steeper than the SHMF one"}.\masc{yes, better the latter! I like the argument but does it go in the 'right direction'?}
%\aastomasc{[[answering your commented comment: it implies more small subhaloes at least. It could imply less large subhaloes but because of the normalization I think it's not the case. I am going to add a couple of sentences:]]}
and the normalization constant $c$ is larger for the SHVF. Moreover, we are applying the Roche criterium only when we repopulate via $M_\mathrm{sub}$, this way getting rid of very bright subhaloes located at relatively close distances to the Earth. 
On the other, as already stated, the SRD when repopulating with subhalo velocities peaks around 100 kpc while the function grows until $R_\mathrm{vir}$ in the mass-driven repopulation case.  One should also note again that subhaloes do not exhibit NFW profiles, especially at the outskirts, due to tidal stripping, which can modify the current 2.163 value commonly relating $R_\mathrm{max}$ and $r_\mathrm{s}$. Using another value can generate differences in the calculation of the J-factor as well.

%\masc{I think that if this is the case then we should conclude something here, i.e. why we think this can still happen. Indeed, why does it happen? I tend to trust more the results via Vmax, but how can we phrase it? } \aastomasc{I see the point. However, I would not like to make such an statement, because  i) this does not go in the conservative direction and ii) most SRDs found in the literature are more similar to the mass case than to the Vmax case. I think that, apart from having different SRDs --letting more Vmax subhaloes being generated closer-- the (cumulative) SHM/VF could have something to do too, since the slope is closer to 3 for Vmax, while closer to 0.9 (rather than 1) for mass. So I think it's more a matter of how we repopulate. I am taking a look at the upper right panel of Figs 7 and 9 and the difference is still there but it's not so large, and in that case it could be explained because of the different integration angle. I don't know... but anyway I think it's more important to focus on the procedure (and on the fact that dark subhaloes are relevant blabla) rather than on which results we should trust more(??)}\masc{fully agree, Ale! I just hope the ref will understand this in the same way we do! It turned to be a tricky issue ;-)}

\section{$V_\mathrm{max} - R_\mathrm{max}$ relation}\label{ap:vmaxrmax}

\begin{figure}%[H]
\centering
\includegraphics[width=\columnwidth]{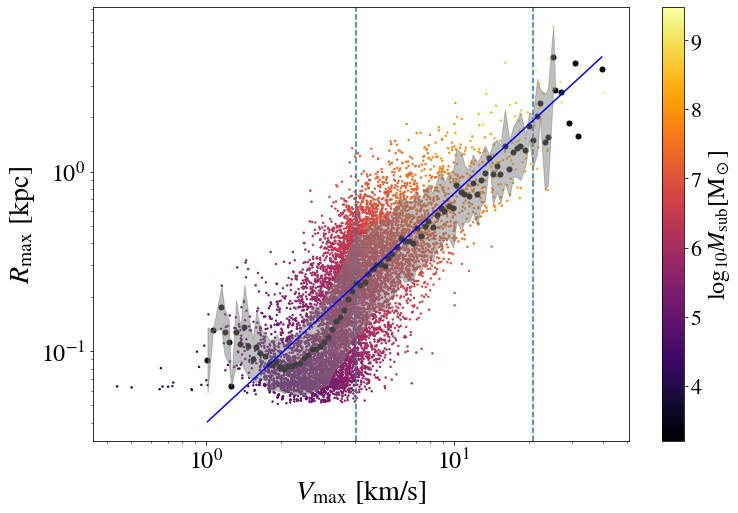}%
\caption{$V_\mathrm{max}-R_\mathrm{max}$ relation for all subhaloes in the original VL-II. The color represents the mass. Black, big dots refer to logarithmic mean values of $R_\mathrm{max}$ binning in $V_\mathrm{max}$; the solid line is a power-law best-fit to the data, with parameters provided in the text of Appendix~\ref{ap:vmaxrmax}. The gray area represents the $1\sigma$ scatter band. The vertical dashed lines indicate the extremes of the range used for the fitting. %
}
\label{fig:vmaxrmax1}
\end{figure}

In this Appendix, we investigate the relation between $V_\mathrm{max}$ and $R_\mathrm{max}$ for subhaloes, which is supposed to behave as a power-law \citep[see e.g.][for the Aquarius simulation]{aquarius}. Such relation is shown in Fig.~\ref{fig:vmaxrmax1} for the original VL-II data. The blue solid line in this Figure represents a power-law fit to the data, whose best-fit parameters are: %
$$ \begin{matrix}
R_\mathrm{max} = 10^c\, V_\mathrm{max}^m \smallskip \\ %
  c = -1.40 \pm 0.06 \\
   m = 1.28 \pm 0.05
 \end{matrix} $$
 For the sake of clarity, we have also plotted mean values (black points) and standard deviations (gray shaded areas). Note that, as expected in $\Lambda$CDM, we find a considerable scatter around mean values. We do not include this scatter in our repopulation work. This is expected to be conservative, since a larger scatter may lead to even higher J-factor values in some cases.

Below $V_\mathrm{max} \sim 4$ km/s %
we obtain lower $R_\mathrm{max}$ values than expected, most likely due to tidal stripping effects. Then, below $V_\mathrm{max} \sim 2$ km/s the resolution becomes too poor, thus the $R_\mathrm{max}$ values artificially increase. %
To stay conservative, %
we will believe/use the results above 4 km/s for our purposes.

%%%%%%%%%%%%%%%%%%%%%%%%%%%%%%%%%%%%%%%%%%%%%%%%%%

% Don't change these lines
\bsp	% typesetting comment
\label{lastpage}
\end{document}